\documentclass[]{dukedissertation2023}

\usepackage[T1]{fontenc}
\usepackage{titlesec}
\usepackage{amsmath, amssymb, amsfonts, amsthm}
\usepackage{bm}
\usepackage{graphicx}
\usepackage[backend=biber,natbib,style=ieee,]{biblatex}
\usepackage[title,titletoc]{appendix}
\usepackage{apptools}
\usepackage{color}
\usepackage{bm}
\usepackage{mathabx} 
\usepackage{multirow}
\usepackage{setspace}

\usepackage{mathpazo} 
\usepackage{indentfirst}
\usepackage{enumitem}
\usepackage{xfrac}
\usepackage{mathtools}
\usepackage{caption}
\usepackage{subcaption}
\usepackage{tabularx}
\usepackage{booktabs}
\usepackage{siunitx}
\usepackage{pdflscape}
\usepackage{listings}
\usepackage{xcolor}
\usepackage{pgfplots}
\usepackage{makecell}
\usepackage[super]{nth}
\usepackage{xr}

\setlist[itemize]{noitemsep,nolistsep}
\setlist[enumerate]{noitemsep,nolistsep}
\usepackage{slashed}
\usepackage{braket}
\usepackage{simpler-wick}
\newcommand{\pv}{{\bf p}}
\newcommand{\Pv}{{\bf P}}
\newcommand{\qv}{{\bf q}}
\newcommand{\lv}{{\bf l}}
\newcommand{\kv}{{\bf k}}
\newcommand{\bv}{{\bf b}}

\newcommand{\zb}{\bar{z}}
\newcommand{\zt}{\tilde{z}}

\newcommand{\as}{{\alpha_s}}
\newcommand{\Otsosing}{\braket{\mathcal{O}^{J/\psi}(^3S_1^{[1]})}}
\newcommand{\Otsooct}{\braket{\mathcal{O}^{J/\psi}(^3S_1^{[8]})}}
\newcommand{\Oosz}{\braket{\mathcal{O}^{J/\psi}(^1S_0^{[8]})}}
\newcommand{\Otpz}{\braket{\mathcal{O}^{J/\psi}(^3P_0^{[8]})}}

\graphicspath{{./Pictures/}}

\author{Reed Hodges}
\title{Studies of $T_{cc}^+$ Decays and Transverse-Momentum-Dependent $J/\psi$ Production Using Effective Field Theory}
\supervisor{Thomas C. Mehen}
\department{Physics} 

\date{2024} 

\copyrighttext{ All rights reserved except the rights granted by the\\
   \href{http://creativecommons.org/licenses/by-nc/3.0/us/}
        {Creative Commons Attribution-Noncommercial Licence}
}

\defensedate{March 25, 2024}
\member{Roxanne Springer}
\member{Mark C. Kruse}
\member{Harold U. Baranger}
\member{Michael A. Troxel}

\usepackage[pdftex, pdfusetitle, plainpages=false, letterpaper, bookmarks, bookmarksnumbered, colorlinks, linkcolor=black, citecolor=black, filecolor=black, urlcolor=black,linktoc=none]{hyperref}
\usepackage[capitalise]{cleveref}
\creflabelformat{equation}{#2#1#3}
\makeatletter

\newcommand\getfontsizeofheading{The current font size is: \fontname\font \f@size pt}
\newcommand\getfontsizeforfontsizeclass[1]{{\string #1 is printed as #1  \fontname\font \f@size pt\par}}
\newcommand\printfontsizeforfontsizeclass[1]{{#1 \string #1 is printed in  \fontname\font \f@size pt\par}}
\makeatother

\AtBeginBibliography{\vspace*{\baselineskip}}
\titleformat{\chapter}
[display] 
{\raggedleft\LARGE\bfseries\usefont{T1}{ppl}{b}{n}\fontsize{18pt}{\baselineskip}} 
{\thechapter} 
	{.5em} 
{\normalbaselines} 

\titleformat{\section}[block]{\fontencoding{T1}\fontfamily{ppl}\fontseries{b}%
  \fontshape{n}\fontsize{14pt}{11}\selectfont}{\thesection}{.5em}{\normalbaselines}
 \titleformat{\subsection}[block]{\fontencoding{T1}\fontfamily{ppl}\fontseries{b}%
  \fontshape{it}\fontsize{12pt}{11}\selectfont}{\thesubsection}{.5em}{\normalbaselines}
 \titleformat{\subsubsection}[block]{\fontencoding{T1}\fontfamily{ppl}\fontseries{b}%
  \fontshape{it}\fontsize{11pt}{11}\selectfont}{\thesubsubsection}{.5em}{\normalbaselines}

\titlespacing\section{0pt}{14pt plus 0pt minus 0pt}{6pt plus 0pt minus 0pt}
\titlespacing\subsection{0pt}{14pt plus 0pt minus 0pt}{6pt plus 0pt minus 0pt}
\titlespacing\subsubsection{0pt}{14pt plus 0pt minus 0pt}{6pt plus 0pt minus 0pt}
\titlespacing\paragraph{0pt}{6pt plus 0pt minus 0pt}{6pt plus 0pt minus 0pt}
\titlespacing\subparagraph{0pt}{6pt plus 0pt minus 0pt}{6pt plus 0pt minus 0pt}

\newcolumntype{L}[1]{>{\raggedright\let\newline\\\arraybackslash\hspace{0pt}}b{#1}}
\newcolumntype{C}[1]{>{\centering\let\newline\\\arraybackslash\hspace{0pt}}b{#1}}
\newcolumntype{R}[1]{>{\raggedleft\let\newline\\\arraybackslash\hspace{0pt}}b{#1}}

\usepackage{subfiles}
\begin{document}

\maketitle

\abstract

We describe the application of effective field theories for quantum chromodynamics (QCD) to two bound states involving heavy quarks: the $T_{cc}^+$ exotic meson and the $J/\psi$.  We study the decay of the $T_{cc}^+$ in an effective theory for hadronic molecules, and find agreement with experiment.  We also use the nonrelativistic QCD (NRQCD) factorization formalism to derive the leading-order transverse momentum dependent fragmentation functions (FFs) for quarks and gluons fragmenting to $J/\psi$.  We then make use of these TMD FFs to compare the $J/\psi$ production mechanisms of light quark fragmentation and photon-gluon fusion, where the conclusions we draw can motivate future experiments at the Electron Ion Collider, shedding light on the inner structure of nucleons and testing ideas from NRQCD factorization.  These results showcase the utility of effective field theories in explaining experiments and testing key concepts in nuclear/particle physics.

\dedication{To my parents.}

\tableofcontents 
\listoftables	
\listoffigures	
\abbreviations


\section*{Abbreviations}

\begin{symbollist}
    \item[BNL] Brookhaven National Laboratory
    \item[CDF] Collider Detector at Fermilab
    \item[CERN] Conseil europ\'een pour la Recherche nucl\'eaire (European Organization for Nuclear Research)
    \item[$\chi$PT] chiral perturbation theory
    \item[DESY] Deutsches Elektronen-Synchrotron (German Electron Synchrotron)
    \item[DIS] deep inelastic scattering
    \item[EFT] effective field theory
    \item[EIC] Electron Ion Collider
    \item[ERL] energy recovery LINAC
    \item[FF] fragmentation function
    \item[HH$\chi$PT] heavy hadron chiral perturbation theory
    \item[IR] infrared
    \item[JLab] Thomas Jefferson National Accelerator Facility (Jefferson Lab)
    \item[LHC] Large Hadron Collider
    \item[LHCb] Large Hadron Collider beauty
    \item[LO] leading-order
    \item[LQF] light quark fragmentation
    \item[LSZ] Lehmann-Symanzik-Zimmermann
    \item[MS] minimal subtraction
    \item[NLO] next-leading-order
    \item[NRQCD] non-relativistic quantum chromodynamics
    \item[PDF] parton distribution function
    \item[PDS] power divergence subtraction
    \item[PGF] photon-gluon fusion
    \item[PHENIX] Pioneering High Energy Nuclear Interaction Experiment
    \item[QCD] quantum chromodynamics
    \item[QED] quantum electrodynamics
    \item[QFT] quantum field theory
    \item[RHIC] Relativistic Heavy Ion Collider
    \item[SIDIS] semi-inclusive deep inelastic scattering
    \item[SLAC] Stanford Linear Accelerator Center
    \item[STAR] Solenoidal Tracker at RHIC
    \item[TMD] transverse momentum dependent
    \item[TMDs] transverse momentum dependent distribution functions
    \item[UV] ultraviolet
    \item[XEFT] effective field theory for the $X(3872)$, a.k.a.~$\chi_{c1}(3872)$
\end{symbollist}

\section*{Notation}

Unless otherwise indicated, this work uses the following notation conventions.

\begin{itemize}
    \item Lorentz four-vectors are indicated with regular typeface; Cartesian three-vectors are indicated in bold face
    \item Lorentz indices are indicated with Greek letters; Cartesian indices are indicated with Latin letters
    \item color indices are indicated with Latin letters; upper case indices enumerate the generators, and lower case correspond are fundamental (quark) indices 
    \item heavy quarks are indicated with an upper case $Q$; light quarks are indicated with a lower case $q$
\end{itemize}

\section*{Units}

\begin{itemize}
    \item $\hbar=c=1$
    \item $g_{\mu\nu} = \text{diag}(1,-1,-1,-1)$
\end{itemize} 

\acknowledgements

First and foremost I would like to thank my parents, without whose love, support, and sacrifices I would not be where I am today.  Thank you as well to my siblings, Will, Bailey, and Chloe, for always being there for me.

Thank you to Kennedy, who has been my partner, confidante, and best friend throughout everything.

My time in grad school would not have been as enjoyable as it was if not for the amazing friends I made along the way, especially Charlie, Drew, Elise, Keith, Matt, and Utsav.  A special thank you goes to Baran for being an exceptional friend and companion as my roommate during a challenging few years.

Thank you to my friends from undergrad, especially Kelvin, Matt, and the staff at Kennedy Hall, for times that I recall fondly every day.  I also appreciate Maxim Durach, my undergraduate research advisor, for his guidance as I started my journey in physics research.

Particular thanks go to my doctoral advisor, Tom Mehen, for his mentorship and advice throughout my studies.  I would also like to thank my other collaborators, especially Sean Fleming, Marston Copeland, Lin Dai, and Abhishek Mohapatra, for making me a better physicist, and Son Nguyen for being my friend and lone classmate in my quantum field theory education.  Lastly, thank you to those who sat on my preliminary exam and dissertation committees: Roxanne Springer, Mark Kruse, Harold Baranger, Dan Scolnic, and Michael Troxel.

%
%
%
\chapter{Introduction}
\label{chap:intro}

\begin{figure*}[t]
\centering
\includegraphics[width=\textwidth]{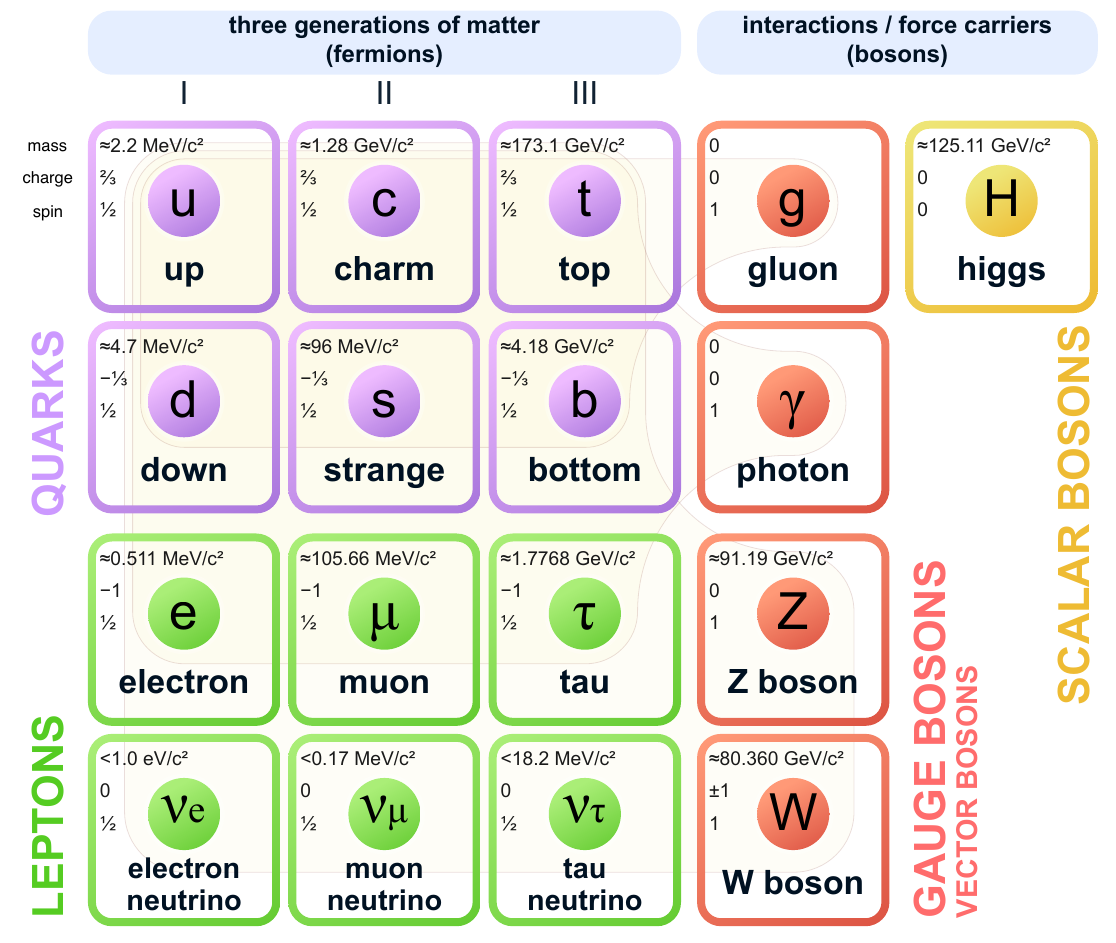}
\caption[Elementary particles of the Standard Model.]{Elementary particles of the Standard Model. Figure from Ref.~\cite{MissMJ}.}
\label{fig: SM particles}
\end{figure*}

All of the known elementary particles in the universe, as well as three of the four fundamental forces describing their interactions, are described by the Standard Model of particle physics.  The matter particles are all spin-$1/2$ fermions, and are divided into two types -- quarks and leptons -- of three generations each.  The interactions are mediated by ``force carriers'', which are bosons of either spin-$0$ or spin-$1$.  Both quarks and leptons participate with the electroweak interaction -- the unified description of electromagnetism and the weak force -- which is mediated by the massive vector bosons $Z$ and $W^\pm$ and the massless photon $\gamma$.  Quarks are the only matter particles that participate in the strong interaction, mediated via the exchange of gluons, which are massless and spin-$1$.  The massive particles get their mass through interactions with the Higgs boson, a massive spin-$0$ particle.  All of these particles are summarized in Fig.~\ref{fig: SM particles}.

The equations of motion of a physical system are generally obtained from the principle of least action $\frac{\delta S}{\delta \phi_i} = 0$.  The action $S[\phi_i]=\int d{\bf s} \, \mathcal{L}\big(\phi_i,\frac{\partial\phi}{\partial s_j},s_j\big)$ is a functional dependent on the degrees of freedom $\phi_i$ of the system, where ${\bf s}$ represent some generalized coordinate(s).  The function $\mathcal{L}$ is the Lagrangian density, or simply the Lagrangian,\footnote{This is a mild abuse of language, as the Lagrangian is technically a separate object equal to the Lagrangian density integrated over the spatial coordinates.} and it contains all of the physics of the system.  The Standard Model is a quantum field theory (QFT), which means the degrees of freedom in its Lagrangian are quantum fields: quantum-operator-valued objects that are a function of space and time, and the generalized coordinates are the coordinates of spacetime.  The fundamental particles are excitations of these quantum fields.  More precisely speaking, the Standard Model is a gauge QFT, which means the Lagrangian is invariant under certain local gauge transformations.  This restriction leads the matter fields, which are described by Dirac spinor fields, to be coupled to gauge fields which are associated with the vector bosons.  The symmetry group of these interactions is $SU(3)\times SU(2)_L\times U(1)$.  The $SU(3)$ portion describes the strong interaction, where the fundamental particles carry a color charge.  The $SU(2)_L\times U(1)$ portion is the electroweak interaction, and this symmetry group is spontaneously broken to a different $U(1)$ symmetry group by the Higgs mechanism, where the coupling to a scalar field for the Higgs boson causes the $Z$ and $W^\pm$ to have nonzero mass.  Electric charge is a quantum number derived from the electroweak sector of the Lagrangian; quarks have fractional electric charge while the leptons have integer electric charge.

The sector of the Standard Model describing the strong interaction is called quantum chromodynamics (QCD).  It is the theory that describes nuclear matter, and therefore is something that underlies all the atoms and molecules that make up the things we encounter in everyday life.  Despite its ubiquitous nature, QCD has many interesting properties and quirks that make it difficult to perform predictive calculations, and physicists are constantly learning new things about systems described by QCD.  This dissertation deals exclusively with such systems. 

Equally important to theoretical descriptions of elementary particles are the experimental endeavors which measure them.  Both aspects can influence and help the other: theorists can predict new phenomena and inspire experimentalists to look for them, and experimentalists can make discoveries that require theorists to revise their explanations or come up with new ones.  Experiments that probe the physics at small distances require enormous amounts of energy.  This is achieved at particle accelerators, which produce beams of particles and accelerate them to high speeds, before smashing them together and detecting the remnants.  Perhaps the most famous particle accelerator is the Large Hadron Collider (LHC), a machine over $26$ kilometers in circumference straddling the border between Switzerland and France, which can achieve collision energies of $13.6$ TeV.  Researchers at the LHC have discovered, among many other things, the $T_{cc}^+$ exotic meson, whose theoretical description is a major topic of this dissertation.  Physicists often aim to build new accelerators that they hope will shed light on aspects of particle physics that are poorly understood.  One such future accelerator is the Electron Ion Collider (EIC), which will be built at Brookhaven National Laboratory over roughly the next decade.  The EIC will hopefully teach us more about the inner structure of protons and neutrons.  This prospect is the primary motivation behind some of the other research projects discussed in this work.  

The outline of this dissertation is as follows.  The current introductory chapter provides the basic information about quantum chromodynamics, and discusses two frameworks which aid in calculation: effective field theory and the parton model.  Chapter \ref{chap:Tcc} discusses an effective field theory which is used to calculate the decay of the exotic meson $T_{cc}^+$.  Chapter \ref{chap:FFs} outlines the derivation of spin- and transverse-momentum-dependent fragmentation functions for $J/\psi$ production, and in Chapter \ref{chap:JPsiProduction} these fragmentation functions are utilized in comparing different $J/\psi$ production mechanisms, which can inform future experiments at the Electron Ion Collider.  Concluding remarks are given in Chap.~\ref{chap:conclusion}.  In places in the Chapters where a particular topic (e.g., a lengthy derivation) would be too distracting from the main discussion, the extra information is provided in an Appendix.  

The research projects presented in this dissertation were initially published in four papers \cite{Fleming:2021wmk,Dai:2023mxm,Copeland:2023wbu,Copeland:2023qed}; the author contributed to the entire analysis and writing of each paper.  Their relevance to the field can be understood in the context of the interplay between theory and experiment.  Using effective field theories can allow us to calculate concrete theoretical predictions for the results of collider experiments.  In doing so, the validity of the theories are tested, and the underlying systems are better understood.

\section{Quantum chromodynamics}

The physics of quarks and gluons is described by a non-Abelian gauge QFT with symmetry group $SU(3)$ called quantum chromodynamics (QCD), whose Lagrangian is:
\begin{equation}
    \mathcal{L}_{\rm QCD} = \sum_q \bar{\psi}_{q,a}(i\gamma^\mu \partial_\mu \delta_{ab}-g_s \gamma^\mu t_{ab}^C A_\mu^C - m_q \delta_{ab})\psi_{q,b} - \frac{1}{4}F_{\mu\nu}^AF^{A\mu\nu} \; .
\label{eq: QCD Lag}
\end{equation}
The $\psi$ fields are Dirac spinors which create and annihilate quarks of flavor $q$ and mass $m_q$; there are six flavors of quarks divided into three generations. The $A_\mu$ fields create and annihilate gluons.  Their kinetic term is written in terms of the field strength tensor $F_{\mu\nu}^A = \partial_\mu A_\nu^A-\partial_\nu A_\mu^A - g_s f_{ABC}A^B_\mu A^C_\nu$. The coupling constant between the quarks and gluons is $g_s$, and the Dirac matrices are denoted by $\gamma^\mu$.

Quarks and gluons both carry what is called color charge, which is analogous to the electric charge of electromagnetism.  The information about color is contained in the color indices on the fields. The quarks transform in the fundamental representation of $SU(3)$, and so their color indices, indicated by $a$ and $b$ in Eq.~(\ref{eq: QCD Lag}), run from 1 to 3.\footnote{The fundamental indices are often suppressed in the notation.}  The gluons transform in the adjoint representation of $SU(3)$, and so their indices, represented by $A$, $B$, and $C$ in Eq.~(\ref{eq: QCD Lag}), run from 1 to 8.  The eight $3\times 3$ matrices represented by $t_{ab}^C$ in the Lagrangian are the generators of $SU(3)$ color.  The commutation relation for the generators is $[t^A,t^B]=if_{ABC}t^C$.  

\subsection{Feynman diagrams}

\begin{figure*}[t]
\centering
\begin{minipage}{0.5\textwidth}
\centering
\subfloat[]{\includegraphics[scale=1.7]{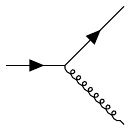}\label{fig: simple tree}}
\end{minipage}%
\begin{minipage}{0.5\textwidth}
\centering
\subfloat[]{\includegraphics[scale=1.7]{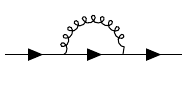}\label{fig: simple loop}}
\end{minipage}
\caption[A simple tree-level and one-loop Feynman diagram.]{A simple (a) tree-level and (b) one-loop Feynman diagram.}
\label{fig: simple feyn diag}
\end{figure*}

Calculations in a quantum field theory usually involve computing a matrix element of an operator.  If one is interested in scattering amplitudes, the LSZ reduction formula \cite{Lehmann:1954rq} relates the $S$-matrix elements to correlation functions, which are vacuum expectation values of a product of field operators.  More generally, we are interested in a matrix element with arbitrary in and out states:
\begin{equation}
    \bra{\text{out}}T\,\mathcal{O}\exp\bigg[-i\int dt\; H_{\rm int}\bigg]\ket{\text{in}} \; ,
\label{eq: matrix element of operator}
\end{equation}
where $T$ indicates time ordering.  The fields in $\mathcal{O}$ are in the interaction picture, and $H_{\rm int}$ is the interaction Hamiltonian.  Wick's theorem \cite{Wick:1950ee} then tells us that the time-ordered product of fields is equal to the normal-ordered sum of all possible contractions of the fields.  In the case of a vacuum expectation value of the time-ordered product of two fields, the statement of the theorem is
\begin{equation}
    T \, \phi(x) \phi(y) = N \, \phi(x) \phi(y) + N \,
    \wick{
        \c1 \phi(x_1) \c1 \phi(x_2)
    } \; .
\end{equation}
The contraction indicates that we replace the fields with the Feynman propagator appropriate for the type of field, and it has a diagrammatic interpretation as a particle propagating from $y$ to $x$.  Matrix elements like in Eq.~(\ref{eq: matrix element of operator}), then, have pictorial representations in the form of Feynman diagrams, examples of which are shown in Fig.~\ref{fig: simple feyn diag}.  Each has an associated mathematical value with it, and scales as a particular power of the coupling, which for QCD is $\alpha_s = g_s^2/4\pi$.  If the coupling is small, then we can utilize perturbation theory and compute the observable to the desired accuracy, where each further order in $\alpha_s$ brings with it more Feynman diagrams.

\subsection{Renormalization}

Beginning at $\mathcal{O}(\alpha_s)$, loop diagrams can contribute, e.g., the quark self-energy correction in Fig.~\ref{fig: simple loop}. In these instances, to evaluate the Feynman diagram one must integrate over the momentum flowing through the loop.  These integrals are often divergent in $d=4$ dimensions.  These divergences can be ultraviolet (UV) divergences, arising when the loop momentum is large, or infrared (IR) divergences, arising when it is small.  A common way to regulate the divergences is to use dimensional regularization, where the loop integral is evaluated in an arbitrary number of spatial dimensions.  One of the simpler integrals that can arise has the following answer in dimensional regularization:
\begin{equation}
    \bigg(\frac{\mu^2 e^{\gamma_E}}{4\pi}\bigg)^{2-\frac{d}{2}} \int \frac{d^dl}{(2\pi)^d}\frac{1}{(l^2-\Delta)^n} = \bigg(\frac{\mu^2 e^{\gamma_E}}{4\pi}\bigg)^{2-\frac{d}{2}} \frac{(-1)^n i}{(4\pi)^{d/2}} \frac{\Gamma(n-\frac{d}{2})}{\Gamma(n)}\bigg(\frac{1}{\Delta}\bigg)^{n-\frac{d}{2}} \; .
\end{equation}
It is common to write the number of spacetime dimensions as $d=4-2\epsilon$, in which case the divergences appear as poles that go as inverse powers of $\epsilon$.  Here $\mu$ is the renormalization scale associated with the modified minimal subtraction ($\overline{\rm MS}$) scheme.

The final answer for a physical observable must not be divergent, so when a Feynman diagram contains a divergence, something must be done to remedy the situation.  The procedure by which these ``infinities'' are absorbed into the theory is called renormalization.  To simplify the discussion, here we turn to the case of quantum electrodynamics (QED), with one fermion and gauge boson.  We contend that the fields and couplings in the Lagrangian are bare quantities, which we now denote with a superscript $(0)$:
\begin{equation}
    \mathcal{L} = i\bar{\psi}^{(0)}\gamma^\mu(\partial_\mu - i g_{\rm em}^{(0)}A_\mu^{(0)})\psi^{(0)} - m^{(0)}\bar{\psi}^{(0)}\psi^{(0)} - \frac{1}{4}F_{\mu\nu}^{(0)}F^{(0)\mu\nu} \; .
\end{equation}
We then introduce a ``renormalized'' version of each of these quantities that is related to the bare by an overall constant: \cite{Manohar:2000dt}
\begin{equation}
    \begin{aligned}
        A_\mu = & \; \frac{1}{\sqrt{Z_A}} A_\mu^{(0)} \; , \qquad & \psi & = \frac{1}{\sqrt{Z_\psi}}\psi^{(0)} \; , \\
        g_{\rm em} = & \; \frac{1}{Z_g}\mu^{-\epsilon/2} g_{\rm em}^{(0)} \; , \qquad & m & = \frac{1}{Z_m}m^{(0)} \; .
    \end{aligned}
\end{equation}
The Lagrangian can then be written to look like the original Lagrangian but in terms of the renormalized fields, plus a set of counterterms.
\begin{equation}
    \mathcal{L} = i\bar{\psi}\gamma^\mu(\partial_\mu - i g_{\rm em}A_\mu)\psi - m \bar{\psi} \psi - \frac{1}{4}F_{\mu\nu}F^{\mu\nu} + \text{counterterms} \; .
\end{equation}
Then, order-by-order in the coupling, whenever there is a divergence arising in a loop diagram involving a renormalized quantity, there is a corresponding diagram associated with a counterterm, and the value of the renormalization constant(s) can be fixed so that the divergence cancels in the final answer.  In this way, the bare quantities are indeed divergent, but infinite values for the renormalization constants are chosen to compensate, yielding finite renormalized quantities, which are the ones that we actually measure.  For example, at $\mathcal{O}(g_{\rm em}^2)$, there is a one-loop diagram for the fermion self energy, like in Fig.~\ref{fig: simple loop}, but with a photon in place of the gluon.  It evaluates to the divergent expression,
\begin{equation}
    \frac{i}{32\pi^2\epsilon}(4g_{\rm em}^2)\bigg(-2m+\frac{1}{2}\gamma \cdot p \bigg) \; ,
\end{equation}
The counterterms contribute a tree-level diagram to this process,
\begin{equation}
    i(Z_\psi - 1)\gamma \cdot p - i (Z_mZ_\psi-1)m \; .
\end{equation}
Adding the two contributions and requiring that the answer be finite fixes two of the renormalization constants:
\begin{equation}
    \begin{aligned}
        Z_\psi = & \; 1 - \frac{g_{\rm em}^2}{16\pi^2 \epsilon} \; , \\
        Z_m = & \; 1 - 3 \frac{g_{\rm em}^2}{16\pi^2 \epsilon} \; .
    \end{aligned}
\end{equation}

\begin{figure*}[t]
\centering
\includegraphics[width=0.7\textwidth]{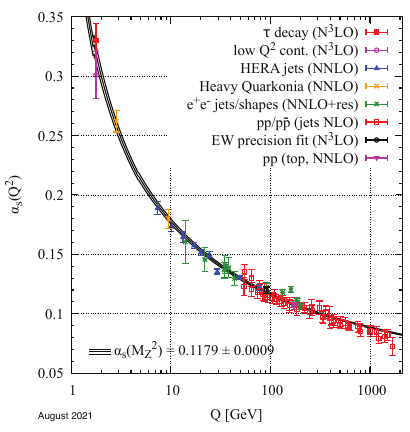}
\caption[Measurements of $\alpha_s$ showing the running coupling.]{Measurements of $\alpha_s$ showing the running coupling.  Figure from Ref.~\cite{ParticleDataGroup:2022pth}.}
\label{fig: alpha s PDG}
\end{figure*}

The bare quantities, which are the original fields and parameters of the theory, must be independent of the subtraction scale $\mu$, since that is an arbitrary parameter in our regularization scheme.  The renormalized quantities, however, do scale with $\mu$.  Returning from QED back to QCD, the evolution of the coupling $g_s$ with $\mu$ is described by the beta function of QCD:
\begin{equation}
    \beta(g_s) \equiv \mu \frac{dg_s}{d\mu} = -\frac{g_s^3}{16\pi^2}\bigg(11-\frac{2}{3}N_f \bigg) + \mathcal{O}(g_s^5) \; .
\label{eq: QCD beta function}
\end{equation}
Here $N_f$ is the number of quark flavors, which is equal to $6$ in our current understanding of the Standard Model.  This means that the beta function is negative, and therefore the coupling gets weaker as the energy scale increases (Fig.~\ref{fig: alpha s PDG}).  This phenomenon is called asymptotic freedom; it was predicted by Gross, Wilczek, and Politzer \cite{Gross:1973id,Politzer:1973fx}, and is one of the key characteristics of QCD.  

We can solve Eq.~(\ref{eq: QCD beta function}) to obtain the value of $\alpha_s=g_s^2/4\pi$ at one scale as a function of its value at another:
\begin{equation}
    \frac{1}{\alpha_s(\mu_2)} = \frac{1}{\alpha_s(\mu_1)} + \beta_0 \ln\bigg(\frac{\mu_2^2}{\mu_1^2}\bigg)  \; ,
\end{equation}
for $\beta_0 = (33-2N_f)/12\pi$.  Defining a scale $\Lambda_{\rm QCD} \equiv \mu e^{-1/[2\beta_0 \alpha_s(\mu)]}$ and taking the logarithm of both sides, we see:
\begin{equation}
    \frac{1}{\alpha_s(\mu)} = \beta_0 \ln \bigg( \frac{\mu^2}{\Lambda_{\rm QCD}^2} \bigg) \, .
\label{eq: alpha s lambda QCD}
\end{equation}
Equation (\ref{eq: alpha s lambda QCD}) implies that $\Lambda_{\rm QCD}$ is the scale at which the coupling constant diverges and thus perturbation theory breaks down.  The experimental value of $\Lambda_{\rm QCD}$, derived from measurements of $\alpha_s$, is about $200$ MeV.

\subsection{The quark model and bound states}

So far the discussion has been about the fundamental particles of QCD -- quarks and gluons -- and their treatment as quantum fields.  However, these particles are not observed individually.  Nature only allows physical states which are color singlets, a fact called color confinement; therefore all the states we observe are bound states of multiple quarks and gluons.  These bound states are called hadrons, and are further divided into mesons and baryons, which are bosons and fermions, respectively.  

All of the hadrons can be neatly organized using the quark model, which utilizes the fact that there are six flavors of quarks along with their antiparticles, and delineates combinations of them by appealing to their quantum numbers.  The multitude of quantum numbers for each flavor are summarized in Table \ref{tab: quantum numbers}.  The electric charge is related to the others by
\begin{equation}
    Q = I_z + \frac{\mathcal{B}+S+C+B+T}{2} \; ,
\end{equation}
for isospin $z$ component $I_z$, baryon number $\mathcal{B}$ ($1$ for baryons and $0$ for mesons), strangeness $S$, charm $C$, bottomness $B$, and topness $T$.  Another quantum number of note is the hypercharge $Y=\mathcal{B} + S - (C-B+T)/3$.

\begin{table}[]
\caption{Quantum numbers of the quarks.}
\centering
\label{tab: quantum numbers}
\begin{tabular}{|c|c|c|c|c|c|c|c|}
\hline
 \textbf{Symbol} & \textbf{Name}  & $d$ & $u$ & $s$ & $c$ & $b$ & $t$ \\ \hline \hline
 $Q$& electric charge  & $-1/3$  & $+2/3$ & $-1/3$ & $+2/3$  & $-1/3$  & $+2/3$  \\ \hline
 $I$& isospin & $1/2$ & $1/2$ & $0$ & $0$ & $0$ & $0$ \\ \hline
 $I_z$& $z$ component of isospin & $-1/2$ &$+1/2$  & $0$ & $0$  & $0$ & $0$ \\ \hline
 $S$& strangeness  & $0$ & $0$  & $-1$ & $0$ & $0$ & $0$ \\ \hline
 $C$& charm & $0$ & $0$ & $0$ & $+1$ & $0$ & $0$ \\ \hline
 $B$& bottomness & $0$ & $0$ & $0$ & $0$ & $-1$ & $0$ \\ \hline
 $T$& topness & $0$ & $0$ & $0$ & $0$ & $0$ & $+1$ \\ \hline
\end{tabular}
\end{table}

\begin{figure*}[t]
\centering
\includegraphics[width=0.5\textwidth]{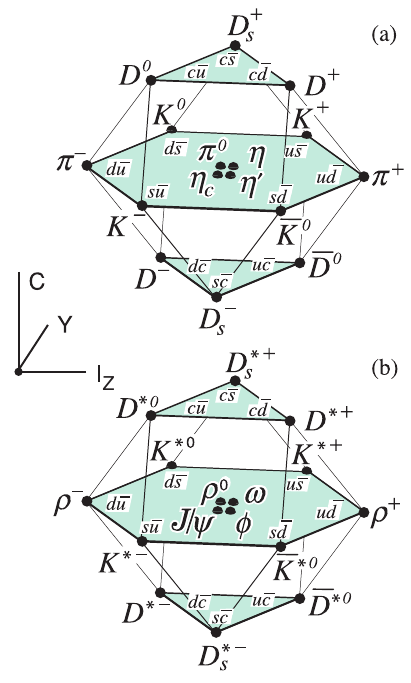}
\caption[16-plets of pseudoscalar and vector mesons.]{Diagram showing the 16-plets of pseudoscalar (a) and vector (b) mesons of $u$, $d$, $s$, and $c$ quarks, organized according to isospin $I_z$, charm $C$, and hypercharge $Y$.  Figure from  Ref.~\cite{ParticleDataGroup:2022pth}.}
\label{fig: meson 16-plet}
\end{figure*}

The quark model is independent of the fact that the quarks can be described by quantum fields, and in fact predates QCD.  In order for the hadron to be colorless, the simplest mesons consist of a quark/antiquark pair $q\bar{q}$, and the simplest baryons consist of three quarks $qqq$. As Chaps.~\ref{chap:FFs} and \ref{chap:JPsiProduction} deal primarily with the $J/\psi$, which is a bound state of $c\bar{c}$, $q\bar{q}$ mesons are the most relevant hadron for this dissertation.  Including up to the charm quark, they can be organized in 16-plets according to their hypercharge, charm, and $z$ component of isospin (Fig.~\ref{fig: meson 16-plet}).  Anything that does not fit into the simplest $q\bar{q}$ or $qqq$ categories of hadrons is called an exotic hadron \cite{ParticleDataGroup:2022pth}.  The $T_{cc}^+$, which is the primary focus of Chap.~\ref{chap:Tcc}, is an exotic meson which is a possible tetraquark state $cc\bar{u}\bar{d}$.

\section{Effective field theory}

Calculations in a QFT often involve physics that is irrelevant to the problem at hand.  It can be of interest, then, to use an approximation to the full theory that only includes the necessary ingredients to solve a particular problem.  This is the idea behind effective field theory (EFT).  The key elements to constructing an EFT Lagrangian are the degrees of freedom and the symmetries.  The degrees of freedom are all of the quantum fields that are applicable to the problem, or, conversely, none of the fields that will have a negligible affect on the dynamics of the system.  The desired symmetries determine the variety of operators in the Lagrangian.  A common maxim is that to write down an EFT, one writes down all the operators involving the relevant degrees of freedom that are consistent with the symmetries of the system.

Another crucial idea behind EFTs is that of power counting.  All EFTs have a parameter (or set of parameters), called the power counting parameter, that is small in the considered energy regime.  The EFT can be thought of as a series expansion in this parameter, and the desired accuracy can determine the number of terms in the Lagrangian, the number of Feynman diagrams calculated, etc.  As an example, if an EFT takes a full theory and removes the physics at the high-energy scale $M$, and keeps the physics at the low-energy scale $m$, the power counting parameter is the ratio $m/M$.  

Broadly speaking, there are two types of EFTs: top-down and bottom-up theories.  In top-down EFTs, the full theory is known, but heavy particles are removed from the theory to simplify the calculations.  This removal is known as ``integrating out'' the heavy particles.  This results in an EFT with different operators than the full theory.  The couplings (also known as Wilson coefficients) associated with the new interaction terms are fixed through a process called matching, where a particular scattering process is studied in both the full theory and the EFT, and the results are forced to agree in the low-energy, or infrared (IR), region where the theories are designed to coincide.  The EFT couplings, then, contain all of the information regarding the high-energy, or ultraviolet (UV), physics.  Non-relativistic QCD, which is an EFT we use in Chaps.~\ref{chap:FFs} and \ref{chap:JPsiProduction}, is a top-down EFT.  

In bottom-up EFTs, the full theory is unknown, or else unsuitable to match onto the low-energy theory, and the effective theory is constructed by writing down all possible terms with the desired degrees of freedom that obey the required symmetries.  Here, the couplings are not matched to any other theory, but rather fixed to experimental or numerical results.  Chiral perturbation theory, which will underly the theories discussed in Chap.~\ref{chap:Tcc}, is a bottom-up EFT.

\subsection{Example: Fermi weak theory}

\begin{figure*}[t]
\centering
\begin{minipage}{0.5\textwidth}
\centering
\subfloat[]{\includegraphics[scale=1.2]{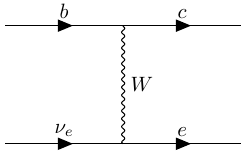}}
\end{minipage}%
\begin{minipage}{0.5\textwidth}
\centering
\subfloat[]{\includegraphics[scale=1.2]{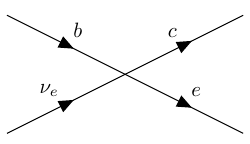}}
\end{minipage}
\caption[Bottom quark decay in the full theory and the effective theory.]{Bottom quark decay in (a) the full theory (the Standard Model) and (b) the effective theory.}
\label{fig: fermi weak}
\end{figure*}

As an example of a top-down EFT, consider an EFT for low-energy weak interactions, called Fermi weak theory, which is a frequent example in introductory material on the subject \cite{Neubert:2005mu,Stewart:2014ln,Manohar:2018aog}.  Here the full theory is the Standard Model, whose $W$ boson interaction terms with quarks and leptons are given by a coupling to a weak current:
\begin{equation}
    j_W^\mu = V_{ij}(\bar{u}_i\gamma^\mu P_L d_j) + (\bar{\nu}_\ell \gamma^\mu P_L \ell) \; ,
\end{equation}
where $u_i\in {u,c,t}$ are the $2/3$ charge quarks, $d_j=d,s,b$ are the $-1/3$ charge quarks, and $V_{ij}$ is the CKM matrix \cite{Cabibbo:1963yz,Kobayashi:1973fv}, which parametrizes the strength of the flavor-changing weak interaction.\footnote{When neglecting neutrino masses, there is no mixing matrix for the leptons.} The amplitude for the decay of a bottom quark to a charm quark through the exchange of a $W$ boson with an electron neutrino, $b + \nu_e \rightarrow c + e^-$, is given by:
\begin{equation}
    \mathcal{A} = \bigg( \frac{-ig}{\sqrt{2}}\bigg)^2 V_{cb}(\bar{c}\gamma^\mu P_L b) (\bar{\ell}\gamma^\nu P_L \nu_\ell)\bigg(\frac{-ig_{\mu\nu}}{p^2-M_W^2}\bigg) \; .
\end{equation}
Fermi weak theory applies at momenta much lower than the $W$ boson mass, $p\ll M_W$.  In this regime, the $W$ boson propagator can be expanded in $p^2/M_W^2$.
\begin{equation}
    \frac{1}{p^2-M_W^2} = -\frac{1}{M_W^2}\sum_{n=0}^\infty \frac{p^{2n}}{M_W^{2n}} \; .
\label{eq: prop expansion}
\end{equation}
Keeping only the $n=0$ constant term for the propagator gives an amplitude that is the same as would be yielded by the effective Lagrangian,
\begin{equation}
    \mathcal{L} = - \frac{g^2}{2M_W^2}V_{cb}(\bar{c}\gamma^\mu P_L b) (\bar{\ell}\gamma^\nu P_L \nu_\ell) + \mathcal{O}\bigg(\frac{1}{M_W^4}\bigg) \; .
\end{equation}
Now the effect of the $W$ boson exchange is approximated by a four-fermion contact operator, as in Fig.~\ref{fig: fermi weak}; the $W$ boson has been integrated out.  The other light particles in the theory have been kept, and the power counting parameters are considered to be the ratios of the light particle masses to that of the $W$, e.g., $m_c/M_W$.  

This was a simple example of a matching calculation: we could have started by writing down the operator $(\bar{c}\gamma^\mu P_L b) (\bar{\ell}\gamma^\nu P_L \nu_\ell)$ with an arbitrary Wilson coefficient, as that operator contains all the degrees of freedom we need for the problem. By expanding the full theory propagator in the limit we are interested in, as in Eq.~(\ref{eq: prop expansion}), we can see that the Wilson coefficient must be $- \frac{g^2}{2M_W^2}V_{cb}$ for the EFT to agree with the results of the full theory in the $p\ll M_W$ regime.

\section{The parton model}

A key goal of nuclear/particle physics is to understand the inner structure of nucleons.  This was pioneered experimentally at the Stanford Linear Accelerator Center (SLAC) with a process called deep inelastic scattering (DIS), $e + p \rightarrow e^\prime + X$ \cite{Bloom:1969kc,Breidenbach:1969kd}.  Here, an electron $e$ scatters off of a proton $p$, causing it to break apart and result in a final state $X$.  The DIS experiments showed that the interactions were as if the electron were scattering off of point-like particles in simple quantum electrodynamics processes \cite{Peskin:1995ev}.  Moreover, rather than detecting a scattered proton in the final state, $X$ included an abundance of hadrons.  This suggests that the proton is a composite particle, made up of smaller constituents collectively known as partons, an idea originally proposed by Feynman \cite{Feynman:1969wa}.  The partons are now identified with quarks and gluons.  Nucleons, then, are composite hadrons of quarks and gluons interacting via QCD.  The quarks are asymptotically free at high energies, which means they almost free particles inside the nucleons.

\begin{figure*}[t]
\centering
\includegraphics[width=\textwidth]{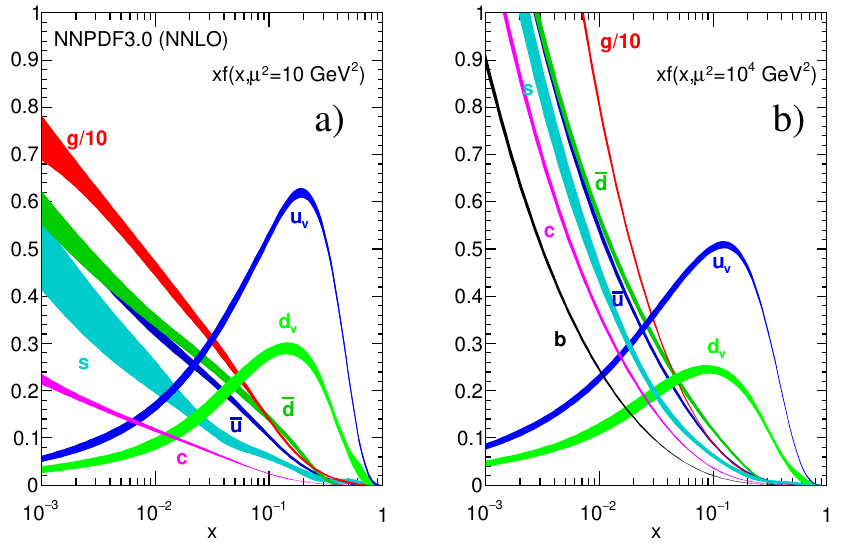}
\caption[Unpolarized parton distribution functions for various parton types, at two different energy scales.]{Unpolarized parton distribution functions for various parton types, at two different energy scales, obtained by the NNPDF collaboration \cite{NNPDF:2014otw}.  Figure from  Ref.~\cite{ParticleDataGroup:2020ssz}.}
\label{fig: PDF plots}
\end{figure*}

If the proton is a collection of quasi-free partons, one might suppose that to calculate some differential cross section $d\sigma_{ep\rightarrow e^\prime X}$ for DIS, we can instead calculate the differential cross sections $d\hat{\sigma}$ for the partonic processes and sum over them, weighted according to some probability distributions $f$ for the partons. 
\begin{equation}
    \begin{aligned}
        \;d\sigma_{ep\rightarrow e^\prime X} \approx \sum_i \int d\xi \; f_{i/p}(\xi) \; d\hat{\sigma}_{ei \rightarrow e^\prime X}  \; .
    \end{aligned}
    \label{eq: DISfact 1}
\end{equation}
This is a statement of the QCD factorization theorem.  The $f$ are known as parton distribution functions (PDFs), and $\xi$ is the fraction of the proton's momentum carried by the parton.  Equation (\ref{eq: DISfact 1}) takes advantage of the separation of scales in the problem.  The partonic processes involve the exchange of a highly virtual photon with $Q^2 \gg \Lambda_{\rm QCD}^2$, where the strong coupling is small and thus the $\hat{\sigma}$ can be calculated in perturbation theory.  All of the nonperturbative physics is contained in the PDFs, which in principle are universal can be extracted from experiment.  See the PDFs for unpolarized partons plotted as a function of the momentum fraction in Fig.~\ref{fig: PDF plots}.  Because it allows us to write down cross sections for otherwise nonperturbative processes, factorization is a powerful tool in understanding the physics of nucleons, and for DIS it has been rigorously proven \cite{Collins:1989gx}.  It will be discussed more explicitly in Sec.~\ref{sec: sidis and factorization} when we apply it to $J/\psi$ production.

\chapter{Decays of the exotic meson $T_{cc}^+$}
\label{chap:Tcc}

\begin{figure*}[t]
\centering
\includegraphics[width=\textwidth]{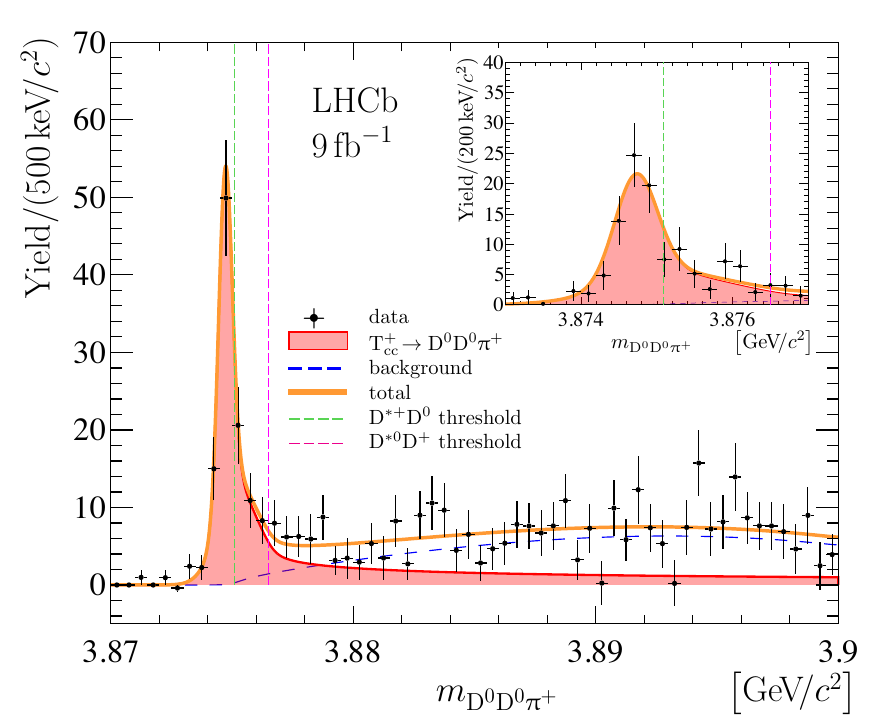}
\caption[Distribution of $D^0D^0\pi^+$ mass, with a resonance corresponding to the $T_{cc}^+$ state.]{Distribution of $D^0D^0\pi^+$ mass, with a resonance corresponding to the $T_{cc}^+$ state.  Note the closeness to the two $D^*D$ thresholds.  Figure from Ref.~\cite{LHCb:2021auc}.}
\label{fig: Tcc resonance}
\end{figure*}

The\footnote{The work presented in this chapter was initially published in Refs.~\cite{Fleming:2021wmk,Dai:2023mxm}. The contributions of each author are listed below. \begin{itemize}
    \item R.~Hodges: analysis and writing for both papers
    \item L.~Dai: checking calculations and editing manuscript (Ref.~\cite{Dai:2023mxm})
    \item S.~Fleming: checking calculations and editing manuscripts for both papers
    \item T.~Mehen: analysis and writing (Ref.~\cite{Fleming:2021wmk}), checking calculations and editing manuscript (Ref.~\cite{Dai:2023mxm})
\end{itemize}} LHCb collaboration announced the discovery of an exotic state in the $D^0D^0\pi^+$ mass spectrum in July 2021 \cite{Muheim,Polyakov,An,LHCb:2021vvq,LHCb:2021auc}.  It was observed as a narrow resonance close to both the $D^{*0}D^+$ and $D^{*+}D^0$ thresholds (Fig.~\ref{fig: Tcc resonance}), consistent with a tetraquark with quark content $cc\bar{u}\bar{d}$ and quantum numbers $J^P=1^+$.  The resonance is named $T_{cc}^+$.  This discovery was similar to that of the $\chi_{c1}(3872)$ \cite{Belle:2003nnu,CDF:2003cab,D0:2004zmu,BaBar:2004oro}, originally known as the $X(3872)$, a possible tetraquark state with quark content $c\bar{c}u\bar{u}$ and quantum numbers $J^{PC}=1^{++}$.  Both resonances lie extremely close to $D^*D$ thresholds and can be regarded as hadronic molecules.  

The decays of the $\chi_{c1}(3872)$ have been studied with an EFT called XEFT (``effective field theory for the X''), initially developed in Ref.~\cite{Fleming:2007rp}.  The similarities of that state with the $T_{cc}^+$, as well as the fact that there are more experimental observables with which to compare to theory, motivated a generalization of XEFT to apply to the $T_{cc}^+$.  In this chapter we discuss this EFT for the $T_{cc}^+$, and its predictions for the total decay width and differential decay width distributions, which are in excellent agreement with the LHCb results and were first presented in Refs.~\cite{Fleming:2021wmk,Dai:2023mxm}.  We begin by reviewing the theories that underpin XEFT.

\section{Background for XEFT}

XEFT, as a theory of the interactions between heavy mesons and pions, is built from heavy hadron chiral perturbation theory \cite{Wise:1992hn,Burdman:1992gh,Yan:1992gz}. This, in turn, is an extension of chiral perturbation theory. 
  
\subsection{Chiral perturbation theory}
\label{sec: ChPT}

The QCD Lagrangian when only considering light quarks is:
\begin{equation}
    \mathcal{L} = -\frac{1}{4}(F^A_{\mu\nu})^2 + \bar{q}(i\slashed{D}-m_q)q \, ,
\end{equation}
where $q=\begin{pmatrix} u & d & s \end{pmatrix}^T$ is a vector of light quark fields, and $m_q = {\rm diag} (m_u,m_d,m_s)$ is the light quark mass matrix.  Each of these masses is small compared to $\Lambda_{\rm QCD}$, so it is interesting to consider the Lagrangian in the chiral limit $m_q\rightarrow 0$.
\begin{equation}
    \mathcal{L}_q \rightarrow \bar{q}i\slashed{D}q = \bar{q}_Li\slashed{D}q_L + \bar{q}_Ri\slashed{D}q_R \, ,
\end{equation}
where $q_L = \frac{1}{2}(1-\gamma_5)q$ and $q_R=\frac{1}{2}(1+\gamma_5)q$. In this limit, there is a chiral symmetry $SU(3)_L \times SU(3)_R$ where the left- and right-handed fields transform independently: $q_L\rightarrow Lq_L$ and $q_R\rightarrow Rq_R$, with $L\in SU(3)_L$ and $R\in SU(3)_R$ \cite{Manohar:2000dt}.  

Let us investigate the vacuum expectation value (vev) of the light quark bilinears in the chiral limit, with the flavor indices $i$ and $j$ explicit: $\bra{0}\bar{q}^iq^j\ket{0} = \bra{0}(\bar{q}_R^iq_L^j + \bar{q}_L^iq_R^j)\ket{0}$. For simplicity consider only one of these terms and  define $v\, \delta^{ij}\equiv\bra{0}\bar{q}_R^i q_L^j \ket{0}$ \cite{Manohar:2000dt}.  Making a $SU(3)_L \times SU(3)_R$ transformation, the vev of the transformed fields is: 
\begin{equation}
    \begin{aligned}
        \bra{0} \bar{q}_R^k (R^\dagger)^{ki}L^{jl} q_L^l \ket{0} = & \; (R^\dagger)^{ki}L^{jl} \bra{0} \bar{q}_R^k q_L^l \ket{0} \; , \\
        = & \; (R^\dagger)^{ki}L^{jl} \, v \, \delta^{kl} \; , \\
        = & \; v \, (R^\dagger)^{ki} L^{jk}  \; , \\
        = & \; v \, (L R^\dagger)^{ji} \; .
    \end{aligned}
\label{eq: vev}
\end{equation}
The vev is unchanged under the transformation if $(LR^\dagger)^{ji} = \delta^{ij}$, i.e., if $L=R$.  Therefore, the vev breaks the chiral symmetry to a single $SU(3)_V$ subgroup.  Each of the eight broken generators has a corresponding Goldstone boson, which can be written in a matrix $M$:
\begin{equation}
    M =
    \begin{pmatrix}
        \frac{1}{\sqrt{2}}\pi^0+\frac{1}{\sqrt{6}}\eta & \pi^+ & K^+ \\
        \pi^- & -\frac{1}{\sqrt{2}}\pi^0+\frac{1}{\sqrt{6}}\eta & K^0 \\
        K^- & \bar{K}^0 & -\sqrt{\frac{2}{3}}\eta
    \end{pmatrix} \, .
\end{equation}
These eight fields are identified with the pions, kaons, and eta mesons.  Chiral perturbation theory ($\chi$PT) attempts to write down an EFT for these fields.  

When only considering the two lightest quarks $q=\begin{pmatrix} u & d \end{pmatrix}^T$, $M$ reduces to a two-by-two matrix of only the pion fields:
\begin{equation}
    M \rightarrow 
    \begin{pmatrix}
        \frac{1}{\sqrt{2}}\pi^0 & \pi^+  \\
        \pi^- & -\frac{1}{\sqrt{2}}\pi^0
    \end{pmatrix} \equiv \pi \; .
\end{equation}
The EFT developed with these fields is $SU(2)$ $\chi$PT.  Its power counting parameters are even smaller than for $SU(3)$ $\chi$PT due to the smaller quark masses.  Since we are not interested in the strange quark for the discussion of XEFT and the $T_{cc}^+$, we will be dealing with $SU(2)$ $\chi$PT moving forward.  

The combination $\Sigma = \exp(2i\pi/f_\pi)$ transforms under the chiral symmetry transformations as $\Sigma\rightarrow L\Sigma R^\dagger$.  As $\Sigma \Sigma^\dagger = 1$, the most general leading-order Lagrangian you can write down that is invariant under this transformation is:
\begin{equation}
    \mathcal{L} = \frac{f_\pi^2}{8}{\rm tr} \; \partial^\mu \Sigma \partial_\mu \Sigma^\dagger \; .
\end{equation}
If we want to include the effects of the light quark masses, we can proceed with a spurion analysis.  The mass terms in the QCD Lagrangian appear as $-\bar{q}_Lm_qq_R-\bar{q}_Rm_q^\dagger q_L$, which is not invariant under $q_L\rightarrow Lq_L$, $q_R \rightarrow Rq_R$.  We have used the fact that the mass matrix is diagonal and real.  Inspecting these operators, one can imagine introducing a fictitious spurion field $S$ which transforms as $S\rightarrow L S R^\dagger$; then the operators $-\bar{q}_LSq_R - \bar{q}_RS^\dagger q_L$ are invariant under $SU(2)_L\times SU(2)_R$.  Replacing $S\rightarrow m_q$ would make this the mass term.  We can add the most general dependence on $S$ that obeys the symmetry to our chiral Lagrangian:
\begin{equation}
    \mathcal{L} = \frac{f_\pi^2}{8}{\rm tr} \; \partial^\mu \Sigma \partial_\mu \Sigma^\dagger - \frac{B f_\pi^2}{8} {\rm tr}(S\Sigma^\dagger + S^\dagger \Sigma)\; .
\end{equation}
Explicitly breaking the symmetry, the dependence on the light quark masses can then be expressed as:
\begin{equation}
    \mathcal{L} = \frac{f_\pi^2}{8}{\rm tr} \; \partial^\mu \Sigma \partial_\mu \Sigma^\dagger - \frac{Bf_\pi^2}{8} \, {\rm tr} (m_q^\dagger \Sigma + m_q \Sigma^\dagger) \; ,
\end{equation}
The interpretation of $B$ can be seen as follows.  Consider a path integral $Z=\int d\phi \; \exp[i\int d^4 x \; \mathcal{L}(\phi)]$, where $\phi$ represents all the fields in the given Lagrangian.  If we consider the derivative $iZ^{-1}\partial Z / \partial (m_q)^{ij}$ in both QCD and $\chi$PT, we see
\begin{equation}
    \begin{aligned}
        \frac{i}{Z_{\rm QCD}} \frac{\partial Z_{\rm QCD}}{\partial(m_q)^{ij}} = & \; \int d^4 x \; \bra{0}\bar{q}^i \bar{q}^j \ket{0} \; , \\
        \frac{i}{Z_{\chi{\rm PT}}} \frac{\partial Z_{\chi{\rm PT}}}{\partial(m_q)^{ij}} = & \; - \frac{Bf_\pi^2}{8} \int d^4 x \; \bra{0}(\Sigma + \Sigma^\dagger)^{ij}\ket{0} \; .
    \end{aligned}
\end{equation}
To lowest order in the pion fields, the expectation value in second line reduces to $\delta^{ij}$, and so we see that the constant $B$ is directly related to the vev discussed in Eq.~(\ref{eq: vev}).  

Breaking the chiral symmetry gives masses to the pions, now making them pseudo-Goldstone bosons.  Expanding out the $\Sigma$ matrices to write the Lagrangian in terms of the pion fields themselves, we get:
\begin{equation}
\begin{aligned}
    {\mathcal L}_\pi ={\rm tr}(\partial^\mu\pi^\dagger \partial_\mu \pi - m_\pi ^2 \pi^\dagger\pi) \, .
\end{aligned}
\end{equation}
This is the relativistic Lagrangian for the propagation of pions that we will use in our EFT for the $T_{cc}^+$.  The power counting is in $p^2/\Lambda_\chi^2$ and $m_\pi^2/\Lambda_\chi^2$, where $p$ is a typical momentum for a pion and $\Lambda_\chi \equiv 4\pi f_\pi$ is the scale of chiral symmetry breaking, which is near the charm quark mass at around $1.6$ GeV.  It can be shown that the masses of the pions are related to the vev by: \cite{Gell-Mann:1968hlm}
\begin{equation}
    m_\pi^2 = \frac{2}{f_\pi^2}(m_u+m_d)\bra{0}\bar{q}q\ket{0} \; .
\end{equation}
%

\subsection{Heavy hadron chiral perturbation theory}

To eventually study the interactions of pions with the $\chi_{c1}(3872)$ and $T_{cc}^+$, we need to couple the pions to heavy meson fields.  Heavy hadron chiral perturbation theory (HH$\chi$PT) \cite{Wise:1992hn,Burdman:1992gh,Yan:1992gz} does this by incorporating into $\chi$PT mesons of the type $Q\bar{q}^a$, where $a$ is a light quark flavor index. In the charm sector these are the pseudoscalar mesons $D^0$, $D^+$, $D_s^+$, and the vector mesons $D^{*0}$, $D^{*+}$, and $D_s^{*+}$.  Again, we are uninterested in strange quarks for the purposes of XEFT and the $T_{cc}^+$, so we neglect $D_s^+$ and $D_s^{*+}$.  

The pseudoscalar and vector mesons are degenerate in the heavy quark symmetry limit $m_Q \rightarrow \infty$ \cite{Manohar:2000dt}.  Therefore, it is convenient to combine the pseudoscalar meson fields $P_a$ and the vector meson fields $P_{a\mu}^*$ into a new field $H_{a,v}$, where $a$ is an index for the light quark in the meson, and $v$ is the four-velocity of the mesons.
\begin{equation}
    H_{a,v} = \frac{1+\slashed{v}}{2}(\slashed{P}_a^*-P_a\gamma_5) \, .
\end{equation}
In the rest frame with $v^\mu=(1,0,0,0)$, this reduces to:
\begin{equation}
    H_{a,v} \rightarrow 
    \begin{pmatrix}
        0 & -{\bf P}_a^* \cdot \boldsymbol{\sigma} - P_a  \\
        0 & 0
    \end{pmatrix} \; .
\end{equation}
Then we can define $2 \times 2$ matrix fields $H_a = {\bf P}_a^* \cdot \boldsymbol{\sigma} + P_a $, which transform under the $SU(2)_L \times SU(2)_R$ chiral symmetry as $H_a^\prime = H_b V_{ba}^\dagger$, where $V$ is a matrix which gives a nonlinear realization of the symmetry, and under the $SU(2)$ heavy quark spin symmetry as $H_a^\prime = SH_a$. The Lagrangian for HH$\chi$PT in this context is then: \cite{Hu:2005gf}
\begin{equation}
    \mathcal{L}={\rm tr}[H_a^\dagger (iD^0)_{ba} H_b] - g \, {\rm tr}[H_a^\dagger H_b \boldsymbol{\sigma}\cdot {\bf A}_{ba}] + \frac{\Delta}{4}{\rm tr}[H_a^\dagger \boldsymbol{\sigma}^i H_a \boldsymbol{\sigma}^i] \, .
\end{equation}
Here $g$ is the axial coupling of the heavy mesons to the axial vector field ${\bf A} = \frac{i}{2}(\xi^\dagger \nabla \xi - \xi \nabla \xi^\dagger)$, where $\xi = \sqrt{\Sigma}=\exp{(i\pi/f_\pi)}$. The scale $\Delta$ is the hyperfine splitting of the heavy mesons; its term in the Lagrangian breaks heavy quark spin symmetry.

We also need to couple to photons, and the interaction terms with the magnetic field $\bf B$ are given by: \cite{Amundson:1992yp}
\begin{equation}
    \mathcal{L}_{\rm EM} = \frac{e\beta}{2}\, {\rm tr}(H_a^\dagger H_b \boldsymbol{\sigma}\cdot {\bf B} Q_{ab}) + \frac{e}{2m_Q}Q^\prime \, {\rm tr}(H_a^\dagger \boldsymbol{\sigma}\cdot {\bf B}H_a) \; .
\end{equation}
The matrix $Q_{ab}={\rm diag}(2/3,-1/3)$ is the light quark charge matrix, and $Q^\prime$ is the heavy quark charge, which is $2/3$ for the charm quark.  The parameter $\beta$ is a coupling with mass dimension $-1$; it needs to be fit to experimental results for partial widths of $D^*$ decays \cite{Amundson:1992yp,Stewart:1998ke}.

\section{XEFT}
\label{sec: XEFT}

XEFT was initially developed in Ref.~\cite{Fleming:2007rp}, and has been further utilized or expanded upon by many authors \cite{Fleming:2008yn,Fleming:2011xa,Mehen:2011ds,Margaryan:2013tta,Braaten:2010mg,Canham:2009zq,Jansen:2013cba,Guo:2014hqa,Jansen:2015lha,Mehen:2015efa,Alhakami:2015uea,Braaten:2015tga,Dai:2019hrf,Braaten:2020iye,Braaten:2020nmc,Braaten:2020iqw}. As a first step, it takes HH$\chi$PT and adapts it to apply to a particular weakly bound molecule of neutral charm mesons, consistent with the properties of the $\chi_{c1}(3872)$. Initially known as the $X(3872)$, the exotic state was discovered in $e^+e^-$ collisions by the Belle Collaboration in 2003 \cite{Belle:2003nnu}, and further confirmed by the CDF, D\O, and BaBar collaborations \cite{CDF:2003cab,D0:2004zmu,BaBar:2004oro}.  Its mass averaged from these experiments is $m_X = 3871.65 \pm 0.06$ MeV \cite{ParticleDataGroup:2022pth}.  This is close to the $D^*D$ threshold: $E_X = m_D + m_{D^*} - m_X = 0.04 \pm 0.09$ MeV, which suggests a molecular bound-state interpretation.  Furthermore, the $D^*$-$D$ hyperfine splitting is only $7$ MeV higher than the neutral pion mass of $135$ MeV, and so any pions produced in a $D^*$ decay are non-relativistic.  These facts invite an effective field theory description, first developed by Fleming et.~al \cite{Fleming:2007rp} to study pion interactions with the bound state.  They consider the resonance a weakly bound molecule of $D^0 \bar{D}^{*0}$ and $D^{*0}\bar{D}^0$, in a state which is positive under charge conjugation: $\ket{DD^*} = (\ket{D^0 \bar{D}^{*0}} + \ket{D^{*0}\bar{D}^0})/\sqrt{2}$.  The starting point for deriving the XEFT Lagrangian is the two-component HH$\chi$PT Lagrangian, together with kinetic terms for pions from the $\chi$PT Lagrangian, both of which were discussed in the previous section.  A key difference between XEFT and HH$\chi$PT is that XEFT includes a kinetic energy term for the heavy mesons.  In HH$\chi$PT this term is subleading in the power counting in $1/m_H$, but in XEFT it is leading in the power counting in the relative velocity of the heavy mesons $v$ \cite{Fleming:2007rp}.  Several field redefinitions then lead to the initially-posited XEFT Lagrangian:
\begin{equation}
    \begin{aligned}
        \mathcal{L}= & \; \boldsymbol{D}^{\dagger}\left(i \partial_0+\frac{\nabla^2}{2 m_{D^*}}\right) \boldsymbol{D}+D^{\dagger}\left(i \partial_0+\frac{\nabla^2}{2 m_D}\right) D \\
        & +\bar{\boldsymbol{D}}^{\dagger}\left(i \partial_0+\frac{\nabla^2}{2 m_{D^*}}\right) \bar{\boldsymbol{D}}+\bar{D}^{\dagger}\left(i \partial_0+\frac{\nabla^2}{2 m_D}\right) \bar{D} \\
        & +\pi^{\dagger}\left(i \partial_0+\frac{\nabla^2}{2 m_\pi}+\delta\right) \pi \\
        & +\left(\frac{g}{\sqrt{2} f_\pi}\right) \frac{1}{\sqrt{2 m_\pi}}\left(D \boldsymbol{D}^{\dagger} \cdot \nabla \pi+ \bar{D}^{\dagger} \bar{\boldsymbol{D}} \cdot \nabla \pi^{\dagger}\right)+\text { H.c. } \\
        & -\frac{C_0}{2}(\bar{\boldsymbol{D}} D+\boldsymbol{D} \bar{D})^{\dagger} \cdot(\bar{\boldsymbol{D}} D+\boldsymbol{D} \bar{D}) \\
        & +\frac{C_2}{16}(\bar{\boldsymbol{D}} D+\boldsymbol{D} \bar{D})^{\dagger} \cdot\left(\bar{\boldsymbol{D}}(\overleftrightarrow{\nabla})^2 D+\boldsymbol{D}(\overleftrightarrow{\nabla})^2 \bar{D}\right)+\text { H.c. } \\
        & +\frac{B_1}{\sqrt{2}} \frac{1}{\sqrt{2 m_\pi}}(\bar{\boldsymbol{D}} D+\boldsymbol{D} \bar{D})^{\dagger} \cdot D \bar{D} \nabla \pi+\text { H.c. }+\cdots .
\end{aligned}
\end{equation}
The fields that annihilate the $D^{*0}$, $\bar{D}^{*0}$, $D^0$, $\bar{D}^0$, and $\pi^0$ are denoted $\boldsymbol{D}$, $\bar{\boldsymbol{D}}$, $D$, $\bar{D}$, and $\pi$,\footnote{This $\pi$ field is a nonrelativistic representation for the pion, and is different from the $\pi$ field in HH$\chi$PT.} respectively.  The residual mass $\delta=7$ MeV is the difference between the $D^*D$ hyperfine splitting and the neutral pion mass.  The coupling to pions $g=0.54$ is the axial coupling from HH$\chi$PT \cite{Wise:1992hn,Burdman:1992gh,Yan:1992gz}, and the pion decay constant is $f_\pi = 132$ MeV. The two-directional derivative is $\overleftrightarrow{\nabla}=\overleftarrow{\nabla}-\overrightarrow{\nabla}$.

\begin{figure*}[t]
\centering
\includegraphics[width=\textwidth]{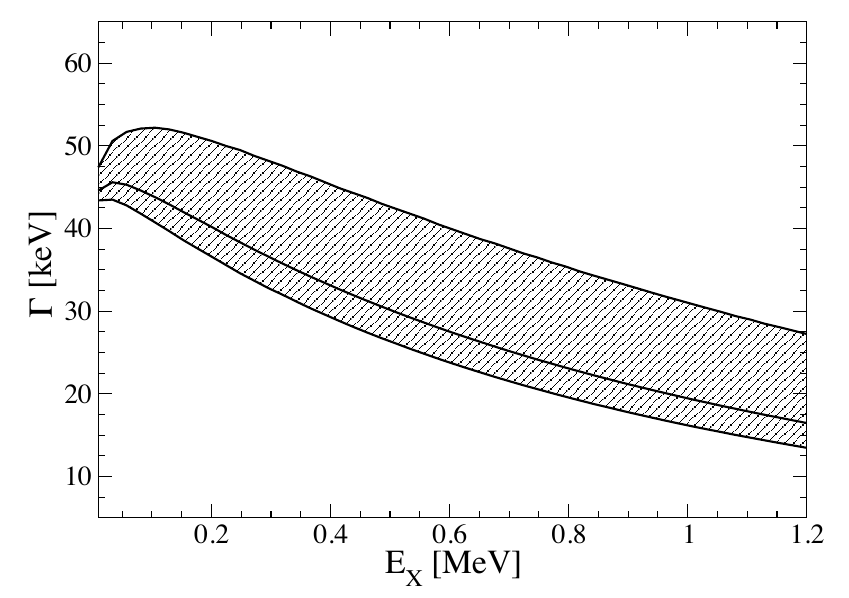}
\caption[Original XEFT prediction for the decay rate of the $\chi_{c1}(3872)$ as a function of the binding energy $E_X$.]{Original XEFT prediction for the decay rate of the $\chi_{c1}(3872)$ as a function of the binding energy $E_X$.  The solid line is the LO prediction, and the shaded band is the range of the NLO prediction.  Figure reused with permission from Ref.~\cite{Fleming:2007rp}.}
\label{fig: XEFT decay rate}
\end{figure*}

Other key additions in XEFT are the contact interactions mediated by $C_0$, $C_2$, and $B_1$, which involve the interpolating field for the bound state, $X^i=(D^0\bar{D}^{0*i} + \bar{D}^0 D^{0*i})/\sqrt{2}$.  We will see that these are the interactions that are responsible for the bound state.  To derive the decay rate for the $\chi_{c1}(3872)$, Fleming et.~al consider a two-point function of the interpolating field.
\begin{equation}\label{twopointG}
    G(E) \delta^{ij} = \int d^4 x \; e^{-iEt} \bra{0}T[X^i(x)X^j(0)]\ket{0} = \delta^{ij}\frac{i Z(-E_X)}{E+E_X+i\Gamma/2} \, .
\end{equation}
We see that this has a pole at the location of the bound state, whose imaginary part is the decay width, and whose residue is the wave function renormalization for the interpolating field.  The $C_0$ interaction is treated nonperturbatively, and in terms of the $C_0$-irreducible graphs contributing to $G(E)$, collectively represented by $-i\Sigma(-E_X)$, this can be written as:
\begin{equation}
\begin{aligned}
    G =& \; \frac{-i\Sigma}{1+C_0\Sigma} \\
    =& \; \frac{-i\,{\rm Re}\,\Sigma+{\rm Im}\,\Sigma}{1+C_0\,{\rm Re}\,\Sigma+iC_0\,{\rm Im}\,\Sigma} \, .
\end{aligned}
\label{eq: G re and im}
\end{equation}
From Eq.~(\ref{twopointG}) we see that the real part of the denominator must vanish at the location of the bound state, so we have $1+C_0\,{\rm Re}\,\Sigma(-E_X)=0$.  The value of $C_0$ is therefore tuned to the location of the bound state.  We can expand the real part about $E+E_X$, but not the imaginary part since it arises from the NLO diagrams and is thus already small.  
\begin{equation}
    G = \frac{i}{C_0^2(E+E_X)\,{\rm Re}\,\Sigma^\prime(-E_X)+iC_0^2\,{\rm Im}\,\Sigma(-E_X)} - \frac{i}{C_0} \, .
\end{equation}
Ignoring the irrelevant overall constant, we can read off:
\begin{equation}
\begin{aligned}
    Z =& \; \frac{1}{C_0^2 \, {\rm Re}\, \Sigma^\prime(-E_X)} \, , \\
    \Gamma =& \; \frac{2\,{\rm Im}\,\Sigma(-E_X)}{{\rm Re}\,\Sigma^\prime(-E_X)} \, .
\end{aligned}
\label{eq: XEFT Gamma}
\end{equation}
All that remains is to compute the decay rate is to calculate the graphs that contribute to ${\rm Re} \, \Sigma^\prime$ and ${\rm Im} \, \Sigma$.  We do not quote the full result here, as it is essentially the $m_u = m_d$ case of the computation for the $T_{cc}^+$ decay rate, which will be discussed in detail later in the chapter.  The original prediction for the decay rate of the $\chi_{c1}(3872)$ as a function of the binding energy is shown in Fig.~\ref{fig: XEFT decay rate}.  Recall that, using the most recent value of the mass of the $\chi_{c1}(3872)$, $E_X = m_D + m_{D^*} - m_X = 0.04 \pm 0.09$ MeV.  

\begin{figure*}[t]
\centering
\includegraphics[width=\textwidth]{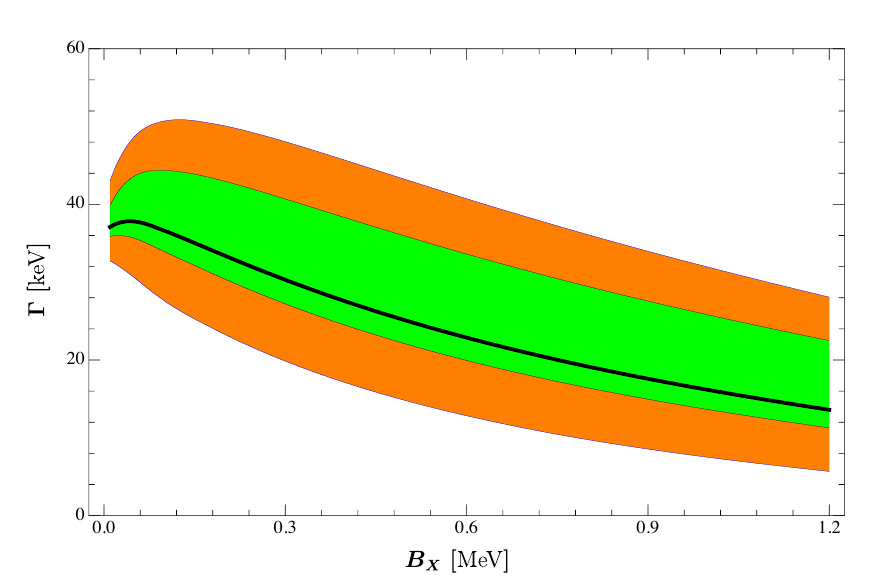}
\caption[Updated XEFT prediction for the decay width of the $\chi_{c1}(3872)$ as a function of the binding energy $B_X$.]{Updated XEFT prediction for the decay width of the $\chi_{c1}(3872)$ as a function of the binding energy $B_X$.  The solid line is the LO prediction, the green band is the original NLO prediction from Ref.~\cite{Fleming:2007rp}, and the orange is the updated NLO prediction including the new final state interaction terms. Figure from Ref.~\cite{Dai:2019hrf}.}
\label{fig: XEFT updated decay rate}
\end{figure*}

\begin{figure*}[t]
\centering
\includegraphics[width=\textwidth]{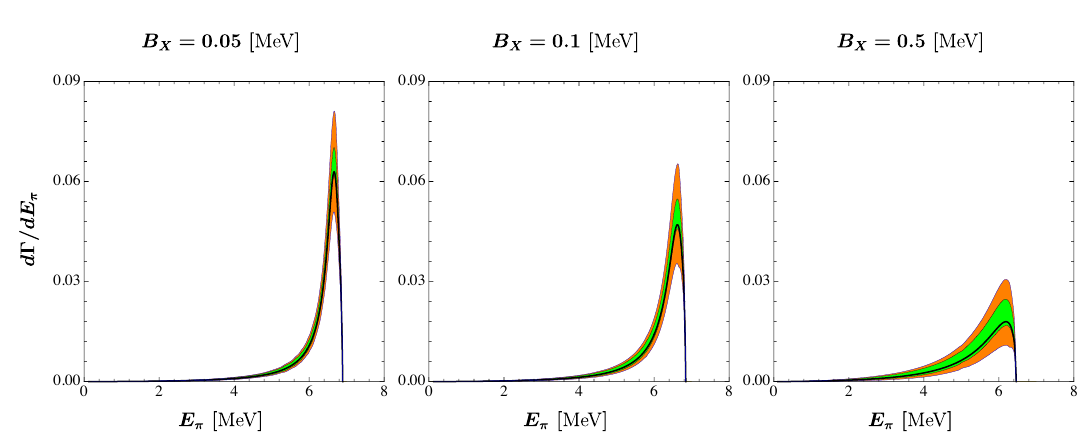}
\caption[Decay width for the $\chi_{c1}(3872)$, differential in the final state pion energy $E_\pi$, for three different binding energies $B_X$.]{Decay width for the $\chi_{c1}(3872)$, differential in the final state pion energy $E_\pi$, for three different binding energies $B_X$.  The band colorings are the same as in Fig.~\ref{fig: XEFT updated decay rate}. Figure from Ref.~\cite{Dai:2019hrf}.}
\label{fig: XEFT diff distr}
\end{figure*}

Reference \cite{Dai:2019hrf} revisits the decay studied in Ref.~\cite{Fleming:2007rp} and includes contributions from final-state rescattering, introducing two new interaction terms at NLO:
\begin{equation}
    \begin{aligned}
        \mathcal{L} = \frac{C_\pi}{2m_{\pi^0}}(D^\dagger \pi^\dagger D\pi + \bar{D}^\dagger \pi^\dagger \bar{D}\pi) + C_{0D} D^\dagger \bar{D}^\dagger D \bar{D} \; .
    \end{aligned}
\end{equation}
These new interaction terms can have a substantial effect on the decay rate, as evidenced in Fig.~\ref{fig: XEFT updated decay rate}.  The authors argue that, since the NLO interactions can change the predicted decay width by as much as about $50\%$, the decay width alone is not a good observable through which to extract the binding energy of the $\chi_{c1}(3872)$.  Instead, the authors look at the decay width differential in the final state pion energy, as in Fig.~\ref{fig: XEFT diff distr}.  The distribution is sharply peaked, and the location of the peak is insensitive to the NLO effects; they only change the overall magnitude.  The magnitude also decreases as the binding energy increases, and so this differential distribution may be a better observable to extract the binding energy.

\begin{figure*}[t]
\centering
\includegraphics[width=\textwidth]{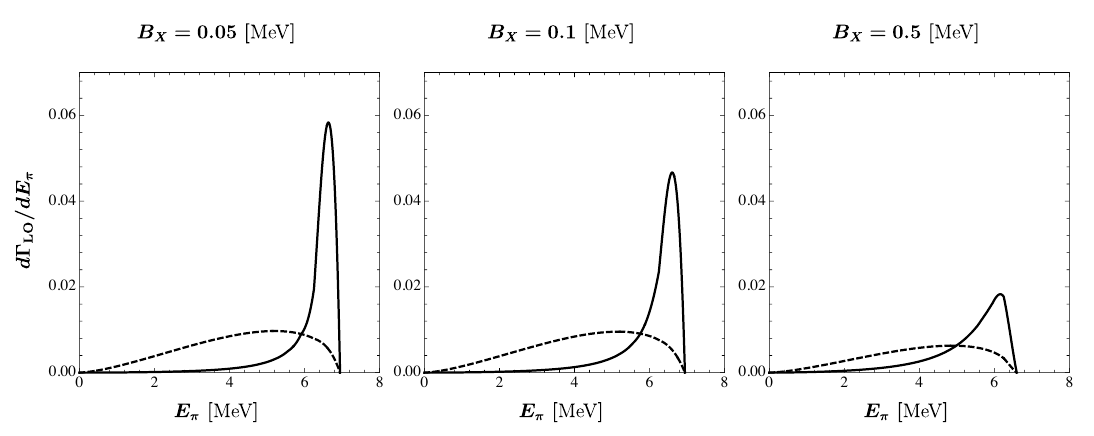}
\caption[Leading-order $\chi_{c1}(3872)$ decay width differential in the final state pion energy $E_\pi$ for different binding energies $B_X$.]{Leading-order $\chi_{c1}(3872)$ decay width differential in the final state pion energy $E_\pi$ for different binding energies $B_X$.  The solid line is the XEFT prediction, and the dotted line is the XEFT prediction with the $D^*$ propagator replaced by a constant, normalized to the same total width. Figure from Ref.~\cite{Dai:2019hrf}.}
\label{fig: XEFT const prop}
\end{figure*}

Another key result of Ref.~\cite{Dai:2019hrf} is that XEFT provides a strong prediction regarding the molecular nature of the $\chi_{c1}(3872)$.  The sharpness of the differential decay distribution is a direct consequence of the virtual $D^*$ propagator, arising in the tree-level decay like in Fig.~\ref{figX1a}, in the decay width.  If the propagator is replaced by a constant, the distribution becomes flat, a fact which holds true for each binding energy considered (Fig.~\ref{fig: XEFT const prop}).  An experimental measurement of the differential distribution yielding a sharp peak thus would support a molecular interpretation for the $\chi_{c1}(3872)$.

\subsection{Summary of the couplings in XEFT}
\label{sec: couplings in XEFT}

There are many couplings in the XEFT Lagrangian, and here we discuss how each are determined.  This will carry over to the corresponding couplings for the $T_{cc}^+$.
\begin{itemize}
    \item {\bf the pion-charm meson coupling} $g$: fit by comparing tree-level $D^*$ decays in XEFT to data from the PDG.
    \item {\bf the contact interaction coupling} $C_0$: tuned to the location of the bound state.
    \item {\bf the derivative contact interaction coupling} $C_2$: the NLO calculation includes a correction to the LO width proportional to $C_2$, and which depends on a cutoff used to regulate the divergent integrals.  In effective range theory, this correction arises from a modification to the normalization of the wave function for the bound state \cite{Fleming:2007rp}; the modification depends on the effective range $r_0$, which is an unknown quantity that must be estimated. The value of $C_2$ is fixed both to cancel the dependence of the cutoff and reproduce the effective range theory prediction.
    \item {\bf the five-point contact coupling} $B_1$: all terms in the NLO calculation dependent on $B_1$ are also proportional to $C_2$, and after the value of $C_2$ is fixed, the value of $B_1$ is fixed accordingly to cancel all cutoff dependence.
    \item {\bf the final state} $D\pi$ {\bf coupling} $C_\pi$: fit to lattice calculations of $D\pi$ scattering lengths \cite{Dai:2019hrf}.
    \item {\bf the final state} $DD$ {\bf coupling} $C_{0D}$: this coupling is proportional to the $D$ meson scattering length \cite{Dai:2019hrf}, about which little is known, so an educated guess is made for its range.
\end{itemize}
We will discuss the explicit values for these couplings in the context of the $T_{cc}^+$, in Sec.~\ref{sec: uncertainties in the nlo couplings}.

\section{An EFT for the $T_{cc}^+$}

Describing the $T_{cc}^+$ with an EFT essentially involves generalizing XEFT to allow for positively charged charm mesons, accounting for the fact that it is now a coupled channel problem because the resonance is near to two thresholds, and dealing with the idiosyncrasies that appear as a result. The publications \cite{Fleming:2021wmk,Dai:2023mxm} that write down this EFT were influenced by the timing of the announcements by the LHCb collaboration, and their fits of the resonance profile.

\subsection{Remarks on the $T_{cc}^+$ experimental fits}
\label{sec: remarks on Tcc fits}

The initial announcement of the observation of the $T_{cc}^+$ resonance was at the European Physical Society Conference on High Energy Physics in July 2021 \cite{Muheim,Polyakov}.  These talks refer to a fit of the resonance profile to a relativistic P-wave two-body Breit-Wigner function with a Blatt-Weisskopf form factor \cite{Blatt:1952ije,VonHippel:1972fg}, which yielded the following values for the binding energy and decay width:
\begin{equation}
\begin{aligned}
    \delta m_{\rm BW} =& \; -273 \pm 61 \pm 5 {}^{+11}_{-14}\, {\rm keV} \, , \\   
    \Gamma_{\rm BW}  =& \;  410 \pm 165 \pm 43 {}^{+18}_{-38}\, {\rm keV} \, .
\end{aligned}
\label{eq: initial BW fit}
\end{equation}
In the weeks following this announcement, a flurry of theory papers \cite{Fleming:2021wmk,Meng:2021jnw,Agaev:2021vur,Wu:2021kbu,Ling:2021bir, Chen:2021vhg,Dong:2021bvy,Feijoo:2021ppq,Yan:2021wdl,Dai:2021wxi,Weng:2021hje,Huang:2021urd,Chen:2021kad,Xin:2021wcr,Albaladejo:2021vln,Du:2021zzh,Jin:2021cxj,Abreu:2021jwm,Dai:2021vgf,Deng:2021gnb,Azizi:2021aib} rose to the task of predicting various properties of the $T_{cc}^+$.  In particular, Refs.~\cite{Meng:2021jnw,Ling:2021bir,Feijoo:2021ppq,Yan:2021wdl,Albaladejo:2021vln,Du:2021zzh} attempted to predict the decay width.  One of these papers was our initial analysis with the extension of XEFT \cite{Fleming:2021wmk}, where we found decay widths an order of magnitude smaller than that in Eq.~(\ref{eq: initial BW fit}).  This motivated us to understand the discrepancy, and we investigated the possibility of a decay to other shallow bound states of two pseudoscalar charm mesons which would enhance the width; this is discussed in Sec.~\ref{sec: decay to another bound state}.

Additional information regarding the fits of the resonance profile was provided in two publications by the LHCb collaboration \cite{LHCb:2021vvq,LHCb:2021auc}, posted on the same day in September 2021.  Reference \cite{LHCb:2021auc} argues that the ``relativistic P-wave two-body Breit-Wigner with a Blatt-Weisskopf form factor \ldots while sufficient to reveal the existence of the state, does not account for the resonance being in close vicinity of the $D^*D$ threshold.''  Instead, they fit the resonance to a unitarized Breit-Wigner profile, which yields a substantially lower value for the decay width:
\begin{equation}
\begin{aligned}
    \delta m_{\rm u} =& \; -360\pm40^{+4}_{-0}\, {\rm keV} \, ,  \\   
    \Gamma_{\rm u}  =& \;  48 \pm 2_{-14}^{+0}\, {\rm keV} \, .
\end{aligned}
\end{equation}
Since this fit accounts for the closeness to the $D^*D$ threshold, we deem it more appropriate for comparison with theoretical results.  


\subsection{Adapting the XEFT Lagrangian}
\label{sec: adapting XEFT to Tcc}

There are several key differences between XEFT and the EFT for the $T_{cc}^+$. Instead of only $D^0$ mesons and their antiparticles, we now also have bound states involving positively charged $D$ mesons, and we put them in an isospin doublet:
\begin{equation}
    H=\begin{pmatrix}D^+ \\ D^0 \end{pmatrix} \; .
\end{equation}
We also need to include the charged pions.  These facts complicate the derivation of the Lagrangian and the decay width, as now we have to deal with matrices of fields and couplings.  The other difference with XEFT is that, for the $T_{cc}^+$, we treat the pions relativistically, since the addition of the charged pions complicates the definition of a residual mass term for the pions.  The interaction terms are written down by constructing isospin invariants out of the charm meson and pion fields.  For a detailed explanation of this process, refer to App.~\ref{app: interaction terms}.

The LO Lagrangian is as follows.
\begin{equation} \label{LagLO}
\begin{aligned}
 {\mathcal L}_{\rm LO} =& \; H^{* i\dagger}\bigg(i\partial^0+\frac{\nabla^2}{2m_{H^*}}-\delta^*\bigg)H^{* i} + H^\dagger\bigg(i\partial^0+\frac{\nabla^2}{2m_H}-\delta\bigg)H  \\
&-C_0^{(0)} (H^{*T}\tau_2 H)^\dagger (H^{*T}\tau_2 H) - C_0^{(1)} (H^{*T}\tau_2 \tau_a H)^\dagger (H^{*T}\tau_2 \tau_a H) \\
& + \frac{g}{f_\pi} H^\dagger \partial^i \pi  H^{*i} + \frac{1}{2}H^\dagger \mu_D {\bf B}^i H^{*i} + \text{H.c.} \, .
\end{aligned}
\end{equation}
Here $\delta$ and $\delta^*$ are diagonal matrices that contain the residual masses of the $D$ mesons; their entries are $\delta_{ii}^{(*)} = m_i^{(*)}-m_0$. The subscripts on the masses indicate the charge of the charm meson.  The magnetic field is denoted by ${\bf B}$.  The matrix $\mu_D$ contains transition magnetic moments, $\mu_D = {\rm diag}(\mu_0,\mu_+)$, where the entries are fixed to give the partial widths $\Gamma(D^{*+}\rightarrow D^+\gamma) = 1.33$ keV and $\Gamma(D^{*0}\rightarrow D^0 \gamma)=19.9$ keV at tree level; refer to App.~\ref{app: transition magnetic moments} for details on these values. 

The NLO interaction terms are constructed from isospin invariant combinations of the $H$ fields and $\pi$ fields; their derivation and their range of values are covered in App.~\ref{app: interaction terms}.  They are:
\begin{equation}\label{range}
\begin{aligned}
    {\mathcal L}_{C_2} = & \; \frac{C_2^{(0)}}{4} (H^{*T}\tau_2 H)^\dagger (H^{*T}\tau_2\overleftrightarrow{\nabla}^2 H) \\
    & + \frac{C_2^{(1)}}{4} (H^{*T}\tau_2 \tau_a H)^\dagger (H^{*T}\tau_2 \tau_a \overleftrightarrow{\nabla}^2H) + \, {\rm H.c.} \, ,  
\end{aligned}
\end{equation}
\begin{equation}  \label{CpiLag}
\begin{aligned}
    {\mathcal L}_{C_\pi} = & \; C_\pi^{(1)}D^{0\dagger}\pi^{0\dagger} D^+ \pi^- - C_\pi^{(1)} D^{+\dagger}\pi^{0\dagger} D^0 \pi^+ + {\rm H.c.}   \\
    & + C_\pi^{(2)}D^{0\dagger}\pi^{0\dagger} D^0 \pi^0 + C_\pi^{(2)} D^{+\dagger}\pi^{0\dagger} D^+ \pi^0  \\
    & + C_\pi^{(3)}D^{0\dagger}\pi^{+\dagger} D^0 \pi^+ \; ,
\end{aligned}
\end{equation}
\begin{equation} \label{LagB12}
\begin{aligned}
    {\mathcal L}_{B_1} = & \; B_1^{(1)}(D^+D^{*0})^\dagger(D^+D^0\nabla\pi^0) +B_1^{(2)}(D^0D^{*+})^\dagger(D^+D^0\nabla\pi^0)  \\
    & +\frac{B_1^{(3)}}{2}(D^0D^{*+})^\dagger(D^0D^0\nabla\pi^+) + \frac{B_1^{(4)}}{2}(D^+D^{*0})^\dagger(D^0D^0\nabla\pi^+) + \, {\rm H.c.}\; ,  \\ 
\end{aligned}
\end{equation}
\begin{equation}
\begin{aligned}
    {\mathcal L}_{C_{0D}} =& \; \frac{C_{0D}^{(1)}}{2}(D^0D^0)^\dagger(D^0D^0) + C_{0D}^{(1)}(D^+ D^0)^\dagger (D^+ D^0) \, .
\end{aligned}
\end{equation}
We will discuss these couplings in more detail once the decay widths have been discussed.  

The relativistic pion kinetic term is the same as discussed in Sec.~\ref{sec: ChPT}:
\begin{equation}
\begin{aligned}
    {\mathcal L}_\pi ={\rm tr}(\partial^\mu\pi^\dagger \partial_\mu \pi - m_\pi ^2 \pi^\dagger\pi) \, .
\end{aligned}
\end{equation}

\section{Leading and next-leading order decay widths}

\begin{figure*}[t]
\centering
\includegraphics[trim=4.23cm 13.97cm 1.83cm 5.08cm,clip,scale=1]{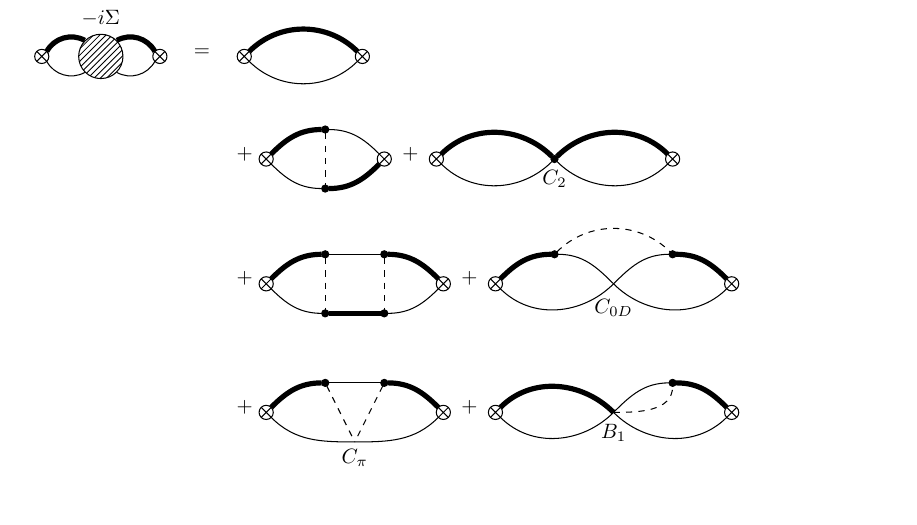}
\caption[The diagrams that contribute to $-i\Sigma$ to NNLO.]{The diagrams that contribute to $-i\Sigma$ to NNLO.  Bold solid lines are $D^*$ mesons, regular solid lines are $D$ mesons, and dotted lines are pions.  NNLO diagrams arising from concatenating two NLO diagrams are not shown.  Figure from Ref.~\cite{Dai:2023mxm}.}
\label{bubbleDiagram}
\end{figure*}

In adapting the XEFT formula for the decay width, Eq.~(\ref{eq: XEFT Gamma}), to the $T_{cc}^+$, we have to account for the coupled channel problem.  The two-point function $G$ in Eq.~(\ref{twopointG}) needs to be generalized to a matrix:
\begin{equation} \label{Geqn}
\begin{aligned}
\hat{G} =&\; \int d^4x\; e^{-iEt} \braket{0|T[X(x)X^T(0)]|0}  \\
& = i\Sigma(1+C_0\Sigma)^{-1} \, ,
\end{aligned}
\end{equation}
where the interpolating field is the appropriate combination of $D^*D$ fields for the $T_{cc}^+$.
\begin{equation}
\begin{aligned}
X=\begin{pmatrix} D^0D^{*+} \\ D^+D^{*0} \end{pmatrix} \, .
\end{aligned}
\end{equation}
The second line of Eq.~(\ref{Geqn}) arises from writing $\hat{G}$ as an infinite sum of the $C_0$-irreducible two-point function $\Sigma$, like in Appendix A of Ref.~\cite{Kaplan:1998sz}, but with $2\times2$ matrices $C_0$ and $\Sigma$.  The diagrammatic representation of $-i\Sigma$ is given to NNLO in Fig. \ref{bubbleDiagram}. The diagonal elements of $-i\Sigma$ contain the graphs where the channel is the same at the initial and final insertion of $C_0$, and the off-diagonal elements are the graphs where the channel swaps.  Any other diagrams contributing to $\Sigma$ that you can write down at this order have a pole structure so that you could evaluate the loop integral $dl_0$ by enclosing only a pion pole.  These are ``radiation pions'', and are their diagrams are suppressed compared to those in Fig.~\ref{bubbleDiagram} \cite{Mehen:1999hz}.

\begin{figure*}[t]
\centering
\begin{minipage}{0.5\textwidth}
\centering
\subfloat[]{\includegraphics[scale=0.99]{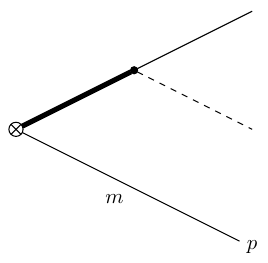}\label{figX1a}}
\end{minipage}%
\begin{minipage}{0.5\textwidth}
\centering
\subfloat[]{\includegraphics[scale=0.99]{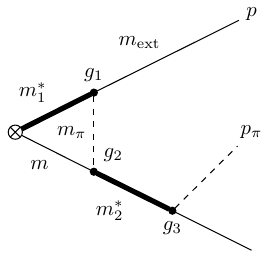}\label{figX1b}}
\end{minipage}%
\\
\begin{minipage}{0.5\textwidth}
\centering
\subfloat[]{\includegraphics[scale=0.99]{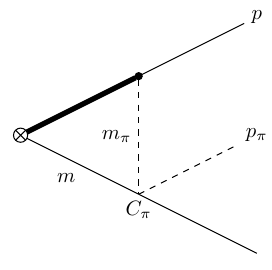}\label{figX1d}}
\end{minipage}%
\bigskip
\begin{minipage}{0.5\textwidth}
\centering
\subfloat[]{\includegraphics[scale=0.99]{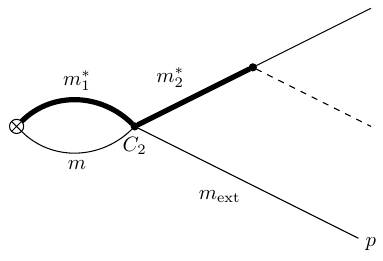}\label{figX1e}}
\end{minipage}%
\\
\begin{minipage}{0.5\textwidth}
\centering
\subfloat[]{\includegraphics[scale=0.99]{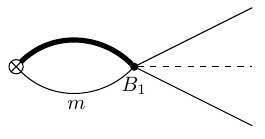}\label{figX1f}}
\end{minipage}%
\begin{minipage}{0.5\textwidth}
\centering
\subfloat[]{\includegraphics[scale=0.99]{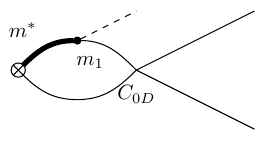}\label{figX1g}}
\end{minipage}
\caption[$T_{cc}^+$ decay diagrams to NLO.]{$T_{cc}^+$ decay diagrams to NLO.  Figures from Ref.~\cite{Dai:2023mxm}.}
\label{DecayDiagrams}
\end{figure*}

A feature of this matrix representation for $\hat{G}$ is that we can project out either the isospin-0 or the isospin-1 channels, each with its own residue and decay width:
\begin{equation} \label{G01}
\begin{aligned}
G_{0/1} =&\; \left(\begin{array}{c} 1 \\ \mp1 \end{array}\right)^T\hat{G}\left(\begin{array}{c} 1 \\ \mp1 \end{array}\right)  \\
 \approx & \; \frac{1}{2}\frac{iZ_{0/1}}{E+E_T + \frac{i\Gamma_{0/1}}{2}} \, ,
\end{aligned}
\end{equation}
The full expression for $G_0$ is:
\begin{equation}
\begin{aligned}
-iG_0 =&\; \frac{-\frac{1}{2}\Sigma_{I=0}-2C_0^{(1)}{\rm det}\,\Sigma}{1+C_0^{(0)}\Sigma_{I=0}+C_0^{(1)}\Sigma_{I=1}+4C_0^{(0)}C_0^{(1)}{\rm det}\,\Sigma} \, ,
\end{aligned}
\end{equation}
where $\Sigma_{I=0/I=1} \equiv \Sigma_{11}+\Sigma_{22}\mp\Sigma_{12}\mp\Sigma_{21}$ is either the isospin-0 or isospin-1 combination of the diagrams. Taking the $T_{cc}^+$ to be an isospin-0 state, we treat $C_0^{(1)}$ perturbatively and expand to LO.
\begin{equation}
\begin{aligned}
-iG_0 \approx & \; \frac{1}{2}\frac{-\Sigma_{I=0}}{1+C_0^{(0)}\Sigma_{I=0}} \, .
\end{aligned}
\end{equation}
Now the expression above is problem is of the same form as Eq.~(\ref{eq: G re and im}), but with the single-channel $\Sigma$ replaced by the isospin-0 combination $\Sigma_{I=0}$ for the coupled channel.  The analysis then proceeds in the same way as in Sec.~\ref{sec: XEFT}, and the wave function renormalization and decay width are:
\begin{equation}
\begin{aligned}
Z_0 =&\; \frac{1}{\big(C_0^{(0)}\big)^2 {\rm Re}\, \Sigma_{I=0}^\prime(-E_T)} \, ,  \\
\Gamma_0 =&\; \frac{2\, {\rm Im}\, \Sigma_{I=0}(-E_T)}{{\rm Re}\, \Sigma_{I=0}^\prime(-E_T)} \, .
\end{aligned}
\end{equation}
We now split $\Sigma_0$ into its LO and NLO contributions, and expand in the NLO ones.  The LO diagrams in $\Sigma_0$ are only on the diagonal, since at LO we cannot swap channels.  
\begin{equation} \label{width2}
\begin{aligned}
\Gamma_0  \approx & \; \Gamma_0^{\rm LO}\bigg(1-\frac{{\rm Re}\, \Sigma_{I=0}^{\prime \rm NLO}(-E_T)}{{\rm Re}\, {\rm tr}\,\Sigma^{\prime \rm LO}(-E_T)}\bigg) + \frac{2\,{\rm Im}\, \Sigma_{I=0}^{\rm NLO}(-E_T)}{{\rm Re}\, {\rm tr}\,\Sigma^{\prime \rm LO}(-E_T)} \; ,
\end{aligned}
\end{equation}
where
\begin{equation}
    \Gamma_0^{\rm LO} = \frac{2 \, {\rm Im}\, \Sigma_{I=0}^{\rm LO}(-E_T)}{{\rm Re}\, {\rm tr}\,\Sigma^{\prime \rm LO}(-E_T)} \; .
\end{equation}
The ${\rm Re}\, \Sigma_{I=0}^{\prime \rm NLO}(-E_T)$ term in Eq.~(\ref{width2}) is a correction to the LO decay width from NLO $D^*D$ self-energy corrections, which are the  diagrams on the second row of Fig.~\ref{bubbleDiagram}.  The self-energy diagrams to NLO are:
\begin{itemize}
    \item the LO self-energy diagram, $-i\Sigma_1$
    \item the one-pion exchange diagram, $-i\Sigma_2$
    \item the $C_2$ contact diagram, $-i\Sigma_3$
\end{itemize}
The ${\rm Im}\, \Sigma_{I=0}^{\rm NLO}(-E_T)$ term in Eq.~(\ref{width2}), by the optical theorem, is related to products of NLO decay diagrams resulting from the cuts of the graphs on the third and fourth rows of Fig.~\ref{bubbleDiagram}.\footnote{${\rm Im}\,\Sigma^{\rm NLO}$ is from self-energy diagrams of one higher order than in ${\rm Re}\,\Sigma^{\rm NLO}$ because the LO self-energy graph has no imaginary part below threshold.} These decay diagrams are shown in Fig.~\ref{DecayDiagrams}.

\subsection{LO calculation}

The LO decay width arises from the tree-level decay in Fig.~\ref{figX1a}, where the dotted line can either be a pion or a photon.  The results for the differential partial widths are:
\begin{equation}
    \begin{aligned}
        \frac{d\Gamma_{\rm LO}[T_{cc}^+ \rightarrow D^0D^0 \pi^+]}{d\pv_1^2 \, d\pv_2^2} = & \; \frac{g^2}{(4\pi f_\pi)^2}\frac{2\gamma_0}{3} \pv_\pi^2 \cos^2{\theta}  \bigg( \frac{1}{\pv_1^2 + \gamma_0^2} + \frac{1}{\pv_2^2 + \gamma_0^2} \bigg) ^2 \; , \\
        \frac{d\Gamma_{\rm LO}[T_{cc}^+ \rightarrow D^+D^0 \pi^0]}{d\pv_+^2 \, d\pv_0^2} = & \; \frac{g^2}{(4\pi f_\pi)^2}\frac{2}{3} \pv_\pi^2 \bigg( \frac{\sqrt{\gamma_0} \cos{\theta}}{\pv_+^2 + \gamma_0^2} - \frac{\sqrt{\gamma_+} \sin{\theta}}{\pv_0^2 + \gamma_+^2} \bigg) ^2 \; , \\
        \frac{d\Gamma_{\rm LO}[T_{cc}^+ \rightarrow D^+D^0 \gamma]}{d\pv_+^2 \, d\pv_0^2} = & \; \frac{E_\gamma^2}{6\pi^2}\bigg( \frac{\sqrt{\gamma_0} \mu_0  \cos{\theta}}{\pv_+^2 + \gamma_0^2} - \frac{\sqrt{\gamma_+} \mu_+ \sin{\theta}}{\pv_0^2 + \gamma_+^2} \bigg) ^2 \; .
    \end{aligned}
\label{eq: LO Tcc widths}
\end{equation}
Here the $\gamma_i$ are the binding momenta and the $\mu_i$ are reduced masses for a particular channel, where the subscript refers to the charm meson which is the pseudoscalar meson in the channel, e.g., $\gamma_0^2 = 2\mu_0(m_0 + m_+^* - m_T)$.  The mass $m_T$ is the mass of the $T_{cc}^+$, which is derived from the binding energy extracted by LHCb. The three-momenta are the momenta of the final state particles, where the $\pv_\pi$ is for a final state pion, $\pv_+$ is for a final state $D^+$, $\pv_0$ is for a final state $D^0$, and $\pv_1$ and $\pv_2$ are for the two final state neutral charm mesons in the first decay.  The angle $\theta$ parametrizes the isospin state of the $T_{cc}^+$; it depends on $\Sigma^\prime(-E_T)$ and for a pure isospin-0 state is $\theta \approx -32.4^\circ $.  Note that if we take the isospin limit $m_0 = m_+$, which makes $\gamma_0 = \gamma_+$ and $\cos{\theta} = -\sin{\theta} = 1/\sqrt{2}$, we get the expression for the decay width of the $\chi_{c1}(3872)$ from Ref.~\cite{Fleming:2007rp}.

\subsection{NLO calculation}

To improve the readability of this section, the expressions for the self-energy diagrams, decay diagrams, and differential decay widths to NLO are provided in full in App.~\ref{app: Tcc diagram expressions}.  Here we discuss one decay diagram, for illustrative purposes.  Consider the one-loop decay with a $C_\pi$ vertex in Fig.~\ref{figX1d}.  Let the mass of the pseudoscalar charm meson be $m$, the vector charm meson be $m^*$, and the pion by $m_\pi$.  Label the momentum of the external $D$ meson which does not have the $C_\pi$ vertex on its line as $p$ with momentum $m^\prime$.  Work in the rest frame of the $T_{cc}^+$, so that $p_T = (E_T, {\bf 0})$, and route the loop momentum $l$ clockwise through the loop.  The diagram has the following components, arising from the Feynman rules derived from the Lagrangian:
\begin{itemize}
    \item the virtual $D^*$ propagator: $\frac{i}{l_0 - \frac{\lv^2}{2m^*}-\delta^* + i \epsilon}$,
    \item the virtual $D$ propagator: $\frac{i}{-l_0 + E_T - \frac{\lv^2}{2m}-\delta+i\epsilon}$,
    \item the virtual pion propagator: $\frac{i}{(l-p)^2-m_\pi^2+i\epsilon}$
    \item the $D^*D\pi$ vertex: $(\pv-\lv)^ig_\pi$,
    \item the $C_\pi$ vertex: $iC_\pi$ 
    \item a polarization vector $\boldsymbol{\varepsilon}_T^i$ for the $T_{cc}^+$, which must contract with the $(\pv-\lv)^ig_\pi$ term
    \item an integration over the loop momentum: $\int \frac{d^d l}{(2\pi)^d}$
\end{itemize}
The appropriate $g_\pi$ and $C_\pi$ will depend on the charges of the particles.  The full integral written out is:
\begin{equation}
\begin{aligned}
    I_{C_\pi} = g_\pi C_\pi \int \frac{d^d l}{(2\pi)^d} & \frac{1}{l_0 - \frac{\lv^2}{2m^*}-\delta^* + i \epsilon} \frac{1}{-l_0 + E_T - \frac{\lv^2}{2m}-\delta+i\epsilon} \\
    & \times \frac{1}{(l-p)^2-m_\pi^2+i\epsilon} (\pv-\lv)\cdot \boldsymbol{\varepsilon}_T \; .
\end{aligned}
\end{equation}
The $l_0$ integral can be evaluated by contour, enclosing the pole at $l_0 = E_T - \frac{\lv^2}{2m}-\delta+i\epsilon$. In terms of the reduced mass $\mu$ and binding momentum $\gamma$ for the diagram, the result is:
\begin{equation}
\begin{aligned}
    I_{C_\pi} = 2i\mu g_\pi C_\pi \int \frac{d^{d-1} \lv}{(2\pi)^{d-1}} & \frac{1}{\lv^2 + \gamma^2 -i\epsilon} \\
    & \times \frac{1}{\bigg(E_T - \frac{\lv^2}{2m}-\delta-p_0\bigg)^2-(\lv-\pv)^2 - m_\pi^2+i\epsilon} (\pv-\lv)\cdot \boldsymbol{\varepsilon}_T \; .
\end{aligned}
\end{equation}
Simplifying the part of pion propagator denominator that is in large parentheses, using $E_T = m_T - 2m_0$, $\delta = m-m_0$, and $p_0 = m^\prime + \frac{\pv^2}{2m^\prime} - m_0$, we have:\footnote{Recall that the energies of the $D$ mesons are measured with respect to the $D^0$ mass $m_0$.}
\begin{equation}
\begin{aligned}
    & \; E_T - \frac{\lv^2}{2m}-\delta-p_0 \\
    = & \; m_T - 2m_0 - \frac{\lv^2}{2m} - (m-m_0) - \bigg(m^\prime + \frac{\pv^2}{2m^\prime} - m_0\bigg) \\
    = & \; m_T - m - m^\prime - \frac{\lv^2}{2m} - \frac{\pv^2}{2m^\prime}
\end{aligned}
\end{equation}
All the terms in the equation above dependent on $\lv$ and $\pv$ will be suppressed by factors of the charm meson mass compared to $(\lv-\pv)^2$, so we drop them.
\begin{equation}
\begin{aligned}
    I_{C_\pi} = 2i\mu g_\pi C_\pi \int \frac{d^{d-1} \lv}{(2\pi)^{d-1}} & \frac{1}{\lv^2 + \gamma^2 -i\epsilon} \\
    & \times \frac{1}{(\lv-\pv)^2-(m_T-m-m^\prime)^2 + m_\pi^2 - i\epsilon} (\lv-\pv)\cdot \boldsymbol{\varepsilon}_T \; .
\end{aligned}
\end{equation}
We now see that these are the basis integrals from App.~\ref{app: PDS scheme}; their evaluation is discussed there.  They are evaluated in the power divergence subtraction (PDS) scheme \cite{Kaplan:1998tg}, which makes linear divergences explicit with the use of a hard cutoff $\Lambda_{\rm PDS}$.  We get:
\begin{equation}
    I_{C_\pi} = 2i\mu g_\pi C_\pi \boldsymbol{\varepsilon}_T \cdot \pv [I^{(1)}(\pv) - I(\pv)] \; .
\end{equation}
where the parameters $c_1$, $c_2$, and $b$ (which are defined in the appendix) are
\begin{equation}
    \begin{aligned}
        c_1 = & \; \gamma^2 \; , \\
        c_2 = & \; \pv^2 - (m_T-m-m^\prime)^2 + m_\pi^2 \; , \\
        b = & \; 1 \; .
    \end{aligned}
\end{equation}
The other diagrams proceed with similar analyses.

\subsection{Uncertainties in the NLO couplings}
\label{sec: uncertainties in the nlo couplings}

Evaluating all of the self-energy and decay diagrams for use in the formula for the NLO decay width yields a lengthy answer with complicated dependence on the NLO couplings.  We discussed the determination of the couplings in XEFT in Sec.~\ref{sec: couplings in XEFT}; the analysis will be similar here.

While the couplings $C_0$, $g$, $\mu_D$, and $C_\pi$ are fixed by either experimental or lattice values, we have to estimate reasonable ranges for the other couplings to see their effects on the decay width.  The coupling $C_{0D}$ is related to the $D$ meson scattering length $a$; in XEFT the bounds on the magnitude are chosen so that $C_{0D} \sim -4\pi a/m_0$ \cite{Dai:2019hrf}, which is about $C_{0D} \sim -(1.3 \; {\rm fm})a$. Assuming the $D$ meson scattering length to be of the order its size, which we take to be no larger than the size of the proton at $\mathcal{O}({\rm fm})$, here we vary $C_{0D}$ in the range $[-1,1]$ fm$^2$, which is also what is done in the XEFT papers.  

The dependence of the decay width on the $C_2$ and $B_1$ couplings is simplified by defining the parameters $\beta_i$, defined in App.~\ref{app: beta expressions}.  To see their significance, we can take the isospin limit $m_0 = m_+$, and we see that:
\begin{equation}
    \beta_2, \beta_4 \rightarrow - \frac{4C_2^{(0)}\mu \gamma}{\pi}(\gamma-\Lambda_{\rm PDS})^2 \; ,
\label{eq: beta limit}
\end{equation}
In XEFT this term is proportional to the effective range in the isospin-0 channel \cite{Fleming:2007rp}, $\beta_2=\beta_4 = - \gamma r_0$.  We pick the value of the effective range for nucleons, $r_0 \approx 1/(100 \, \rm MeV)$, as an upper bound of that for the charm mesons.  For the binding momentum $\gamma$, we plot the results for both $\gamma = \gamma_0$ and $\gamma = \gamma_+$ in Sec.~\ref{sec: Tcc plots}.  These values are used to provide error bound estimates on $\beta_2$ and $\beta_4$.  The values for $B_1$ are then chosen to cancel the $\Lambda_{\rm PDS}$ dependence in the decay width after the $C_{2}$ have been fixed by Eq.~(\ref{eq: beta limit}).  The values for the partial widths in Table~\ref{tab:widths} are with the couplings chosen to give upper and lower bounds.

In the same isospin limit, the rest of the $\beta_i$ parameters become:
\begin{equation}
\begin{aligned}
    \beta_1 = \beta_3=\beta_5 = & \; \frac{1}{\pi}(\gamma-\Lambda_{\rm PDS})\bigg(\frac{b_0f_\pi}{g}-2C_2^{(0)}\mu\bigg) \; .
\end{aligned}
\end{equation}
This is where the dependence on $B_1$ appears, as $b_0$ is an isospin-0 simplification of $B_1$, introduced in App.~\ref{app: interaction terms}.  Here you can see that the $B_1$ couplings must be fixed to cancel the $\Lambda_{\rm PDS}$ dependence in $\beta_{1,3,5}$.  The effective range can again inform an appropriate range through which to vary these $\beta$s, however we find that the differential distributions are negligibly affected in the range, so we will not show plots regarding them.

\subsection{Results for partial widths}

\def\arraystretch{1.5}
\begin{table*}
\centering
\caption{\label{tab:widths}$T_{cc}^+$ partial widths to NLO. The central value is the LO result, and the error bars give bounds to the NLO corrections.}
\begin{tabular}{|c|c|c|c|}
\hline
& Partial width (keV)  \\ \hline
$\Gamma[T_{cc}^+\to D^0 D^0 \pi^+]$ & $28_{-7}^{+16}$  \\ \hline
$\Gamma[T_{cc}^+\to D^+ D^0 \pi^0]$ & $13_{-5.2}^{+8}$  \\ \hline
$\Gamma_{\rm strong}[T_{cc}^+]$ & $41_{-12}^{+25}$  \\ \hline
$\Gamma_{\rm strong}[T_{cc}^+]+\Gamma_{\rm EM}^{\rm LO}[T_{cc}^+]$ & $47_{-12}^{+25}$  \\ \hline
\end{tabular}
\end{table*}

We can integrate the LO and NLO differential distributions over phase space to get the partial widths (Table \ref{tab:widths}).  The total predicted LO decay width of $47$ keV is already in excellent agreement with the experimental value $\Gamma_{\rm u} = 48 \pm 2^{+0}_{-14}$ keV.  There are no free parameters at LO; the coupling $C_0$ is merely tuned so that the two-point function $G_0$ has a pole at the location of the bound state, and the rest of the couplings are fixed from $D^*$ decays.  The NLO contributions can add significant uncertainty, yielding $\Gamma = 47^{+53\%}_{-25\%}$ keV, which arises due to the uncertainties in the NLO couplings.

\subsection{Decay to another bound state}
\label{sec: decay to another bound state}

\begin{figure*}[t]
\centering
\includegraphics[scale=0.7]{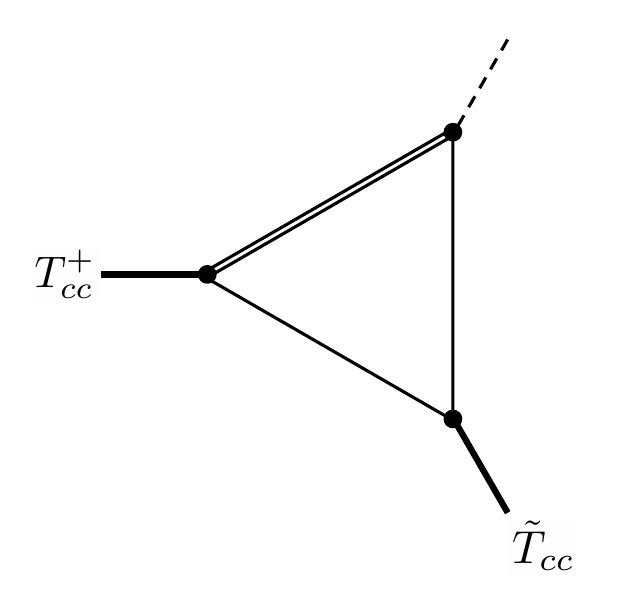}
\caption[Triangle diagram for the decay of the $T_{cc}^+$ to a final state with a $\tilde{T}_{cc}$.]{Triangle diagram for the decay of the $T_{cc}^+$ to a final state with a $\tilde{T}_{cc}$.  Figure from Ref.~\cite{Fleming:2021wmk}.}
\label{fig: decay to another state}
\end{figure*}

We mentioned in Sec.~\ref{sec: remarks on Tcc fits} that the initial $T_{cc}^+$ decay width quoted by the LHCb collaboration ($\Gamma_{\rm BW}$) was orders of magnitude larger than the value obtained by the early theory calculations, including our initial value of $\Gamma=52$ keV.  To attempt to provide an explanation for this discrepancy, in Ref.~\cite{Fleming:2021wmk} we explored the possibility of a decay of the $T_{cc}^+$ to a hypothetical bound state $\tilde{T}_{cc}$ of two pseudoscalar $D$ mesons, either $D^0D^0$ or $D^+D^0$.  Predictions for such $I=1$, $J^P=0^+$ doubly-charm tetraquarks are anywhere from a few to a few hundred MeV above the $DD$ threshold \cite{LHCb:2021vvq,Tan_2020,PhysRevD.80.114023}, and given uncertainties it is reasonable to investigate the existence of these $\tilde{T}_{cc}$ states.   

\begin{figure*}[t]
\centering
\begin{minipage}{0.5\textwidth}
\centering
\includegraphics[width = 0.95 \textwidth]{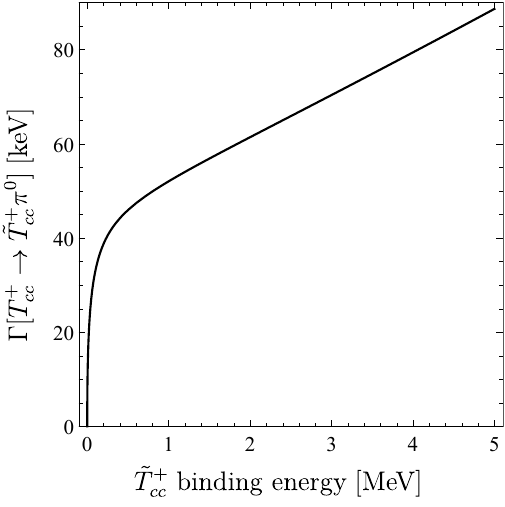}
\label{LoopPi01}
\end{minipage}%
\begin{minipage}{0.5\textwidth}
\centering
\includegraphics[width = 0.95 \textwidth]{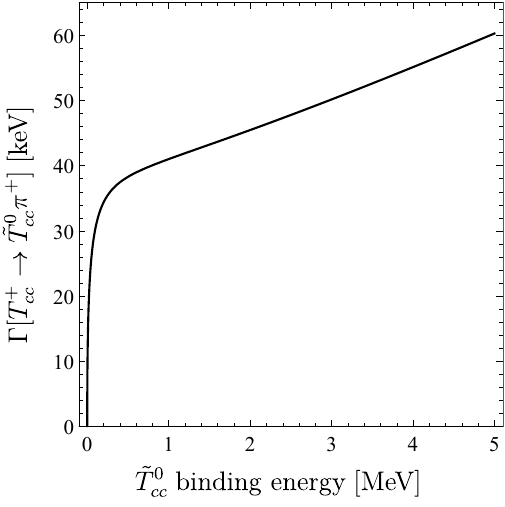}
\label{LoopPiPlus1}
\end{minipage}%
\\
\begin{minipage}{0.5\textwidth}
\centering
\includegraphics[width = 0.95 \textwidth]{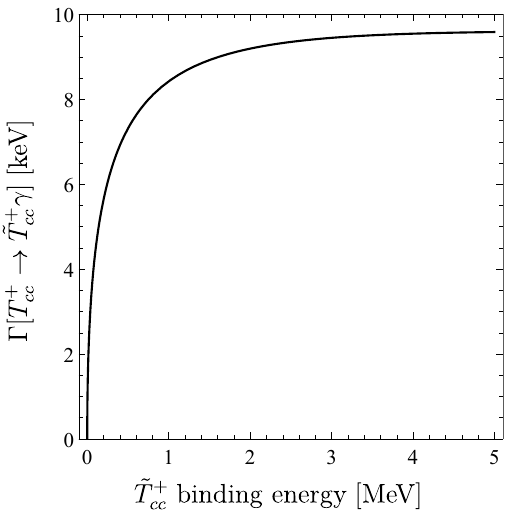}
\label{LoopPhoton1}
\end{minipage}
\caption[Partial widths for the $T_{cc}^+$ decay to a $\tilde{T}_{cc}$ and a pion or photon.]{Partial widths for the $T_{cc}^+$ decay to a $\tilde{T}_{cc}$ and a pion or photon.  Figures from Ref.~\cite{Fleming:2021wmk}.}
\label{fig: plots for another bound state}
\end{figure*}

We consider three decays at one-loop level: $T_{cc}^+ \rightarrow \tilde{T}_{cc}^+ \pi^0$,  $\tilde{T}_{cc}^0 \pi^+$, and $\tilde{T}_{cc}^+ \gamma$.  The corresponding triangle diagram is shown in Fig.~\ref{fig: decay to another state}.  The partial widths can be written in terms of an integral of the form of $I(\pv)$ from App.~\ref{app: PDS scheme}; we quote the result here with a parametrization specific to these decays.
\begin{equation}
    \begin{aligned}
        F(m_1,m_2,m_3,m_{\tilde{T}}) = & \; \sqrt{\frac{\gamma_{12}\gamma_{13}}{b^2}}\bigg[ \tan^{-1}\bigg(\frac{c_2-c_1}{2\sqrt{c_2b^2{\bf p}_{\pi}^2}}\bigg) \\
        & + \tan^{-1} \bigg(\frac{2b^2{\bf p}_\pi^2+c_1-c_2}{2\sqrt{b^2{\bf p}_\pi^2(c_2-b^2{\bf p}_\pi^2)}}\bigg)\bigg] \;,
    \end{aligned}
\end{equation}
where the parameters are:
\begin{equation}
    \begin{aligned}
        \gamma_{12} &= \; \sqrt{-2\mu_{12}(m_T-m_1-m_2)}, \quad
        \gamma_{13} =\sqrt{-2\mu_{13}(E_{\tilde{T}}-m_1-m_3)}, \\
        \mu_{ij}^{-1} &= \; m_i^{-1}+m_j^{-1}, \quad
        c_1 = \gamma^2_{12}, \quad
        c_2 = \frac{\mu_{13}}{m_3}{\bf p}_\pi^2+ \gamma_{13}^2,  \quad
        b =\mu_{13}/m_3 \; . \quad
    \end{aligned}
\end{equation}
In terms of the unknown mass $m_{\tilde{T}}$ of the hypothetical $\tilde{T}_{cc}$ state, the partial widths are then:
\begin{equation}
    \begin{aligned}
        \Gamma[T_{cc}^+ \rightarrow \tilde{T}_{cc}^+ \pi^0] =& \; 
        \frac{| {\bf p}_\pi|m_{\tilde{T}}}{6 \pi m_T}
        \bigg(\frac{g}{\sqrt{2}f_\pi}\bigg)^2
        \big|\cos{\theta} \, F(m_0,m_{D^{*+}},m_+,m_{\tilde{T}}) \\
        & - \sin{\theta} \, F(m_+,m_0^*,m_0,m_{\tilde{T}})\big|^2 \; , \\
        \Gamma[T_{cc}^+ \rightarrow \tilde{T}_{cc}^0 \pi^+] =& \; \frac{| {\bf p}_\pi|m_{\tilde{T}}}{6 \pi m_T}\bigg(\frac{g}{f_\pi}\bigg)^2\big|\cos{\theta} \, F(m_0,m_{D^{*+}},m_0,m_{\tilde{T}})\big|^2 \; , \\
        \Gamma[T_{cc}^+ \rightarrow \tilde{T}_{cc}^+ \gamma] =& \; 
        \frac{| {\bf p}_\gamma|m_{\tilde{T}}}{3 \pi m_T}
        \big|\mu_+\cos{\theta} \, F(m_0,m_{D^{*+}},m_+,m_{\tilde{T}}) \\
        & + \mu_0\sin{\theta} \, F(m_+,m_0^*,m_0,m_{\tilde{T}})\big|^2 \; .
    \end{aligned}
\end{equation}
We can then plot these partial widths in Fig.~\ref{fig: plots for another bound state} as a function of the $\tilde{T}_{cc}$ binding energy from $0$ to $5$ MeV.  The pion decays have widths that increase linearly for most of the domain, and the photon decay has a width that approaches $10$ keV as the binding energy increases.  In all, these processes could increase the total width of the $T_{cc}^+$ by as much as $150$ keV for a $\tilde{T}_{cc}$ binding energy of $5$ MeV, and by about $80$ keV if the binding energy is the same as that of the $T_{cc}^+$.  This is far above the experimental value $\Gamma_{\rm u} \approx 48$ keV; this is evidence that the $\tilde{T}_{cc}$ bound states do not exist.

\section{Differential decay distributions}
\label{sec: Tcc plots}

\begin{figure*}[t]
\centering
\includegraphics[width=\textwidth]{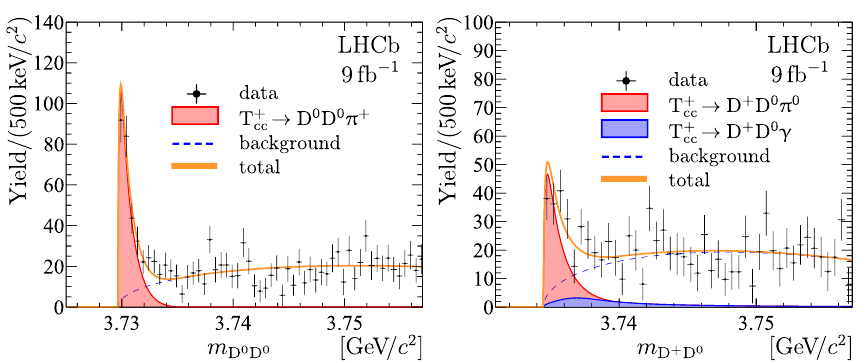}
\caption[LHCb fits to data for the total yield as a function of the invariant mass of the final state charm meson pair.]{LHCb fits to data for the total yield as a function of the invariant mass of the final state charm meson pair.  Figure from Ref.~\cite{LHCb:2021auc}.}
\label{fig: lhcb two plots}
\end{figure*}

In this section we plot the differential distribution $d\Gamma / dm_{DD}$, where $m_{DD}$ is the invariant mass of the charm mesons in the final state, $m_{DD} = (p_1 + p_2)^2$, and compare to the LHCb experimental data for the total yield, which is shown in Fig.~\ref{fig: lhcb two plots} for comparison.\footnote{Thanks to M.~Mikhasenko for providing the data for use in our plots.}  The normalizations of the curves are fixed by performing a least-squares fit of the LO distribution to the data.

In the plots, the solid lines represent the LO curves, and dotted or dashed lines represent bounds on the NLO contributions.  The LHCb data is overlaid with the background subtracted.  We show three plots that include a subset of the NLO terms:
\begin{itemize}
    \item Figure \ref{ContributionsNAandSE}: includes non-$C_2$-dependent NLO self-energy corrections (the first diagram on the second line of Fig.~\ref{bubbleDiagram}), and decays like Fig.~\ref{figX1b}.  These increase the magnitude of the differential distribution by a small, but noticeable amount.
    \item Figure \ref{ContributionsC0D}: the contributions from $C_{0D}$ are shown, varying it in two ranges: $[-1,1]$ fm$^2$ and $[-0.25,0.25]$ fm$^2$.  The change in the neutral pion decay is twice that of the charged pion decay, due to the coupling of charged pions to charm mesons being larger by a factor of $\sqrt{2}$.  The magnitude of the peak is sensitive to the couplings magnitude; for $|C_{0D}| = 1$ fm$^2$ the peak is either too high or too low for the neutral pion decay.
    \item Figure \ref{ContributionsBeta}: the parameters $\beta_2$ and $\beta_4$ are varied together so as to maximize the effect on the distribution.  The magnitudes of the peaks are even more sensitive to these; for $\beta_2=\beta_4 = -0.59$ the peaks are far too high.  Clearly the appropriate choice for the binding momentum and the effective range are important for obtaining reasonable values for these parameters.  Both choices for $\beta_2=\beta_4$ serve to increase the partial widths.
\end{itemize}
The $C_\pi$ diagrams, as well as terms proportional to $\beta_1$, $\beta_3$, and $\beta_5$, contribute negligibly to the distributions, so we do not plot their contribution individually.

The total NLO contributions to the differential distribution are plotted in Fig.~\ref{ContributionsAll}.  The parameters are varied between $-1 \; {\rm fm}^2 \le C_{0D} \le 0.25 \; {\rm fm}^2$ and $-0.25 \le {\beta_2,\beta_4} \le 0$.  An important observation is that while the magnitude of the distributions can vary significantly with the choices for the NLO couplings, the qualitative nature of the plots remains the same: the peak stays at the same invariant mass, and the charged pion decay remains larger than the neutral pion decay.  As in XEFT, the shape of the differential distributions are a result of the amplitudes being dependent on a virtual $D^*$ propagator, $1/(\pv^2 + \gamma^2)$.  The binding momentum is a small number, and so the distributions are sharply peaked at small $\pv$.  If we were to instead replace the propagator with a constant, the curves would be much flatter (Fig.~\ref{constantProp}) and would not be in good agreement with the LHCb data.  

An interesting future direction would be to perform a careful fit of the experimental data to obtain values for the NLO couplings, since there is not much known about several of them at this time.  This would increase the predictive power of the EFT, putting tighter constraints on the couplings.  Such an analysis could be further improved by having even more precise experimental data to which to fit.

\begin{landscape}

\begin{figure*}[t]
\centering
\includegraphics[width=0.9\linewidth]{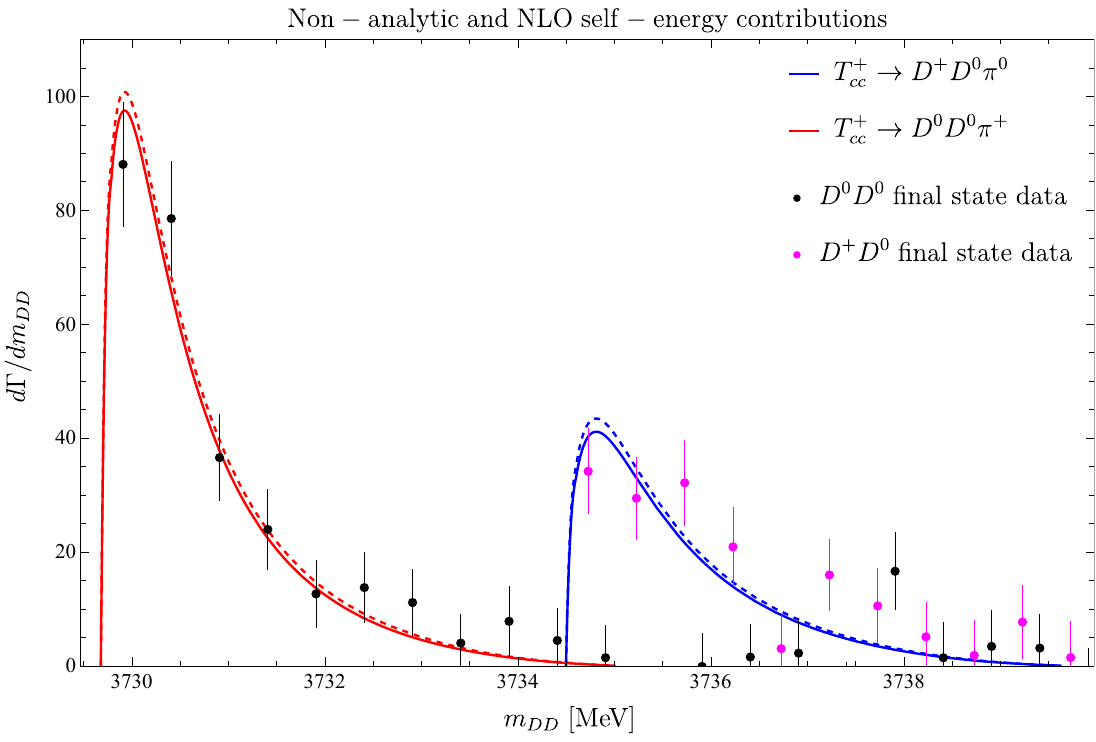}
\caption[Differential decay width at LO, and NLO contributions from non-$C_2$-dependent self-energy terms and Fig.~\ref{figX1b}.]{Differential decay width at LO (solid lines), and NLO contributions from non-$C_2$-dependent self-energy terms and Fig.~\ref{figX1b}.  Figure from Ref.~\cite{Dai:2023mxm}.}
\label{ContributionsNAandSE}
\end{figure*}

\begin{figure*}[t]
\centering
\includegraphics[width=0.9\linewidth]{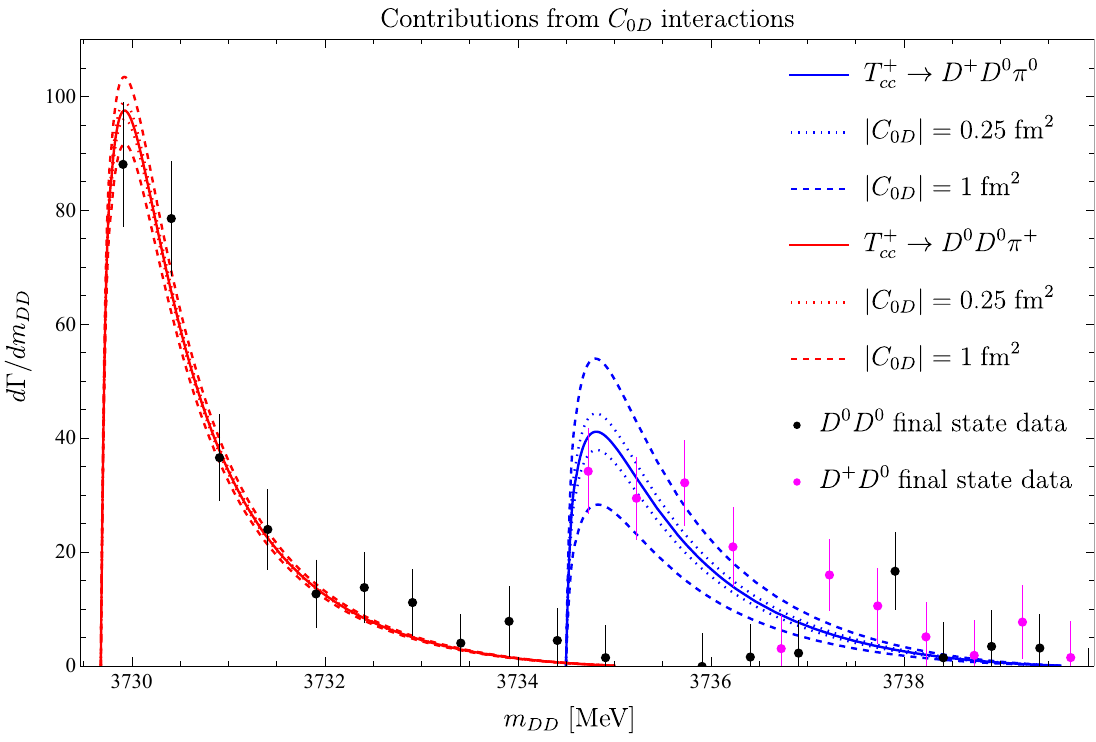}
\caption[Differential decay width at LO, and NLO contributions from $C_{0D}$.]{Differential decay width at LO (solid lines), and NLO contributions from $C_{0D}$. The legend shows the absolute value of $C_{0D}$; the positive values increase the width and the negative values decrease it.  Figure from Ref.~\cite{Dai:2023mxm}.}
\label{ContributionsC0D}
\end{figure*}

\begin{figure*}[t]
\centering
\includegraphics[width=0.9\linewidth]{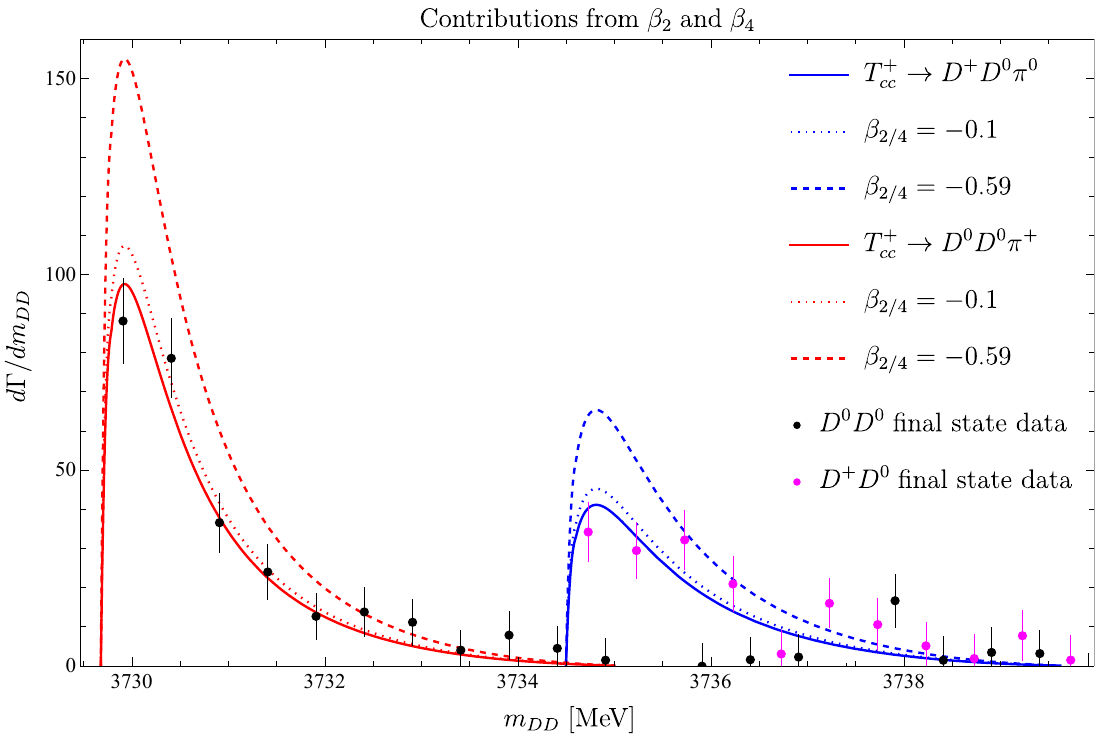}
\caption[Differential decay width at LO, and NLO contributions from the parameters $\beta_{2}$ and $\beta_4$.]{Differential decay width at LO (solid lines), and NLO contributions from the parameters $\beta_{2}$ and $\beta_4$, which are set to be equal in these plots.  Figure from Ref.~\cite{Dai:2023mxm}.}
\label{ContributionsBeta}
\end{figure*}

\begin{figure*}[t]
\centering
\includegraphics[width=0.9\linewidth]{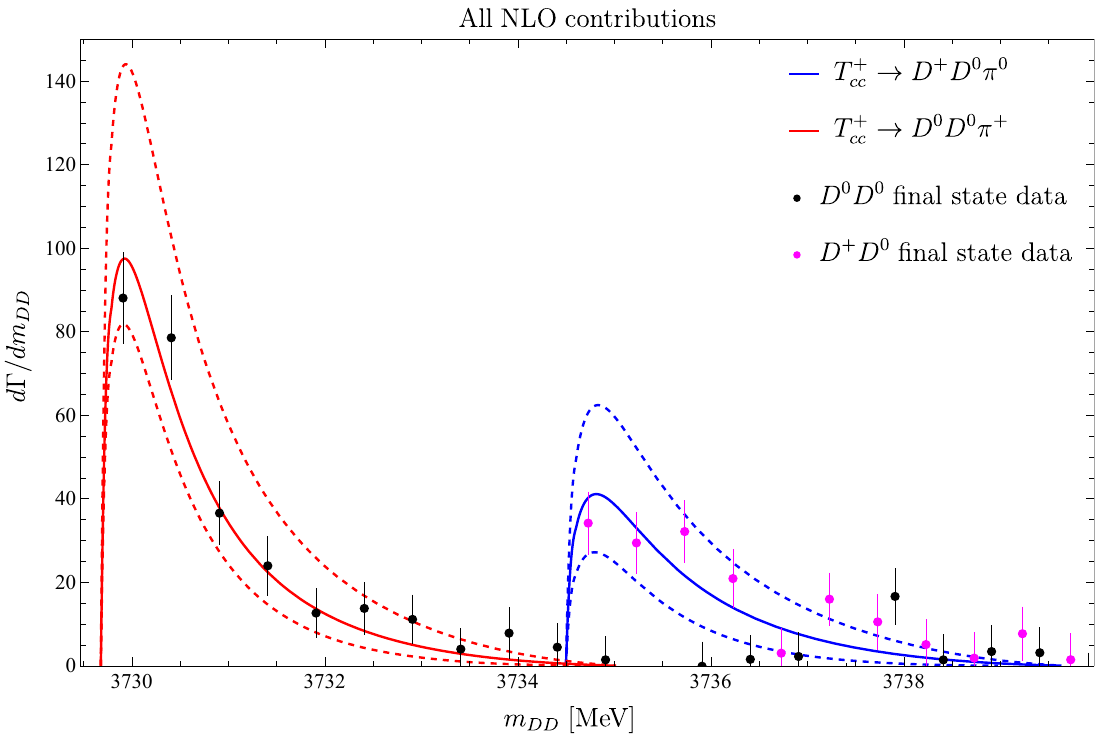}
\caption[Differential decay width at LO, and all NLO contributions with $-1 \; {\rm fm}^2 \leq C_{0D} \leq 0.25 \; {\rm fm}^2$ and $-0.26\leq \beta_{2/4} \leq 0$.]{Differential decay width at LO (solid lines), and all NLO contributions with $-1 \; {\rm fm}^2 \leq C_{0D} \leq 0.25 \; {\rm fm}^2$ and $-0.26\leq \beta_{2/4} \leq 0$. Figure from Ref.~\cite{Dai:2023mxm}.}
\label{ContributionsAll}
\end{figure*}

\begin{figure*}[t]
\centering
\includegraphics[width=0.9\linewidth]{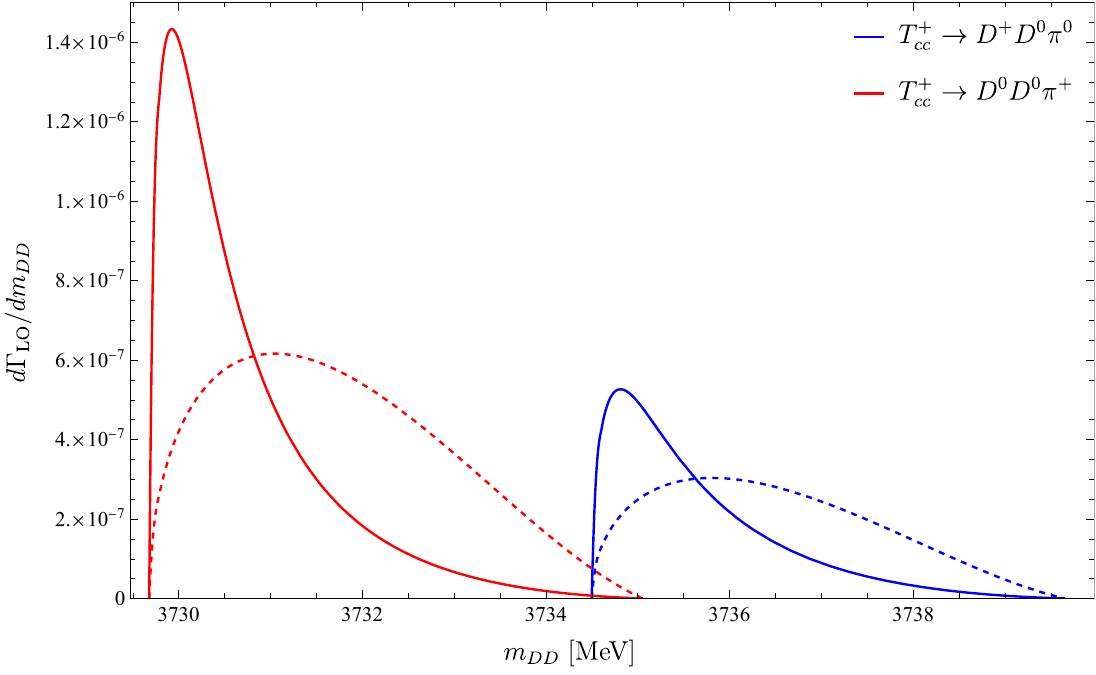}
\caption[Comparing the LO differential decay width to one with constant virtual $D^*$ propagators.]{Comparing the LO differential decay width to one with constant virtual $D^*$ propagators.  The normalization is fixed to give the curves the same partial width. Figure from Ref.~\cite{Dai:2023mxm}.}
\label{constantProp}
\end{figure*}

\end{landscape}
\chapter{Spin-dependent TMD fragmentation functions for $J/\psi$ production}
\label{chap:FFs}

In\footnote{The work presented in this chapter was initially published in Ref.~\cite{Copeland:2023wbu}. The contributions of each author are listed below. \begin{itemize}
    \item R.~Hodges, M.~Copeland, S.~Fleming: analysis and writing
    \item R.~Gupta: analysis
    \item T.~Mehen: checking calculations and editing manuscript
\end{itemize}} this chapter we derive the spin-dependent TMD fragmentation functions (FFs) for $J/\psi$ production in the framework of NRQCD.  These FFs are fundamental objects describing the hadronization of a hard parton, and can be utilized in different scattering processes.  Before discussing the FFs themselves, we review the properties of the $J/\psi$, as well as the theoretical background for the production mechanism we consider, semi-inclusive deep inelastic scattering. 

\section{$J/\psi$ meson}

\begin{figure*}[t]
\centering
\includegraphics[width=0.7\textwidth]{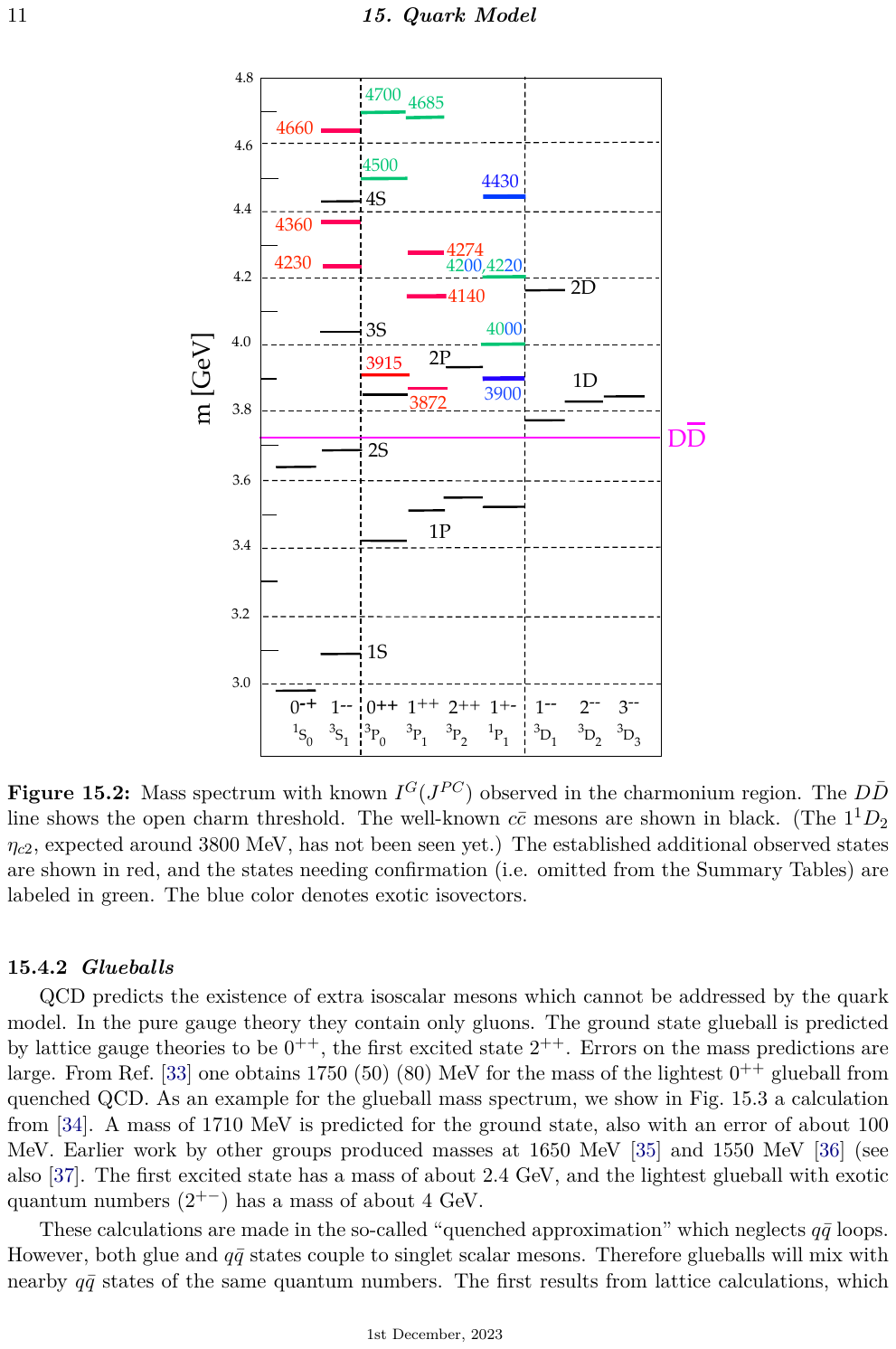}
\caption[Mass spectrum of charmonium states.]{Mass spectrum of charmonium states.  Black indicates conventional quarkonium states, red indicates exotic isoscalars, blue indicates exotic isovectors, and green indicates candidates for exotic states.  The $D\bar{D}$ threshold is shown in pink. Figure from Ref.~\cite{ParticleDataGroup:2022pth}.}
\label{fig: charmonium spectrum}
\end{figure*}

The $J/\psi$ meson was discovered independently by Ting et al.~\cite{E598:1974sol} at Brookhaven National Lab (BNL) and Richter et al.~\cite{SLAC-SP-017:1974ind} at the Stanford Linear Accelerator Center (SLAC) in 1974.   The discovery confirmed the existence of the charm quark, posited by Bj{\o}rken and Glashow \cite{Bjorken:1964gz} in 1964. The $J/\psi$ was the first of so-called charmonium states to be observed.  The charmonia are bound states of charm-anticharm with different quantum numbers $^{2S+1}L_J$, where $S$ is the spin, $L$ is the orbital angular momentum, and $J$ is the total angular momentum.  They can also be described in terms of $J^{PC}$, for parity $P=(-1)^{L+1}$ and charge conjugation parity $C=(-1)^{L+S}$.  The $J/\psi$ has quantum numbers $^3S_1$ and $1^{--}$ and a mass of $3.1$ GeV.  Many other charmonium states have been measured as well (Fig.~\ref{fig: charmonium spectrum}). 

Consider the $J/\psi$ decay modes in Table \ref{tab: J psi decays}.  The strong decays, in particular the $3g$ decay, are more predominant than the leptonic deacys.  The relative proportion of $3g$ decays to leptonic decays can be understood by looking at the powers in the couplings: $3g$ decays will go as $\alpha_s^3 \sim 10^{-3}$, whereas the leptonic decays will go as $\alpha_{\rm em}^2 \sim 10^{-4}$.  However, the lepton decays have considerably higher branching fractions than for other hadrons, for example, corresponding branching fractions for the neutral rho meson $\rho^0$ are smaller by a factor of $10^{-5}$ \cite{ParticleDataGroup:2022pth}. Indeed, the initial BNL discovery was in decays to $e^+e^-$, and the SLAC discovery in decays to $\mu^+\mu^-$.  This can be attributed to the fact that the $J/\psi$ lies below the $D\bar{D}$ threshold of about $m_{D\bar{D}}=3.74$ GeV, so there are fewer hadronic decays that are kinematically allowed.  

The relatively high proportion of leptonic decays makes the $J/\psi$ an attractive subject for experiment; muons in particular are readily measured, since they are charged particles that are not absorbed as often as electrons, and obviously do not lead to hadronization, so their momentum can be resolved without many other complications.  It is important to have theoretical predictions for $J/\psi$ production to go along with experimental observations.  In this work we consider $J/\psi$ production via a process called semi-inclusive deep-inelastic scattering.

\begin{table}[t]
\renewcommand{\arraystretch}{1.5}
\centering
\caption[$J/\psi$ decay modes from the Particle Data Group.]{$J/\psi$ decay modes from the Particle Data Group \cite{ParticleDataGroup:2022pth}.}
\label{tab: J psi decays}
\begin{tabular}{|c|c|}
\hline
Mode & Fraction \\ \hline \hline
$J/\psi \rightarrow 3g$ & $(64.1 \pm 1.0)\%$ \\ \hline
$J/\psi \rightarrow \gamma gg$& $(8.8 \pm 1.1)\%$ \\ \hline
$J/\psi \rightarrow \gamma^* \rightarrow \text{hadrons}$& $(13.50 \pm 0.30)\%$ \\ \hline
$J/\psi \rightarrow \gamma^* \rightarrow e^+ e^-$ & $(5.971 \pm 0.032)\%$ \\ \hline
$J/\psi \rightarrow \gamma^* \rightarrow \mu^+ \mu^-$ & $(5.961 \pm 0.033)\%$ \\ \hline
\end{tabular}
\end{table}

\section{SIDIS and factorization}
\label{sec: sidis and factorization}

In Chap.~\ref{chap:intro}, we discussed deep inelastic scattering (DIS), $e(l) + h(p) \rightarrow e^\prime(l^\prime) + X(p_X)$, where an electron $e$ scatters off of a hadron $h$ (usually a proton), which hadronizes to a final state $X$.  The momentum transfer of the electron to the proton is parametrized by the virtuality of the exchanged photon, $q^2 = (l^\prime-l)^2 = -Q^2$.  DIS is best described in a frame where the proton moves fast, and so the partons move in a mostly parallel direction with momentum $k \approx \xi p \sim Q$ \cite{Boussarie:2023izj}.  The transverse momentum is small, $k_T \ll Q$.  The DIS cross section can be approximated using QCD factorization:
\begin{equation}
    \begin{aligned}
        E^\prime \frac{d\sigma_{eh\rightarrow e^\prime X}}{d^3 l^\prime} \approx \sum_i \int d\xi \; f_{i/h}(\xi) \; E^\prime \frac{d\hat{\sigma}_{ei \rightarrow e^\prime X}}{d^3 l^\prime} \equiv \sum_i f_{i/h} \otimes \hat{\sigma}_i \; .
    \end{aligned}
    \label{eq: DISfact}
\end{equation}
Here, the full cross section $\sigma$ is factorized into a product of a probability distribution $f_{i/h}(\xi)$, which describes the probability density to find a parton of type $i$ with momentum fraction $\xi$ inside the hadron, and a partonic cross section $\hat{\sigma}$, which is the cross section for the corresponding partonic process.  The distribution $f_{i/h}$ is called a parton distribution function (PDF), and is extracted from experiment.  

\begin{figure*}
\centering
    \includegraphics[scale=1.3]{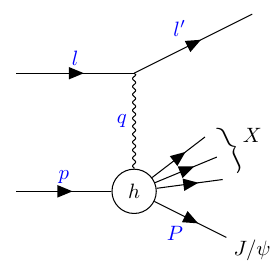}
    \caption[Diagram for semi-inclusive deep inelastic scattering.]{Diagram for semi-inclusive deep inelastic scattering. Figure from Ref.~\cite{Copeland:2023wbu}.}
\end{figure*}

Consider now that we will measure a hadron $H$ of momentum $P$ in the final state along with the electron; this is semi-inclusive DIS (SIDIS).  We can define the kinematic variables
\begin{equation}
    x = \frac{Q^2}{2p\cdot q} \; , \qquad y = \frac{p\cdot q}{p\cdot l} \; , \qquad z = \frac{p\cdot P}{p\cdot q} \; .
\end{equation}
The parameter $x$ is the usual DIS Bjorken scaling variable, and in the proton rest frame, $y$ is the fractional energy loss of the electron, and $z$ is the ratio of the energy of the hadron to that of the photon \cite{Boussarie:2023izj}.  At leading order, we can glean the significance of Bjorken $x$ by considering the $\mathcal{O}(\alpha_{\rm em})$ partonic process where the photon is absorbed by a quasi-free quark inside the proton of mass $m_q$ and initial momentum $k$.  We then have
\begin{equation}
    m_q^2 + 2k \cdot q + q^2 = m_q^2  \quad \Rightarrow  \quad 1=\frac{Q^2}{2k \cdot q} \approx \frac{Q^2}{2\xi p \cdot Q} \; .
\label{eq: mom conservation parton}
\end{equation}
This implies that at LO, $x\approx\xi$, i.e., Bjorken $x$ is the fraction of the proton's momentum carried by the parton.  This need not be true at higher orders where more particles are involved in the partonic process and thus Eq.~(\ref{eq: mom conservation parton}) does not hold.

The cross section for SIDIS still factorizes: \cite{Collins:2011zzd,Echevarria:2011epo,Echevarria:2012js,Chiu:2012ir}
\begin{equation}
    \frac{d\sigma}{dx \, dy \, dz} = \sum_{i,j} \int_x^1 d\xi \int_{z}^1 d\zeta \; f_{i/h}(\xi) \, D_{H/j}(\zeta) \, \frac{d\hat{\sigma}_{ij}(\xi, \zeta)}{dx \, dy \, dz} \bigg[1 + \mathcal{O}\bigg(\frac{\Lambda_{\rm QCD}^2}{Q^2}\bigg) \bigg]
\end{equation}
Here the partonic process is for an electron and a parton of type $i$ going to an electron and a parton of type $j$: $e^-(l) + i(k) \rightarrow e^-(l^\prime) + j(k^\prime) + X$. The parameter $\xi$ is still the fraction of the hadron $h$'s longitudinal momentum carried by the initial parton, and $\zeta$ is the fraction of the scattered parton's momentum carried by the hadron $H$, $k^\prime = P/\zeta$.  The function $D_{H/j}$ is a fragmentation function (FF), and it describes the likelihood for a parton of type $j$ to fragment to a hadron $H$.  

If we also want to be sensitive to a small transverse momentum of the final state hadron, $\Lambda_{\rm QCD} \lesssim P_T \ll Q$, then we get the transverse-momentum-dependent (TMD) version of the SIDIS cross section: \cite{Boussarie:2023izj}
\begin{equation}
    \begin{aligned}
        \frac{d\sigma}{dx \, dy \, dz \, d^2 \Pv_T} = & \; \sum_i d\hat{\sigma}_{ii}^{\rm TMD}(Q,x,y) \int d^2 \pv_T \, d^2 \kv_T \; \delta^{(2)}(\Pv_T - z \kv_T - \pv_T) \\
        & \; \times f_{i/h}(x, \kv_T) \, D_{H/i}(z, \pv_T) \, \bigg[1 + \mathcal{O}\bigg(\frac{P_T^2}{Q^2},\frac{\Lambda_{\rm QCD}^2}{Q^2}\bigg) \bigg]
    \end{aligned}
\label{eq: TMD fact}
\end{equation}
The integrations over $\xi$ and $\zeta$ have been performed, and the parton is now the same type after exchanging a photon with the electron \cite{Boussarie:2023izj}.  The procedure to obtain a SIDIS cross section, then, is to perturbatively calculate the partonic cross section $\hat{\sigma}^{\rm TMD}$, and utilize it with expressions for the TMD PDF and FF in Eq.~(\ref{eq: TMD fact}).  In practice, we can use experimental fits for the PDFs, and for hadrons with light quarks the FFs must also be determined experimentally.  However, for quarkonium, the FFs can be calculated.  We will use nonrelativistic QCD to do this.

\section{Non-relativistic quantum chromodynamics}

Non-relativistic quantum chromodynamics (NRQCD) is an effective field theory developed by Bodwin, Braaten, and Lepage \cite{Bodwin:1994jh}, to provide rigorous predictions for quarkonium decay and production cross sections.  The EFT has a power counting in $v\ll 1$, the relative velocity of the heavy quarks in the bound state.  For charmonium, $v^2 \approx 0.3$.  There are three energy scales involving this velocity that are relevant: the heavy quark mass $m_Q$, its three-momentum $m_Qv$, and its kinetic energy $m_Qv^2$.  The size of the quarkonium system is of the order $1/m_Qv$, and its binding energy is of order $m_Qv^2$.  The QCD scale $\Lambda_{\rm QCD}$ is taken to be on the order of $m_Qv^2$.  

The NRQCD of Bodwin, Braaten, and Lepage separates the physics of the large scale $m_Q$ from the smaller scales $m_Qv$, $m_Qv^2$, and $\Lambda_{\rm QCD}$.  As is standard in EFT, the effects of high energy are incorporated into the Wilson coefficients of the operators; since $\alpha_s(m_Q) \approx 0.24 \ll 1$ for charmonium, the coefficients can be calculated perturbatively.  The operators themselves scale with $v$, and so the cross sections are double expansions in $\alpha_s$ and $v$. 

\subsection{Lagrangian and power counting}

The NRQCD Lagrangian can be derived by starting with full QCD and diagonalizing the theory with the transformation $Q \rightarrow e^{-im_Q t}(\psi \quad \chi)^T$, to decouple the heavy quark and antiquark fields.\footnote{This can also be done with a Foldy-Wouthuysen-Tani transformation \cite{Tani:1951trl,Foldy:1949wa}, $Q\rightarrow \exp{(-i\boldsymbol{\gamma}\cdot {\bf D}/2m_Q)}Q$ \cite{Braaten:1996ix}.}
\begin{equation}
    \begin{aligned}
        \mathcal{L} = & \; \bar{Q}(i\slashed{D}-m_Q)Q \\
        \rightarrow & \; \psi^\dagger i D_0 \psi + \chi^\dagger(iD_0 + 2m_Q)\chi - \psi^\dagger i \boldsymbol{\sigma}\cdot {\bf D}\chi - \chi^\dagger i \boldsymbol{\sigma} \cdot {\bf D}\psi \; .
    \end{aligned}
\end{equation}
The equation of motion for the antiquark field is:
\begin{equation}
    (iD_0 + 2m_Q)\chi = i \boldsymbol{\sigma}\cdot {\bf D}\psi \; ,
\end{equation}
which can be solved for $\chi$ to integrate out the antiquark field from the Lagrangian.
\begin{equation}
    \mathcal{L} = \psi^\dagger iD_0 \psi - \psi^\dagger i \boldsymbol{\sigma} \cdot {\bf D} \frac{1}{iD_0 + 2m_Q}i \boldsymbol{\sigma}\cdot {\bf D}\psi \; .
\end{equation}
By assumption the heavy quark mass is large, so we can expand in powers of $1/m_Q$.  
\begin{equation}
    \mathcal{L} = \psi^\dagger \bigg(iD_0 + \frac{{\bf D}^2}{2m_Q}\bigg) \psi + \mathcal{O}\bigg(\frac{1}{m_Q^2}\bigg) \; .
\end{equation}
Doing this procedure with the equations of motion for the quark field $\psi$ would give an analogous kinetic term for the antiquarks, leaving us with:
\begin{equation}
    \mathcal{L}_{\rm heavy} = \psi^\dagger\bigg(iD_0 + \frac{{\bf D}^2}{2m_Q}\bigg)\psi + \chi^\dagger \bigg(iD_0 - \frac{{\bf D}^2}{2m_Q}\bigg) \chi \; .
\label{eq: NRQCD kinetic term}
\end{equation}
The light degrees of freedom are described by full QCD in the massless limit:
\begin{equation}
    \mathcal{L}_{\rm light} = -\frac{1}{2}{\rm tr}\, F_{\mu \nu}F^{\mu \nu} + \sum_q \bar{q} i \slashed{D} q \; .
\end{equation}
The interaction terms relevant to quarkonium production are: \cite{Bodwin:1994jh}
\begin{equation}
    \begin{aligned}
        \mathcal{L}_{\rm int} = & \; \frac{c_1}{8m_Q^3} \big[ \psi^\dagger ({\bf D}^2)^2 \psi - (\psi \Rightarrow \chi) \big] \\
        & \; + \frac{c_2}{8m_Q^2} \big[ \psi^\dagger({\bf D} \cdot g {\bf E} - g {\bf E}\cdot {\bf D})\psi + (\psi \Rightarrow \chi) ] \\
        & \; + \frac{c_3}{8m_Q^2} \big[ \psi^\dagger (i {\bf D}\times g {\bf E} - g {\bf E} \times i {\bf D}\cdot \boldsymbol{\sigma}\psi + (\psi \Rightarrow \chi) \big] \\
        & \; + \frac{c_4}{2m_Q}\big[\psi^\dagger(g{\bf B}\cdot \boldsymbol{\sigma})\psi - (\psi \Rightarrow \chi) \big] \; .
    \end{aligned}
\end{equation}
The chromomagnetic and chromoelectric fields are $E^i = F^{0i}$ and $B^i = \frac{1}{2}\epsilon^{ijk}F^{jk}$.  

We can infer the $v$ scaling of the operators using several observations: \cite{Lepage:1992tx,Braaten:1996ix}
\begin{itemize}
    \item the normalization of a quarkonium bound state $\ket{H}$ is unity,
    \item the expectation value of the quark number operator, $N = \int d^3x \, \psi^\dagger \psi$, for the quarkonium bound state is $\mathcal{O}(1)$,
    \item the size of the bound state is $\int d^3 x \sim (m_Qv)^{-3}$.
\end{itemize}
From these we can immediately deduce that the quark field scales as $\psi \sim (m_Qv)^{3/2}$.  The kinetic term of the Lagrangian should scale as $m_Qv^2$, so from that we know $D_0 \sim m_Qv^2$ and ${\bf D} \sim m_Qv$.  The latter aligns with the intuition that ${\bf D}$ should be associated with the three-momentum.  We summarize these scalings, along with the fields associated with the gluons, in Table \ref{tab: v scaling}.

\begin{table}[t]
\renewcommand{\arraystretch}{1.5}
\centering
\caption{Power counting for the 
NRQCD fields.}
\label{tab: v scaling}
\begin{tabular}{|c|c|}
\hline
Operator & Velocity scaling \\ \hline \hline
$\psi$, $\chi$ & $(m_Qv)^{3/2}$ \\ \hline
$D_0$ & $m_Qv^2$ \\ \hline
${\bf D}$ & $m_Qv$ \\ \hline
$g {\bf E}$ & $m_Q^2 v^3$ \\ \hline
$g {\bf B}$ & $m_Q^2 v^4$ \\ \hline
$g A_0$ (Coulomb gauge) & $m_Q^2 v^2$ \\ \hline
$g {\bf A}$ & $m_Q^2 v^3$ \\ \hline
\end{tabular}
\end{table}

An erratum to the original Bodwin, Braaten, and Lepage formulation of NRQCD \cite{Bodwin:1994jh}, which is the formulation discussed up to this point, attempts to clarify some issues with the velocity scaling of the gluon fields, and how that pertains to quarkonium production.  Rather than delve into this erratum, we instead turn to vNRQCD, which is a formulation that has more clarity in the velocity scaling.

\subsection{vNRQCD}

vNRQCD was developed by Luke, Manohar, and Rothstein \cite{Luke:1999kz}, and arises from accounting for the following $v$ scalings for a four-momentum $k = (k_0, \kv)$ to $\mathcal{O}(v^2)$.
\begin{equation}
    \begin{aligned}
        \text{hard:} \quad & (k_0, \kv) \sim (m_Q, m_Q) \; , \\
        \text{soft:} \quad &  (k_0, \kv) \sim (m_Qv, m_Qv) \; , \\
        \text{potential:} \quad &  (k_0, \kv) \sim (m_Qv^2, m_Qv) \; , \\
        \text{ultrasoft:} \quad &  (k_0, \kv) \sim (m_Qv^2, m_Qv^2) \; .
    \end{aligned}
\end{equation}
The hard modes are integrated out of the full theory.  The remaining modes are the scalings that could put a virtual quark or gluon on-shell. Looking at the kinetic term in Eq.~(\ref{eq: NRQCD kinetic term}), it is clear that on-shell fluctuations of quarks must be in the potential regime.  The soft momenta are treated as discrete variables, and the ultrasoft as continuous, so that the quark three-momentum $\Pv = \pv + \kv$ has a discrete component $\pv$ and continuous component $\kv$.  Therefore the soft momenta are indicated as indices on the two-spinor fields: $\psi_\pv(x)$.  

The decomposition $\Pv = \pv + \kv$ is invariant under the redefinition $\kv \rightarrow \kv + {\bf q}$, $\pv \rightarrow \pv - {\bf q}$, for ${\bf q} \sim m_Q v^2$, which is equivalent to $\psi_\pv(x) \rightarrow e^{i{\bf q}\cdot {\bf x}}\psi_{\pv -{\bf q}}(x)$ \cite{Luke:1999kz}.  This is known as reparametrization invariance.  Derivatives on this redefined field act as ${\bf D} e^{i{\bf q}\cdot {\bf x}}\psi_{\pv -{\bf q}}(x) = e^{i{\bf q}\cdot {\bf x}}(i{\bf q}+{\bf D})\psi_{\pv - {\bf q}}(x)$, so for the Lagrangian to be reparametrization invariant, the derivatives must be of the form $(i\pv + {\bf D})$ \cite{Luke:1992cs}. The kinetic term is therefore
\begin{equation}
    \mathcal{L}_{\rm heavy} = \sum_\pv \psi^\dagger_\pv \bigg[ i D_0 - \frac{(\pv-i{\bf D)}^2}{2m_Q}\bigg]\psi_\pv + (\psi \Rightarrow \chi)\; .
\end{equation}

Gluons can be either soft or ultrasoft, and vNRQCD divides the gluons into two degrees of freedom, one for each scaling.  The covariant derivative $D^\mu$ only involves the ultrasoft gluons, to preserve the scaling of the quark kinetic term.  The full effective Lagrangian involving all the fields is:
\begin{equation}
    \begin{aligned}
        \mathcal{L} = & \; -\frac{1}{4}F^{\mu \nu}F_{\mu \nu}  + \sum_p |p^\mu A_p^\nu - p^\nu A_p^\mu |^2 +  \sum_\pv \psi^\dagger_\pv \bigg[ i D_0 - \frac{(\pv-i{\bf D)}^2}{2m_Q}\bigg]\psi_\pv   \\
        & \; -4\pi\alpha_s \sum_{q,q^\prime,\pv, \pv^\prime}\bigg\{ \frac{1}{q_0} \psi^\dagger_{\pv^\prime}[A_{q^\prime}^0, A_q^0]\psi_\pv \\
        & \; + \frac{g^{\nu 0}(q^\prime-p+p^\prime)^\mu - g^{\mu 0}(q-p+p^\prime)^\nu + g^{\mu\nu}(q-q^\prime)^0}{(\pv^\prime-\pv)^2}\psi^\dagger_{\pv^\prime}[A_{q^\prime}^\mu, A_q^\mu]\psi_\pv \bigg\} \\
        & \; + (\psi \Rightarrow \chi, T \Rightarrow \bar{T}) \\
        & \; + \sum_{\pv, {\bf q}} \frac{4\pi \alpha_s}{(\pv - {\bf q})^2}\psi^\dagger_{\bf q}T^A \psi_\pv \chi^\dagger_{- {\bf q}}\bar{T}^A \chi_{-\pv} + \cdots \; .
    \end{aligned}
\end{equation}

The new considerations regarding the velocity scaling means the operators in vNRQCD have a different scaling than in the NRQCD of Bodwin, Braaten, and Lepage.  The new counting is summarized in Table \ref{tab: vNRQCD v scaling}.

\begin{table}[t]
\renewcommand{\arraystretch}{1.5}
\centering
\caption{Power counting in for the vNRQCD fields.}
\label{tab: vNRQCD v scaling}
\begin{tabular}{|c|c|}
\hline
Operator & Velocity scaling \\ \hline \hline
$\psi_\pv$, $\chi_\pv$ & $v^{3/2}$ \\ \hline
$A_p^\mu$ & $v$ \\ \hline
$D_0$ & $v^2$ \\ \hline
${\bf D}$ & $v^2$ \\ \hline
$A^\mu$ & $v^2$ \\ \hline
${\bf E}$ & $v^4$ \\ \hline
${\bf B}$ & $v^4$ \\ \hline
\end{tabular}
\end{table}

\subsection{Quarkonium production}
\label{sec: quarkonium production}

The production of quarkonium $H$ in NRQCD is governed by a 4-fermion production operator\footnote{At this stage we return to the Bodwin, Braaten, and Lepage notation where the heavy quark spinors do not have soft momentum labels, while retaining the knowledge of the $v$ scaling of the fields gained from vNRQCD.}, with the generic form
\begin{equation}
    \begin{aligned}
        \mathcal{O}_n^H = & \; \chi^\dagger \mathcal{K}_n \psi \bigg( \sum_X \sum_{\lambda} \ket{H(\lambda)+X}\bra{H(\lambda)+X} \bigg) \psi^\dagger \mathcal{K}_n^\prime \chi \\
        \equiv & \; \chi^\dagger \mathcal{K}_n \psi (a^\dagger_H a_H) \psi^\dagger \mathcal{K}_n^\prime \chi \; .
    \end{aligned}
\label{eq: NRQCD 4 fermion operator}
\end{equation}
The sums are over other particles $X$ in the final state, which are soft in the quarkonium rest frame, as well as over the spin states $\lambda$.  The operators $\mathcal{K}_n$ and $\mathcal{K}_n^\prime$ are products of color matrices, spin matrices, and/or derivatives. 

The $v$ scaling of the operators depends on how many chromoelectric (${\bf A}\cdot \nabla$) or chromomagnetic (${\bf B}\cdot \boldsymbol{\sigma}$) dipole transitions must be inserted into the matrix element for its quantum numbers to match those of the relevant hadron \cite{Bodwin:1994jh}.  The former operator arises from the ${\bf D}^2$ term in the Lagrangian, and scales as $v^3$ since the gluon field in the covariant derivative is ultrasoft, while the gradient acts on the quark fields and is thus soft.  The latter operator scales as $v^4$ since it involves ultrasoft gluons, and the curl acting on an ultrasoft field scales as $v^2$. 

The $J/\psi$ is a $^3S_1^{[1]}$ configuration, which according to the power counting in Table \ref{tab: vNRQCD v scaling} will have a matrix element that scales as $(v^{3/2})^4v^{-3} = v^3$: a factor of $v^{3/2}$ for each of the four spinors, and a factor of $v^{-3}$ for the normalization to the volume of the hadron.  For a matrix element with orbital angular momentum number $L$ and requiring $E$ chromoelectric dipole transitions and $M$ chromomagnetic dipole transitions, the scaling of the matrix element goes as $v^{3+2L+2E+4M}$.  This can be broken down for a single bilinear $\chi^\dagger \mathcal{K}_n \psi$ as follows:
\begin{itemize}
    \item a factor of $(v^{3/2})^2=v^3$ from the two heavy quark spinors,
    \item a factor of $v$ for each increment in the angular momentum quantum number, since they involve a derivative ${\bf D}$ acting on a heavy quark field,
    \item each chromoelectric dipole transition contributes $v^3v^{-2} = v$; the $v^3$ is from the chromoelectric dipole operator, and the $v^{-2}$ is from the necessary virtual heavy quark propagator,
    \item each chromomagnetic dipole transition contributes $v^4v^{-2}=v^2$; the $v^4$ is from the scaling of the chromomagnetic field, and the $v^{-2}$ is again from the virtual heavy quark propagator.
\end{itemize}
Then since there are two bilinears in a matrix element this cumulative $v$ scaling is squared, and combined with the factor of $v^{-3}$ for the normalization to the volume of the hadron, leading to the $v^{3+2L+2E+4M}$ rule.  The NRQCD operator associated with $c\bar{c}$ production via light quark fragmentation (LQF) at leading order in $\alpha_s$ is in a color octet $^3S_1$, with $\mathcal{K}_n = \sigma_i T^a$.  Two chromoelectric operators must be inserted to bring the state to a color singlet \cite{Bodwin:1994jh,Braaten:1996jt,Fleming:2000ib}, which in the process takes it to a $P$-wave and then back to an $S$-wave.  This means the $^3S_1^{[8]}$ matrix element scales as $v^7$.  Beyond $^3S_1^{[1]}$ and $^3S_1^{[8]}$, two other matrix elements that appear at leading order in $J/\psi$ production are for $^1S_0^{[8]}$ and $^3P_0^{[8]}$; these will be covered in the next chapter when a different production mechanism is discussed.

In terms of these NRQCD matrix elements, the quarkonium production cross sections have a factorized form of:\footnote{This equation is valid in the collinear limit; in the TMD regime it is an approximation.  This will be discussed further in Sec.~\ref{sec: TMD fact in NRQCD}.}
\begin{equation}
    \begin{aligned}
        \sigma(H) = & \; \sum_n \frac{F_n(\Lambda)}{m_Q^{d_\mathcal{O} -4}} \bra{0}\mathcal{O}_n^H(\Lambda)\ket{0} \\
        \equiv & \; \sum_{L,s,c} \frac{F_n}{m_Q^{d_\mathcal{O} -4}} \braket{\mathcal{O}^H(^{2s+1}L_J^{[c]})} \; ,
    \end{aligned}
\label{eq: NRQCD fact}
\end{equation}
where $d_\mathcal{O}$ is the mass dimension of the operator.  This is a sum of terms involving perturbative short-distance coefficients $F_n$, and non-perturbative NRQCD long distance matrix elements (LDMEs) $\braket{\mathcal{O}^H(^{2s+1}L_J^{[c]})}$.

The general procedure for utilizing this NRQCD factorization approach is as follows:
\begin{enumerate}
    \item Write down the full QCD Feynman diagrams appropriate for the production problem at hand.
    \item Perform a nonrelativistic expansion on the Dirac spinors $u$ and $v$ for the heavy quarks, taking dependence on $u$ and $v$ and turning it into dependence on the two-spinors $\xi$ and $\eta$.
    \item Match the color and spin structure of the resulting two-spinor bilinear to the appropriate 4-fermion operator in NRQCD, of the form in Eq.~(\ref{eq: NRQCD 4 fermion operator}).
    \item Be sure that you are accounting for all relevant terms at the order in $\alpha_s$ and $v$ you are considering.
\end{enumerate}
The matching from full QCD Dirac spinors onto NRQCD LDMEs for the most relevant bilinears is outlined in App.~\ref{app: NRQCD matching}.  We will be using this procedure to calculate the $J/\psi$ FFs using NRQCD.

\subsection{TMD factorization in NRQCD}
\label{sec: TMD fact in NRQCD}

As we are considering the transverse momentum in the problem, we will be using a factorization conjecture analagous to Eq.~(\ref{eq: NRQCD fact}).
\begin{equation}
    \begin{aligned}
        \Delta_{i\rightarrow J/\psi}(z,\kv_\perp) \rightarrow \sum_{L,s,c,m} d_{i\rightarrow c\bar{c}}^{(m)}(z,\kv_\perp) \braket{\mathcal{O}^{J/\psi}(^{2s+1}L_J^{[c]})} \, ,
    \end{aligned}
\label{eq: FF NRQCD fact}
\end{equation}
The matching coefficient describes the production of a $c\bar{c}$ pair by a parton $i$.  Its superscript $(m)$ refers to the power in the strong coupling, $d^{(m)}\sim \alpha_s^m$.  

\begin{figure*}
    \centering
    \includegraphics[scale=1.2]{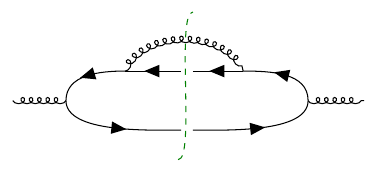}
    \caption[A soft gluon exchange in a $c\bar{c}$ production diagram, which can alter the transverse momentum $p_\perp \sim m_c v$ of the $c\bar{c}$.]{A soft gluon exchange in a $c\bar{c}$ production diagram, which can alter the transverse momentum $p_\perp \sim m_c v$ of the $c\bar{c}$. Figure from Ref.~\cite{Copeland:2023wbu}.}
    \label{fig: gluon across cut}
\end{figure*}

Reference \cite{Copeland:2023wbu} argues that Eq.~(\ref{eq: FF NRQCD fact}) is an approximation, and that the LDME should in fact be replaced by an operator that itself is transverse-momentum dependent.  This is because it is possible for the emission of soft gluons to alter the transverse momentum of the produced $c\bar{c}$ pair.  The collinear momentum, with $P^+\gg m_c$, is unchanged, but soft gluon emissions like in Fig.~\ref{fig: gluon across cut} could change the transverse momentum $p_\perp \sim m_c v$ of the quarks by $\mathcal{O}(m_c v)$, giving the $J/\psi$ a different transverse momentum than the $c\bar{c}$.  In this work we will use Eq.~(\ref{eq: FF NRQCD fact}) as an approximation, as has been done in a previous study \cite{Echevarria:2020qjk} on $J/\psi$ production via TMD fragmentation, since a full TMD treatment of the NRQCD factorization formalism has yet to be developed.

\section{Light quark fragmentation}

We now have the theoretical background to derive the fragmentation functions.  In this chapter, we will be working in a reference frame where the $J/\psi$ has no transverse momentum; all of the transverse momentum dependence will be in the fragmenting partons.
\begin{equation}
    \begin{aligned}
        P^\mu = P^+ \bar{n}^\mu + \frac{M^2}{2P^+}n^\mu \; ,
    \end{aligned}
\end{equation}
where the light-cone coordinate vectors are defined to be:
\begin{equation}
    \begin{aligned}
        n^\mu = & \; \frac{1}{\sqrt{2}}(1,0,0,-1) \, , \\
        \bar{n}^\mu = & \; \frac{1}{\sqrt{2}}(1,0,0,1) \, .
    \end{aligned}
\end{equation}
Using these, any vector $v$ can be written as $v^\mu = v^+\bar{n}^\mu + v^-n^\mu + v_T^\mu$, for $v^+ \equiv n\cdot v$ and $v^- \equiv \bar{n}\cdot v$. 

In position space, the field-theoretic definition of the quark TMD FF is: \cite{Boussarie:2023izj}
\begin{equation}
\begin{aligned}
\label{eq: q TMDFF}
\tilde{\Delta}_{q\to J/\psi} (z, \bv_T, P^+/z) = \frac{1}{2 z N_c}{\rm Tr}\int \frac{d b^-}{2 \pi} & e^{ib^-P^+/z}\sum_X \Gamma_{\alpha \alpha'} \\
& \times \bra{0} W{\scalebox{1.7}{$\lrcorner$}} \psi_i^{ \alpha}(b) \ket{J/\psi (P) , X} \\
& \times \bra{J/\psi (P), X} \bar{\psi}_i^{\alpha'} (0) W_{\scalebox{1.7}{$\urcorner$}} \ket{0} \; ,
\end{aligned}
\end{equation}
Here we have allowed the parton to have an arbitrary polarization.  At leading twist, the initial parton can be unpolarized, longitudinally polarized, or transversely polarized, corresponding to projection operators $\Gamma \in \{\gamma^+/2, \gamma^+\gamma_5/2, i\sigma^{\beta +}\gamma_5/2\}$, respectively. The $W$ objects are Wilson lines -- objects inserted into an operator to preserve gauge invariance.  They are
\begin{equation}
\begin{aligned}
    W{\scalebox{1.7}{$\lrcorner$}} = &\; W_{\hat{b}_T}(+\infty n;b_T,+\infty) W_{n}(b;0,+\infty) \\
    W_{\scalebox{1.7}{$\urcorner$}} = &\; W_{n}(0;+\infty,0) W_{\hat{b}_T}(+\infty n;+\infty,0) \;.
\end{aligned}
\end{equation}
where the Wilson line along a single direction is
\begin{equation}
    W_v (x^\mu; a, b) =  {\cal P} \, {\rm exp}\bigg[i g_s \int_a^b ds~ v\cdot A^c (x + s v) t^c\bigg].
\end{equation}
In the TMDFF, these connect the quark fields that are separated by $b=(0,b^-,{\bf b}_T)$, with a transverse separation ${\bf b}_T\equiv b_T \hat{b}_T$.  Refer to App.~\ref{app: wilson lines} for a more detailed discussion of these Wilson lines.

\subsection{Enumerating the diagrams}

\begin{figure*}[t]
\centering
\begin{minipage}{0.5\textwidth}
\centering
\subfloat[]{\includegraphics[scale=1]{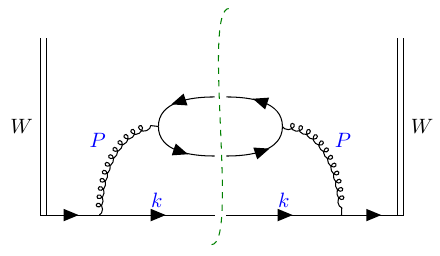}\label{quarkfig1}}
\end{minipage}%
\begin{minipage}{0.5\textwidth}
\centering
\subfloat[]{\includegraphics[scale=1]{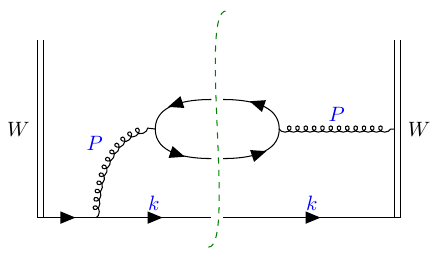}\label{quarkfig2}}
\end{minipage} 
\\ 
\begin{minipage}{0.5\textwidth}
\centering
\subfloat[]{\includegraphics[scale=1]{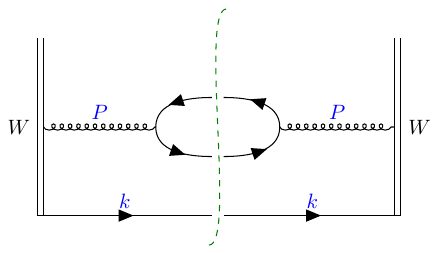}\label{quarkfig3}}
\end{minipage}
\caption[Diagrams at leading order in $\alpha_s$ contributing to the quark TMDFF.]{Diagrams at leading order in $\alpha_s$ contributing to the quark TMDFF. The double lines represent the Wilson lines. Diagram (b) has a mirror diagram.  Figures from Ref.~\cite{Copeland:2023wbu}.}
\label{fig: QuarkFeynDiag}
\end{figure*}

To see how to write down the Feynman diagrams contributing to the quark TMD FF, look more closely at one of the matrix elements in the definition, $\bra{0} W_\lrcorner \psi(b) \ket{J/\psi, X}$.  In NRQCD factorization, the FF will be broken down into a piece to produce a $c\bar{c}$ pair and an LDME, so we can instead work with $\bra{0} W_\lrcorner \psi(b) \ket{c\bar{c}, X}$.  The Wick contractions will be done in the interaction picture, where the matrix element is:
\begin{equation}
    \bra{0} W_\lrcorner \psi_i^{ \alpha}(b) \exp{\bigg(-i \int d^4 x \, \mathcal{H}_{\rm int}\bigg)} \ket{c\bar{c}, X} \; .
\end{equation}
The interaction Hamiltonian density for QCD is $\mathcal{H}_{\rm int} = g_s \bar{\psi} \gamma^\mu A_\mu^a T^a \psi$.  Expanding the time evolution operator and the Wilson line in $g_s$, one of the resluting terms has the following Wick contraction contribute to the matrix element:
\begin{equation}
    g_s^2 \int d^4x_1 \int d^4 x_2\, \bra{0} 
    \wick{
        \c1 \psi(b) \bar{\c1 \psi}(x_1) \gamma^\mu \c2 A_\mu^a(x_1) T^a \c3 \psi(x_1)
        \bar{\psi}\c6 (x_2) \gamma^\nu \c2 A_\nu^b T^b \c5 \psi(x_2) \ket{\c5 c\bar{\c6 c}, \c3 X} \; .
    }
\end{equation}
This clearly corresponds to the left half of Fig.~\ref{quarkfig1}.  The only other contribution of the matrix element $\bra{0} W_\lrcorner \psi(b) \ket{c\bar{c}, X}$ at this order in $\alpha_s$ corresponds to the left half of Fig.~\ref{quarkfig3}; combining two such matrix elements in the full expression for the TMD FF yields four possible diagrams contributing to $\Delta_{q\rightarrow c\bar{c}}$, one of which is a mirror image of another.

\subsection{Quark diagram amplitudes}

Applying the replacement rules for the contractions, the expressions we get for the diagrams in Fig.~\ref{fig: QuarkFeynDiag} are:
\begin{equation}
\begin{aligned}
d_{(a)} =& \; \frac{g_s^4}{4z M^4 N_c}\int\frac{d^Dk}{(2\pi)^D} \int db^- e^{ib^-P^+/z}e^{-ib(k+P)}\\
&\times{\rm Tr}\left[\slashed{k}\gamma^\mu \frac{\slashed{k} + \slashed{P}}{(k+P)^2+i\epsilon} \left(\Gamma  \chi_{\mu \nu}\right) \frac{\slashed{k} + \slashed{P}}{(k+P)^2+i\epsilon} \gamma^\nu\right] \delta(k^2)\\
\end{aligned}
\label{DA}
\end{equation}
\begin{equation}
\begin{aligned}
d_{(b) + {\rm mirror}} = & \; \frac{g_s^4}{4z M^4 N_c}\int\frac{d^Dk}{(2\pi)^D} \int db^- e^{ib^-P^+/z}e^{-ib(k+P)}\\
&\times{\rm Tr}\bigg[\slashed{k}\bigg(\frac{n^\mu}{P^+-i\epsilon}\Gamma \chi_{\mu\nu} \frac{\slashed{k} + \slashed{P}}{(k+P)^2+i\epsilon}\gamma^\nu \\ & - \gamma^\mu \frac{\slashed{k} + \slashed{P}}{(k+P)^2+i\epsilon} \Gamma \chi_{\mu\nu}\frac{n^\nu}{P^++i\epsilon}\bigg)\bigg]\delta(k^2)\\
\end{aligned}
\label{DB}
\end{equation}
\begin{equation}
\begin{aligned}
d_{(c)} =& \; \frac{g^4}{4z M^4 N_c}\int\frac{d^Dk}{(2\pi)^D} \int db^- e^{ib^-P^+/z}e^{-ib(k+P)}\\
&\times{\rm Tr}\left[\slashed{k}\frac{n^\mu}{P^+-i\epsilon}\left(\Gamma  \chi_{\mu \nu}\right) \frac{n^\nu}{P^++i\epsilon}\right] \delta(k^2)\\
\end{aligned}
\label{DC}
\end{equation}
The delta function $\delta(k^2)$ arises from the cuts of the light quark propagator in the diagrams, where we have taken its mass to be zero.  Here the spinor structure is
\begin{equation}
\begin{aligned}
    \chi_{\mu\nu} = & \; \bar{u}(p) \gamma_\mu T^a v(p') \bar{v}(p')\gamma_\nu T^a u(p). 
\label{eq: chi}
\end{aligned}
\end{equation}
This has the following nonrelativistic expansion.
\begin{equation}
\begin{aligned}
    \chi_{\mu\nu} \approx& \; M^2 {\Lambda^{\mu}}_i {\Lambda^{\nu}}_j  [\xi^\dagger \sigma^i T^a\eta][ \eta^\dagger \sigma^j T^a \xi] \; .
\label{eq: NR spinors}
\end{aligned}
\end{equation}

The integration over $d^D k$ is trivial due to the delta function arising from the $b^-$ integral. At this stage we begin to match onto NRQCD operators.  

\subsection{NRQCD matching and $J/\psi$ polarization}

The nonrelativistic expansion of the QCD spinors $u$ and $v$ yields answers in terms of the two-spinors $\xi$ and $\eta$, which match onto the NRQCD heavy quark and antiquark fields $\psi$ and $\chi$.  A projection operator $\mathcal{P}_{J/\psi(\lambda)}$ projects out the $J/\psi$ bound state with a particular helicity $\lambda$.  In this instance we have a color octet $^3S_1$ state, where the matching is:
\begin{equation}
    \begin{aligned}
        M^2 \eta^{\prime\dagger} \sigma_i T^a \xi^\prime \xi^\dagger \sigma_j T^a \eta &\leftrightarrow \braket{\chi^\dagger \sigma_i T^a \psi \, \mathcal{P}_{J/\psi (\lambda)} \, \psi^\dagger \sigma_j T^a \chi}  \, . \\
    \end{aligned}
\end{equation}
Spin symmetry implies that the matrix element on the right-hand side be proportional to $\epsilon^*_{\lambda i}\epsilon_{j\lambda}$ (no sum over $\lambda$), where $\epsilon$ is the polarization three-vector of the $J/\psi$ \cite{Braaten:1997rg}.  The helicity can take values $\lambda = \pm1$, $0$.  The coefficient of this tensor structure is then defined to be the NRQCD LDME:
\begin{equation}
    \begin{aligned}
        \braket{\chi^\dagger \sigma_i T^a \psi \, \mathcal{P}_{J/\psi (\lambda)} \, \psi^\dagger \sigma_j  T^a \chi} &= \frac{2M}{3} \epsilon^*_{\lambda i} \epsilon_{j \lambda} \braket{\mathcal{O}^{J/\psi}(^3S_1^{[8]})} \, , 
\label{eq: 3S18 LDME}
    \end{aligned}
\end{equation}
The polarization tensor for the $J/\psi$ can be decomposed in terms of a parametrization appropriate for a spin-1 particle \cite{Bacchetta:2000jk, Kumano:2020gfk}.
\begin{equation}
     \epsilon^*_{\lambda i} \epsilon_{j \lambda} = \frac13 \delta_{ij} + \frac{i}{2} \epsilon_{ijk}S_k - T_{ij} \, ,
\end{equation}
where the vector polarized piece is in terms of:
\begin{equation}
\label{eq: S vec}
\begin{aligned}
    {\bf S}= {\rm Im}\left(\epsilon_\lambda^* \times \epsilon_\lambda\right) = (S^x_T, S_T^y, S_L) \; ,
\end{aligned}
\end{equation}
and the tensor polarized piece is:
\begin{equation}
\label{eq: Tij}
\begin{aligned}
    T_{ij} = & \; \frac13 \delta_{ij} - {\rm Re}\left( \epsilon^*_{\lambda i} \epsilon_{j \lambda}\right) \\
     \equiv & \; \frac12
\begin{pmatrix}
-\frac23 S_{LL} + S_{TT}^{xx} & S_{TT}^{xy} & S_{LT}^x\\
S_{TT}^{yx} & -\frac23 S_{LL} - S_{TT}^{xx} & S_{LT}^y\\
S_{LT}^x & S_{LT}^y & \frac43 S_{LL}
\end{pmatrix} \; .
\end{aligned}
\end{equation}
The choice of a particular helicity $\lambda$ fixes the values of the parameters in ${\bf S}$ and $T_{ij}$. Refer to App.~\ref{app: pol par interpretation} for interpretations of these parameters.

Boosting to a frame where the $J/\psi$ has arbitrary momentum $P$ yields:\footnote{The boost matrices are discussed in App.~\ref{app: NRQCD matching}.}
\begin{equation}
    {\Lambda^\mu}_i {\Lambda^\nu}_j \epsilon^{*}_i \epsilon_{j} = -\frac13 \left( g^{\mu \nu} -\frac{P^\mu P^\nu}{M^2} \right) + \frac{i}{2M} \epsilon^{\mu\nu\alpha\beta}P_\alpha S_\beta - T^{\mu\nu} \, ,
\label{eq: Jpsi Pol}
\end{equation}
where
\begin{equation}
    S_{\beta} = \left(P^+ \bar{n}_\beta - P^- n_\beta\right) \frac{S_L}{M} + S_{T\beta} \; ,
\label{eq: spin vector}
\end{equation}
\begin{equation}
\begin{aligned}
    T^{\mu \nu} = \frac12\bigg\{\left[\frac43 \frac{(P^+)^2}{M^2} \bar{n}^\mu \bar{n}^\nu - \frac23 \bar{n}^{\{\mu}\bar{n}^{\nu\}} + \frac13\frac{M^2}{(P^+)^2} n^{\mu}n^{\nu}\right]S_{LL}\\
    -\frac{1}{M}\bigg(P^+ \bar{n} - \frac{M^2}{2P^+}n\bigg)^{\{\mu} S^{~\nu\}}_{LT} +\frac23 S_{LL}g^{\mu\nu}_T + S_{TT}^{\mu\nu}\bigg\} \; ,
\end{aligned}
\end{equation}
We consider the production of unpolarized, longitudinally polarized, and transversely polarized $J/\psi$.  Experimentally, this is measured by looking at the angular distribution of the lepton pair to which the $J/\psi$ decays; we will look at this distribution in more detail in Sec.~\ref{sec: polarization puzzle}. The polarization tensor in the unpolarized case is the usual result:
\begin{equation}
    \sum_{\rm pol}\epsilon_U^{*\mu}\epsilon_U^\nu = -g^{\mu \nu} + \frac{P^\mu P^\nu}{M^2} \; .
\end{equation}
The polarization vector for a longitudinally polarized $J/\psi$ is simple to deduce.  It needs to have a Cartesian component that lies along the three-momentum of the $J/\psi$, $\boldsymbol{\epsilon} \, \propto \, {\bf P}$.  Combined with the constraints $\epsilon_L \cdot P = 0$ and $\epsilon_L \cdot \epsilon_L = -1$, the result must be
\begin{equation}
    \epsilon_L^\mu = \frac{1}{M}\left(|{\bf P}|, P^0 \hat{\bf P} \right) \; .
\label{eq: long pol vector}
\end{equation}
The transversely polarized $J/\psi$ tensor can be obtained from the relation $\epsilon_U^{*\mu}\epsilon_U^\nu = \epsilon_L^{*\mu}\epsilon_L^\nu + \epsilon_T^{*\mu}\epsilon_T^\nu$, so it is sufficient to study only the unpolarized and longitudinally polarized case.  Comparing Eq.~(\ref{eq: long pol vector}) with Eq.~(\ref{eq: Jpsi Pol}) reveals that longitudinal polarization corresponds to $S_{LL}=-1$, with all other parameters vanishing.

After matching onto NRQCD and plugging in an arbitrary polarization for the $J/\psi$, the full expression for $\Delta_{q\rightarrow J/\psi}$ is quite large.  However, its terms can be classified according to their dependence on the polarization parameters of $S^\mu$ and $T^{\mu\nu}$, and the polarization of the initial light quark \cite{Bacchetta:2000jk}.  The coefficients of these structures are also referred to as fragmentation functions. The possible FFs depending on the quark and hadron polarization are given in Table \ref{tab: quark TMDFFs}.  Their relations to the full FF $\Delta_{q\rightarrow J/\psi}$ are provided in App.~\ref{app: projection operators}.    

\begin{table*}[t]
\centering
\renewcommand{\arraystretch}{1.5}
\caption{Quark TMD Fs for different parton polarizations, organized according to the $J/\psi$ polarization parametrization.}
\label{tab: quark TMDFFs}
\begin{tabular}{|m{0.7cm}|m{0.7cm}||m{3.8cm}|m{3cm}|m{4cm}|}
 \cline{3-5} 
 \multicolumn{2}{c||}{\multirow{2}{*}{}} & \multicolumn{3}{c|}{Quark polarization} \\ \cline{3-5}
 \multicolumn{2}{c||}{} & Unpolarized & Longitudinal & Transverse \\ \hline \hline
 \multirow{6}{0.5cm}{\rotatebox[origin=c]{90}{Hadron polarization }} & U & $D_1$ (unpolarized) & & $H_1^\perp$ (Collins) \\ \cline{2-5}
 & L &  & $G_1$ (helicity) & $H_{1L}^\perp$ \\ \cline{2-5}
 & T & $D_{1T}^\perp$ (polarizing FF) & $G_{1T}^\perp$ & $H_1,\, H_{1T}^\perp$ \\ \cline{2-5} 
 & LL & $D_{1LL}$ & & $H_{1LL}^\perp$ \\ \cline{2-5}
 & LT & $D_{1LT}$ & $G_{1LT}$ & $H_{1LT}^\perp , \, H_{1LT}^\prime$ \\ \cline{2-5}
 & TT & $D_{1TT}$ & $G_{1TT}$ & $H_{1TT}^\perp$ (transversity), $H_{1TT}^\prime $ \\ \hline
\end{tabular}
\end{table*}

At this order in $\alpha_s$, the only nonzero TMD FFs are:
\begin{equation}
    \begin{aligned}
        D_1(z, \kv_T ; \mu) = & \; \frac{2\alpha_s^2(\mu)}{9\pi N_c M^3 z} \frac{\kv_T^2 z^2(z^2-2z+2)+2M^2(z-1)^2}{[z^2 \kv_T^2+M^2(1-z)]^{2}} \braket{\mathcal{O}^{J/\psi}(^3S_1^{[8]})}\; , \\
        D_{1LL}(z, \kv_T ; \mu) =  & \; \frac{2\alpha_s^2(\mu)}{9\pi N_c M^3 z}\frac{\kv_T^2z^2(z^2-2z+2)-4 M^2(z-1)^2}{[z^2\kv_T^2+M^2(1-z)]^{2}}\braket{\mathcal{O}^{J/\psi}(^3S_1^{[8]})}\; , \\
        D_{1LT}(z, \kv_T ; \mu) = & \; \frac{2\alpha_s^2(\mu)}{3\pi N_c M } \frac{(2-z)(1-z) }{[z^2\kv_T^2+M^2(1-z)]^{2}} \braket{\mathcal{O}^{J/\psi}(^3S_1^{[8]})} \; , \\
        D_{1TT}(z, \kv_T ; \mu) = & \; \frac{2\alpha_s^2(\mu)}{3\pi N_c M} \frac{z(z-1)}{[z^2\kv_T^2+M^2(1-z)]^{2}} \braket{\mathcal{O}^{J/\psi}(^3S_1^{[8]})}\; , \\
        G_{1L}(z, \kv_T ; \mu) = & \; \frac{\alpha_s^2(\mu)}{3\pi N_c M^3} \frac{ \kv_T^2 z^2(2-z)}{[z^2\kv_T^2+M^2(1-z)]^{2}}\braket{\mathcal{O}^{J/\psi}(^3S_1^{[8]})} \; , \\
        G_{1T}^\perp(z, \kv_T ; \mu) = & \;  \frac{2\alpha_s^2(\mu)}{3\pi N_c M} \frac{z(z-1)}{[z^2\kv_T^2+M^2(1-z)]^{2}} \braket{\mathcal{O}^{J/\psi}(^3S_1^{[8]})}\, . \\
    \end{aligned}
\end{equation}

The function $D_1$ was first calculated in Ref.~\cite{Echevarria:2020qjk}; the rest are results new to Ref.~\cite{Copeland:2023wbu}.  These are ready to be utilized in an expression for a $J/\psi$ production cross section.

\section{Gluon fragmentation}

\begin{figure*}
    \centering
    \includegraphics[scale=1.2]{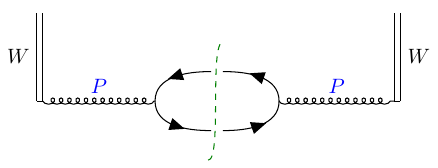}
    \caption[The lone diagram at LO in $\as$ contributing to the gluon TMDFF.]{The lone diagram at LO in $\as$ contributing to the gluon TMDFF.  Figure from Ref.~\cite{Copeland:2023wbu}.}
    \label{gluondiag}
\end{figure*}

The procedure for obtaining the gluon TMD FFs are the same as for the quark.  In terms of Wilson lines in the adjoint representation, $\mathcal{W}$, the gluon TMD FF is: \cite{Boussarie:2023izj}
\begin{equation}
\begin{aligned}
\tilde{\Delta}^{\alpha \alpha'}_{ g \rightarrow J/\psi}(z, \bv_T, P^+z ) = \frac{1}{2 z^2 P^+} \int\frac{d b^-}{2 \pi} & e^{i b^- P^+/z} \sum_X  \bra{0} G^{+ \alpha}(b) {\cal W}{\scalebox{1.7}{$\lrcorner$}} \ket{J/\psi(P), X} \\ 
\times & \bra{J/\psi(P), X} G^{+ \alpha'}(0)  {\cal W}_{\scalebox{1.7}{$\urcorner$}} \ket {0} \; .
\end{aligned}
\label{eq: g TMDFF}
\end{equation}\\
The spin indices are left free.  There is only one diagram at leading order in $\alpha_s$, shown in Fig.~\ref{gluondiag}.  Its amplitude is:
\begin{equation}
\begin{aligned}
d^{\alpha \alpha'}_g = \frac{g^2}{2z^2P^+(N_c^2-1)}\int & \frac{db^-}{(2\pi)} e^{i(b^-P^+/z - P\cdot b)}\left(P^\alpha \frac{n^\mu}{P^2} - P^+ \frac{g^{\alpha\mu}}{P^2}\right)\chi_{\mu \nu}\\
&\times\left(P^{\alpha'} \frac{n^\nu}{P^2} - P^+ \frac{g^{\alpha'\nu}}{P^2}\right) \; .
\label{eq: gTMDFF amp}
\end{aligned}
\end{equation}
Fourier transforming to $\kv_T$ space yields two delta functions:
\begin{equation}
    \int \frac{db^-}{(2\pi)} \frac{d^2{\bf b}_T}{(2\pi)^2} e^{-i(\kv_T \cdot {\bf b_T})} e^{i(b^-P^+ /z - b^-P^+ + \Pv_T \cdot {\bf b}_T)} = \frac{(z^2)_{z=1}}{P^+}\delta(1-z) \delta^{(2)}(\kv_T) \; .
\end{equation}
The amplitude thus reduces to:
\begin{equation}
\begin{aligned}
d^{\alpha \alpha'}_g = \frac{g^2}{2(P^+ M^2)^2(N_c^2-1)}\delta(1-z) \delta^{(2)}(\kv_T)\big(P^\alpha n_\mu - P_+ g^{\alpha}_{ \mu}\big)\chi^{\mu \nu}\big(P^{\alpha'} n_\nu - P_+ g^{\alpha'}_{\nu}\big) \; .
\label{eq: gTMDFF amp 2}
\end{aligned}
\end{equation}
We then need to match onto NRQCD and project out the various polarizations of the gluon and $J/\psi$.  The resulting TMD FFs can be organized in a table like in the quark case (Table \ref{tab:gluonTMDFFs}). 

\begin{table*}[t]
\renewcommand{\arraystretch}{1.5}
\centering
\caption{\label{tab:gluonTMDFFs}Gluon TMD FFs for the unpolarized, antisymmetric, and symmetric combinations of gluons, along with the parametrization of the $J/\psi$ polarization.}
\begin{tabular}{|m{0.7cm}|m{0.7cm}||m{3.5cm}|m{2.5cm}|m{4cm}|}
 \cline{3-5}
 \multicolumn{2}{c||}{\multirow{2}{*}{}} & \multicolumn{3}{c|}{Gluon operator polarization} \\ \cline{3-5}
 \multicolumn{2}{c||}{} & Unpolarized & Helicity 0 antisymmetric & Helicity 2 \\ \hline \hline
 \multirow{6}{0.5cm}{\rotatebox[origin=c]{90}{ Hadron polarization }} & U & $D_1^g$ (unpolarized) & & $H_1^{\perp g}$ (linearly polarized) \\ \cline{2-5}
 & L &  & $G_{1L}^g$ (helicity) & $H_{1L}^{\perp g}$ \\ \cline{2-5}
 & T & $D_{1T}^{\perp g} $ & $G_{1T}^{\perp g}$ & $H_{1T}^g$ (transversity), $H_{1T}^{\perp g}$ \\ \cline{2-5}
 & LL & $D_{1LL}^g $ & & $H_{1LL}^{\perp g}$ \\ \cline{2-5}
 & LT & $D_{1LT}^g $ & $G_{1LT}^g$ & $H_{1LT}^{\perp g} , \, H_{1LT}^{\prime g}$ \\ \cline{2-5}
 & TT & $D_{1TT}^g $ & $G_{1TT}^g$ & $H_{1TT}^{\perp g}, \, H_{1TT}^{\prime g}$ \\ \hline
\end{tabular}
\end{table*}

The nonvanishing TMDFFs are:
\begin{equation}
    \begin{aligned}
        D^g_1(z, \kv_T ; \mu) = & \; \frac{\pi \alpha_s(\mu)}{9M^3}\braket{\mathcal{O}^{J/\psi}(^3S_1^{[8]})}\delta(1-z) \delta^{(2)}(\kv_T)\;  , \\
        D^g_{1LL}(z, \kv_T ; \mu)  = & \; \frac{\pi \alpha_s(\mu)}{9M^3}\braket{\mathcal{O}^{J/\psi}(^3S_1^{[8]})}\delta(1-z) \delta^{(2)}(\kv_T) \; , \\
        G^g_{1L}(z, \kv_T ; \mu)  =  & \; -\frac{\pi \alpha_s(\mu)}{6M^3}\braket{\mathcal{O}^{J/\psi}(^3S_1^{[8]})}\delta(1-z) \delta^{(2)}(\kv_T) \; , \\
        H^g_{1TT}(z, \kv_T ; \mu)  = & \; -\frac{\pi \alpha_s(\mu)}{6M^3}\braket{\mathcal{O}^{J/\psi}(^3S_1^{[8]})}\delta(1-z) \delta^{(2)}(\kv_T) \; . 
    \end{aligned}
\end{equation}
Notice that the transverse momentum dependence is trivial.  Nontrivial $\kv_T$ dependence appears at one loop, and has been calculated recently for the unpolarized case by Echevarria et al.~\cite{Echevarria:2023dme}.

\section{Phenomenology}

We can now turn to making some predictions utilizing the derived TMD FFs.  As their TMD FFs have more interesting transverse momentum dependence at this order in $\alpha_s$, as well as the fact that quark PDFs are better understood, we focus only on LQF.

\subsection{Cross section expressions}

We established in Sec.~\ref{sec: sidis and factorization} that the SIDIS cross section can be factorized into a convolution of a PDF, a FF, and a partonic cross section.  Here we write the expression more explicitly.  The full QCD cross section is given by: \cite{Bacchetta:2000jk,Bacchetta:2006tn,Echevarria:2020qjk}
\begin{equation}
   \frac{d\sigma}{dx \, dz \, dQ^2 \, d{\bf P}_\perp^2} = \frac{\alpha_{\rm em}^2 M}{2Q^2xzs} L^{\mu\nu}W_{\mu\nu} \; ,
\end{equation}
for the kinematic variables:
\begin{equation}
    Q^2 = -q^2 = -(l-l')^2, ~~x = \frac{Q^2}{2p\cdot q}, ~~y= \frac{p\cdot q}{p\cdot l}, ~~z = \frac{p\cdot P}{p\cdot q} \; ,
\end{equation}
and the hadronic and leptonic tensors:
\begin{equation}
    L_{\mu \nu} = e^{-2} \bra{l'} J_\mu(0)\ket{l}\bra{l} J^\dagger_\nu (0) \ket{l'} \, .
\end{equation}
\begin{equation}
W_{\mu \nu} = e^{-2}\int \frac{d^4 x} {(2\pi)^4} e^{-iq\cdot x} \sum_X \bra{p} J_\mu^\dagger(x) \ket{P,X}\bra{P,X} J_\nu(0) \ket{p} 
\end{equation}
The lepton charge is $e$, $J_\mu=e\bar{\psi}A_\mu^{\rm em}\psi$ is the electromagnetic current, and $s=Q^2/xy$ is the center-of-mass energy.  At leading twist, the hadronic tensor factorizes into a convolution of a TMD PDF $\Phi$ and a TMD FF $\Delta$: \cite{Bacchetta:2000jk}
\begin{equation}
    W^{\mu \nu} = 2 z \int d^2\kv_T \, d^2\pv_T ~\delta^{(2)}\left(\pv_T - \kv_T + \frac{\Pv_\perp}{z} \right){\rm Tr} \left[\gamma^\mu \Phi(\pv_T, x) \gamma^{\nu} \Delta(\kv_T, z) \right] \, 
\label{eq: Factorized W}
\end{equation}
We saw in the previous section that the full expression for the TMD FF, for a polarized parton and polarized $J/\psi$, can be organized according to those polarizations to yield multiple coefficient functions. The same can be done for the TMD PDFs in terms of the nucleon and parton polarizations.  The resulting coefficient functions, also called PDFs, are summarized in Table~\ref{tab: quark TMDPDFs} for quarks and Table~\ref{tab: gluon TMDPDFs} for gluons.

\begin{table*}[t]
\centering
\renewcommand{\arraystretch}{3}
\caption{Quark TMD PDFs organized according to quark and nucleon polarization.}
\label{tab: quark TMDPDFs}
\begin{tabular}{|m{0.7cm}|m{0.7cm}||m{3cm}|m{3cm}|m{3cm}|}
 \cline{3-5} 
 \multicolumn{2}{c||}{\multirow{2}{*}{}} & \multicolumn{3}{c|}{Quark polarization} \\ \cline{3-5}
 \multicolumn{2}{c||}{} & Unpolarized (U) & Longitudinal (L) & Transverse (T) \\ \hline \hline
 \multirow{3}{0.5cm}{\rotatebox[origin=c]{90}{Nucleon polarization}} & U & $f_1$ (unpolarized) & & $h_1^\perp$ (Boer-Mulders) \\ \cline{2-5}
 & L &  & $g_1$ (helicity) & $h_{1L}^\perp$ (Worm-gear) \\ \cline{2-5}
 & T & $f_{1T}^\perp$ (Sivers) & $g_{1T}^\perp$ (Worm-gear)& $h_1$ (transversity),  $h_{1T}^\perp$ (pretzelosity) \\ \hline
\end{tabular}
\end{table*}

\begin{table*}[t]
\centering
\renewcommand{\arraystretch}{3}
\caption{Gluon TMD PDFs organized according to gluon and nucleon polarization.}
\label{tab: gluon TMDPDFs}
\begin{tabular}{|m{0.7cm}|m{0.7cm}||m{3cm}|m{2.5cm}|m{3.5cm}|}
 \cline{3-5} 
 \multicolumn{2}{c||}{\multirow{2}{*}{}} & \multicolumn{3}{c|}{Gluon operator polarization} \\ \cline{3-5}
 \multicolumn{2}{c||}{} & Unpolarized & Helicity 0 antisymmetric & Helicity 2 \\ \hline \hline
 \multirow{3}{0.5cm}{\rotatebox[origin=c]{90}{Nucleon polarization}} & U & $f_1^g$ (unpolarized) & & $h_1^{\perp g}$ (linearly polarized) \\ \cline{2-5}
 & L &  & $g_{1L}^g$ (helicity) & $h_{1L}^{\perp g}$ \\ \cline{2-5}
 & T & $f_{1T}^{\perp g}$ & $g_{1T}^{\perp g}$ & $h_{1T}^g$ (transversity),  $h_{1T}^{\perp g}$ \\ \hline
\end{tabular}
\end{table*}

Consider the case of LQF with an unpolarized lepton and unpolarized target.  There are only two surviving convolutions,
\begin{equation}
\begin{aligned}
\label{eq: UU fac cross}
    \frac{d \sigma_{UU}(l + H \to l' + J/\psi + X)}{dx ~dz ~dy ~d^2 {\bf P_\perp}} 
    = &\frac{4\pi \alpha^2 s}{Q^4} \left(1 - y +\frac{y^2}{2}\right) \bigg\{{\bf I}[f_1 D_1] + S_{LL} {\bf  I}[f_1 D_{1LL}]\bigg\} \; ,
\end{aligned}
\end{equation}
where we have defined the convolution integral
\begin{equation}
    {\bf  I}[f~ D] = 2 z \int d^2\kv_T \, d^2\pv_T ~\delta^{(2)}\left(\pv_T - \kv_T + \frac{\Pv_\perp}{z} \right) f(\pv_T) D(\kv_T) \; .
\end{equation}
This clearly has an unpolarized parton; the polarization of the $J/\psi$ is obtained by summing over appropriate values of the parameter $S_{LL}$.  For unpolarized $J/\psi$, the sum is over $S_{LL}\in \{1/2,1/2,-1\}$, and for longitudinally polarized $J/\psi$, it is over $S_{LL}\in \{-1\}$. Transverse polarization can be obtained by taking the difference between the unpolarized and longitudinally polarized cases.

Similarly, for a polarized lepton beam and longitudinally polarized target, the cross section is: \cite{Bacchetta:2000jk}
\begin{equation}
    \frac{d \sigma_{LL}(l + H \to l' + J/\psi + X)}{dx ~dz ~dy ~d^2 {\bf P_\perp}} = \frac{4\pi \alpha^2 s}{Q^4}2 \lambda_e S_{qL} ~ y\bigg(1- \frac{y}{2}\bigg)x\bigg\{{\bf I}[g_{1L}D_1] + S_{LL} {\bf I}[g_{1L}D_{1LL}]\bigg\}.
\label{eq: LL cross section}
\end{equation}
The parameter $\lambda_e$ is the beam helicity, and $S_{qL}$ is the polarization parameter for the quark.  This is not a physical cross section; rather, it is instead a difference of physical cross sections.  In terms of a basis where superscripts represent the helicities of the target and subscripts represent the helicities of the virtual photon, $d\sigma_{LL}$ is: \cite{Bacchetta:2006tn}
\begin{equation}
    d\sigma_{LL} = \frac12(d\sigma^{++}_{++} - d\sigma^{--}_{++}).
\end{equation}
%

\subsection{Transverse momentum dependence of the PDFs}

In plotting the cross sections, we use numerical values for the PDFs that have been fit to experiment.  The TMD PDFs are poorly constrained and are a topic of ongoing research in the field \cite{Boussarie:2023izj, Musch:2007ya, Anselmino:2013lza, Sun:2014dqm, Scimemi:2019cmh, Bertone:2019nxa}.  It is common to use a parametrization where the transverse momentum dependence is in the form of a Gaussian:
\begin{equation}
    \Phi_{i/N}(x, \pv_T) = \frac{1}{\pi \braket{p^2_T}}\Phi_{i/N}(x)e^{-{\pv_T^2}/\braket{p^2_T}} \; ,
\end{equation}
where $\braket{p^2_T}$ is a small parameter and $\Phi_{i/N}(x)$ is the collinear PDF.  However, when applied to our cross sections in Eqs.~(\ref{eq: UU fac cross}) and (\ref{eq: LL cross section}), using the Gaussian in the range $\braket{p^2_T} \in [0.2, 0.8]\; {\rm GeV}^2$ is not sufficiently different from simply using the first-order result in the TMD expansion:
\begin{equation}
\begin{aligned}
    \Phi_{i/N}(x,\pv_T) &\approx \Phi_{i/N}(x) \delta^{(2)}(\pv_T)
    \end{aligned}
\label{eq: delta fn approx}
\end{equation}
Using a delta function for the transverse momentum dependence also avoids computational difficulties associated with highly oscillatory integrands.  Therefore, here we use Eq.~(\ref{eq: delta fn approx}) to produce the plots of the cross sections.

\subsection{Discussion of the phase and parameter spaces}

We plot the cross sections in eight regions of phase space, corresponding to dividing the variables $x$, $z$, $Q$, into two bins each and integrating over those subsets of the domains, and plotting as a function of $\Pv_\perp$.  This is done for the sake of comparison with the plots in Ref.~\cite{Echevarria:2020qjk}, who were the first to derive the TMD FFs for $J/\psi$ production in the case of unpolarized partons and $J/\psi$.  The bins are  $x \in [0.1, 0.5] ~\&~ [0.5,1], z \in [0.1, 0.4] ~\&~ [0.4,0.8], $ and $Q $ (GeV) $ \in [10, 30]~\&~ [30, 50]$, and $\Pv_\perp$ is constrained to be in the TMD regime $\Pv_\perp \in [0, z_{\rm [bin~ min]} Q_{\rm [bin ~min]}/2]$ \cite{Echevarria:2020qjk}.  

There are several parameters in the full expressions for the cross sections.  Here we list them and justify the values we choose for the plots.
\begin{itemize}
    \item {\bf the center-of-mass energy} $s$: we choose $\sqrt{s} = 63$ GeV to be within the capabilities of the future Electron Ion Collider \cite{Accardi:2012qut}, as well as to be consistent with Ref.~\cite{Echevarria:2020qjk},
    \item {\bf the PDF energy scale} $\mu$: the PDFs are associated with an energy exchange of $Q$, so we choose $\mu = 30\; {\rm GeV} \sim Q$,
    \item {\bf the collinear PDFs}: we choose the numerical fits of Ref.~\cite{Bastami:2018xqd}; we have to evolve the polarized PDF $g_{1L}$ to the scale $\mu=30$ GeV,
    \item {\bf the TMD FF renormalization scale} $\mu$: as the matching of the TMD FF from full QCD to NRQCD is done at the scale $\sim M$, the FFs are evaluated at $\mu=M=3.1$ GeV,
    \item {\bf the color-octet} $^3S_1$ {\bf LDME} $\braket{\mathcal{O}^{J/\psi}(^3S_1^{[8]})}$: chosen to be the fit from Ref.~\cite{Chao:2012iv}, $\braket{\mathcal{O}^{J/\psi}(^3S_1^{[8]})}$ = $0.3 \times 10^{-2}$ GeV,\footnote{Choosing an LDME fit is non-trivial; this will be discussed in detail in the next chapter, where a greater variety of LDMEs are utilized.}
    \item {\bf the beam helicity} $\lambda_e$: we use $\lambda_e=-1$, corresponding to a purely left-handed beam,
    \item {\bf the quark polarization parameter} $S_{qL}$: we use $S_{qL}=-1$, which means the target spin is parallel to the photon momentum \cite{Bacchetta:2006tn, Diehl:2005pc},
    \item {\bf an arbitrary normalization factor} $\mathcal{N}$: chosen by hand to allow the different plots to be shown with the same vertical axis.
\end{itemize}

\subsection{Plots}

We plot $d\sigma_{UU}$ for the case of transversely and longitudinally polarized $J/\psi$, in Figs.~\ref{fig: T cross sections} and \ref{fig: L cross sections}, respectively.  In each case we show the full differential cross section, as well as the individual contributions from the two TMD FFs $D_1$ and $D_{1LL}$.  For transversely polarized $J/\psi$, the $D_{1LL}$ contribution suppresses the total cross section and is of roughly equal magnitude to the $D_1$ contribution, such that near $|\Pv_\perp|=0$ the total cross section is essentially zero in all bins.  It follows that for longitudinally polarized $J/\psi$, the $D_{1LL}$ contribution enhances the total cross section, since the $S_{LL}$ parameter has a sign flip.  

The two polarization cases are plotted alongside the unpolarized case in Fig.~\ref{fig: A cross sections}.  Longitudinal $J/\psi$ polarization dominates at smaller $\Pv_\perp$, and transverse polarization takes over around $|\Pv_\perp|=2$ GeV.  The same is true when considering the cross section $d\sigma_{LL}$.  Here, in the larger $x$ bin, all contributions to the cross section are negative, which occurs because the polarized PDF $g_{1L}$ goes negative.  In light of Eq.~(\ref{eq: LL cross section}), this suggests that the negative helicity photon cross section dominates for higher $x$.

\begin{landscape}

\begin{figure}
    \centering
    \includegraphics[width = \linewidth ]{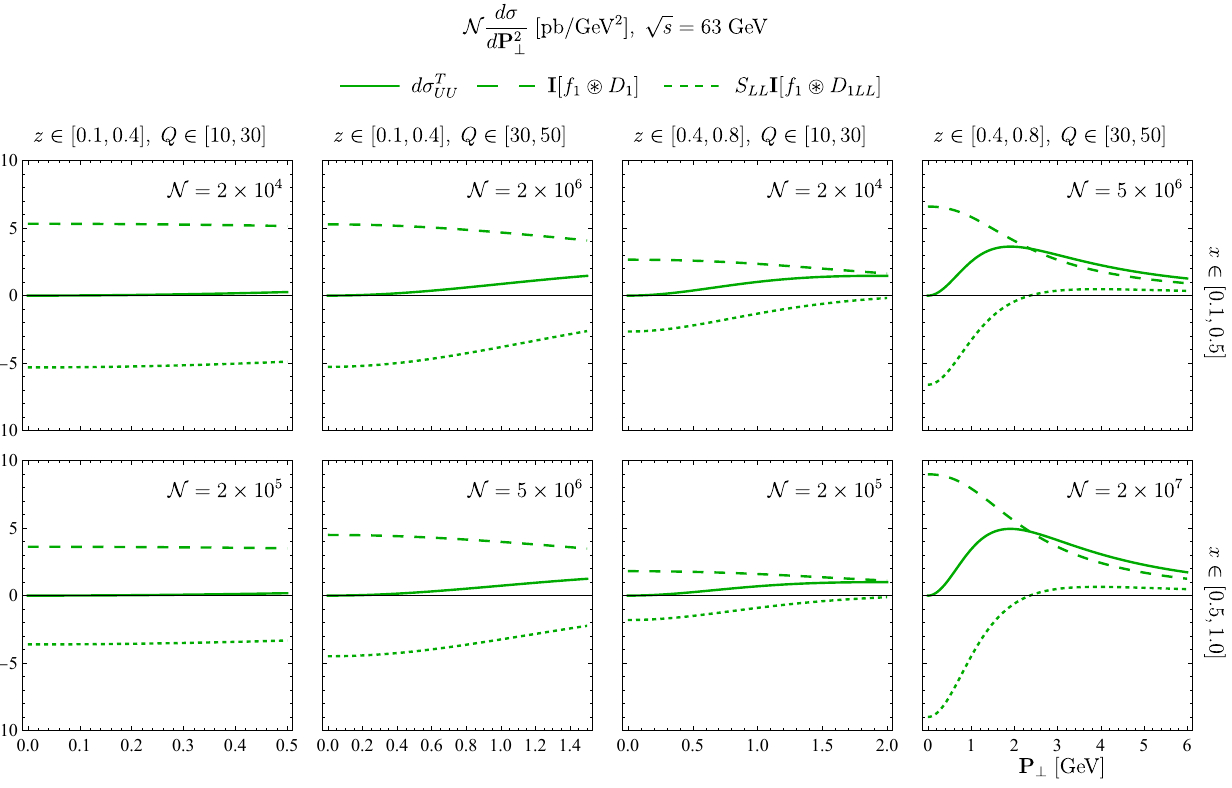}
    \caption[SIDIS cross section for an unpolarized beam and target, producing transversely polarized $J/\psi$.]{SIDIS cross section for an unpolarized beam and target, producing transversely polarized $J/\psi$.  Figure from Ref.~\cite{Copeland:2023wbu}.}
    \label{fig: T cross sections}
\end{figure}

\begin{figure}
    \centering
    \includegraphics[width = \linewidth]{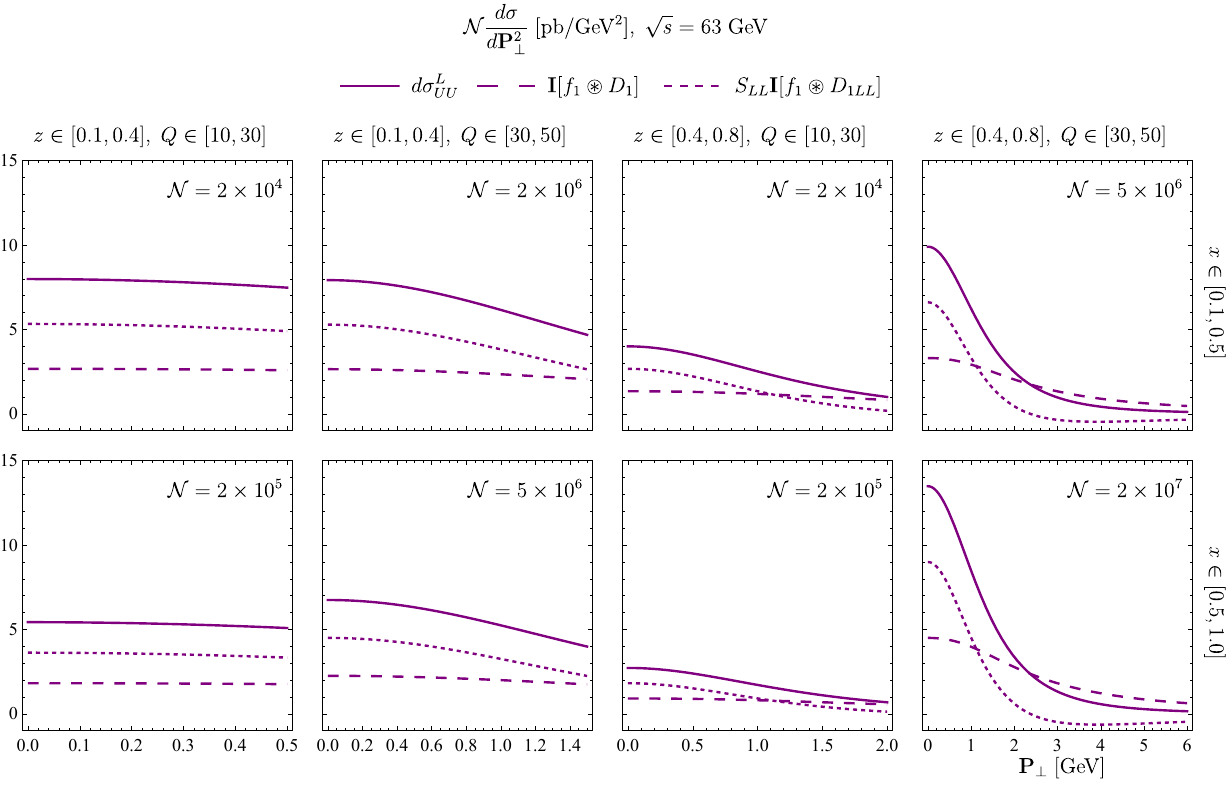}
    \caption[SIDIS cross section for an unpolarized beam and target, producing longitudinally polarized $J/\psi$.]{SIDIS cross section for an unpolarized beam and target, producing longitudinally polarized $J/\psi$.  Figure from Ref.~\cite{Copeland:2023wbu}.}
    \label{fig: L cross sections}
\end{figure}

\begin{figure}
    \centering
    \includegraphics[width = \linewidth]{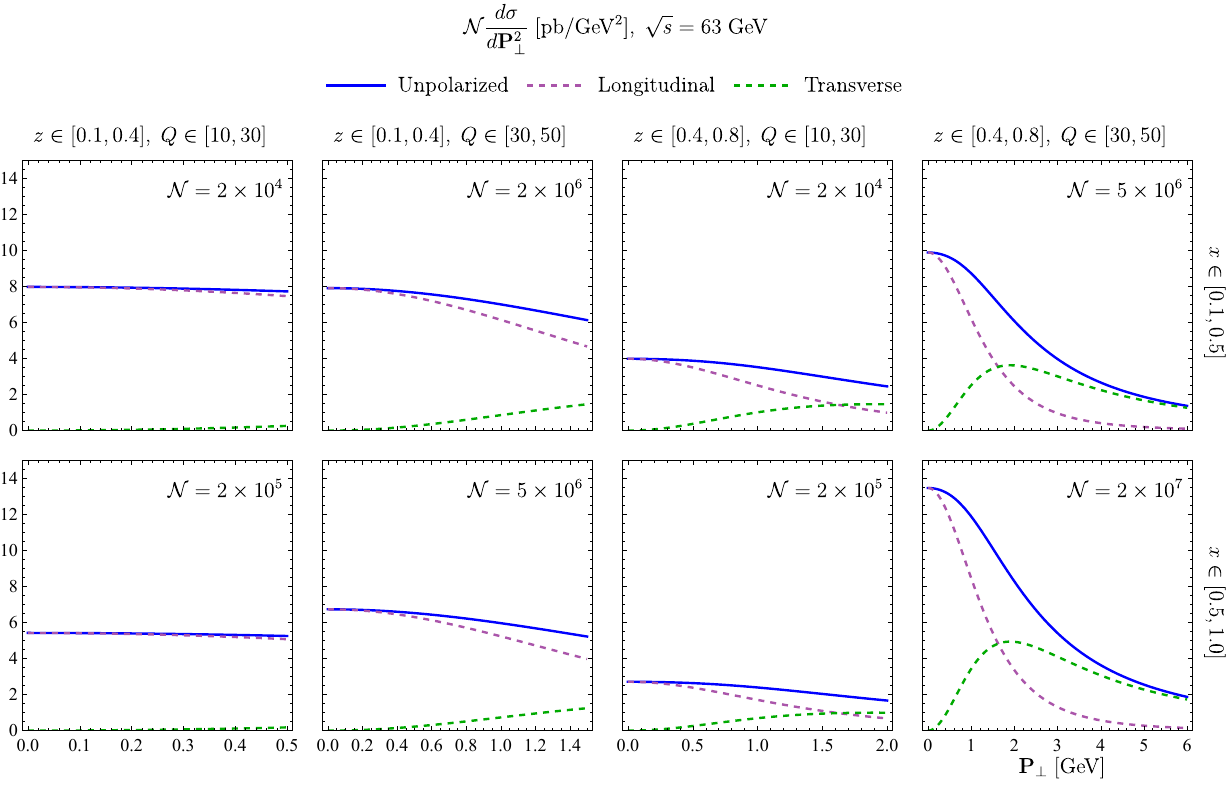}
    \caption[SIDIS cross section for an unpolarized beam and target, producing different polarizations of $J/\psi$.]{SIDIS cross section for an unpolarized beam and target, producing different polarizations of $J/\psi$.  Figure from Ref.~\cite{Copeland:2023wbu}.}
    \label{fig: A cross sections}
\end{figure}

\begin{figure}
    \centering
    \includegraphics[width = \linewidth]{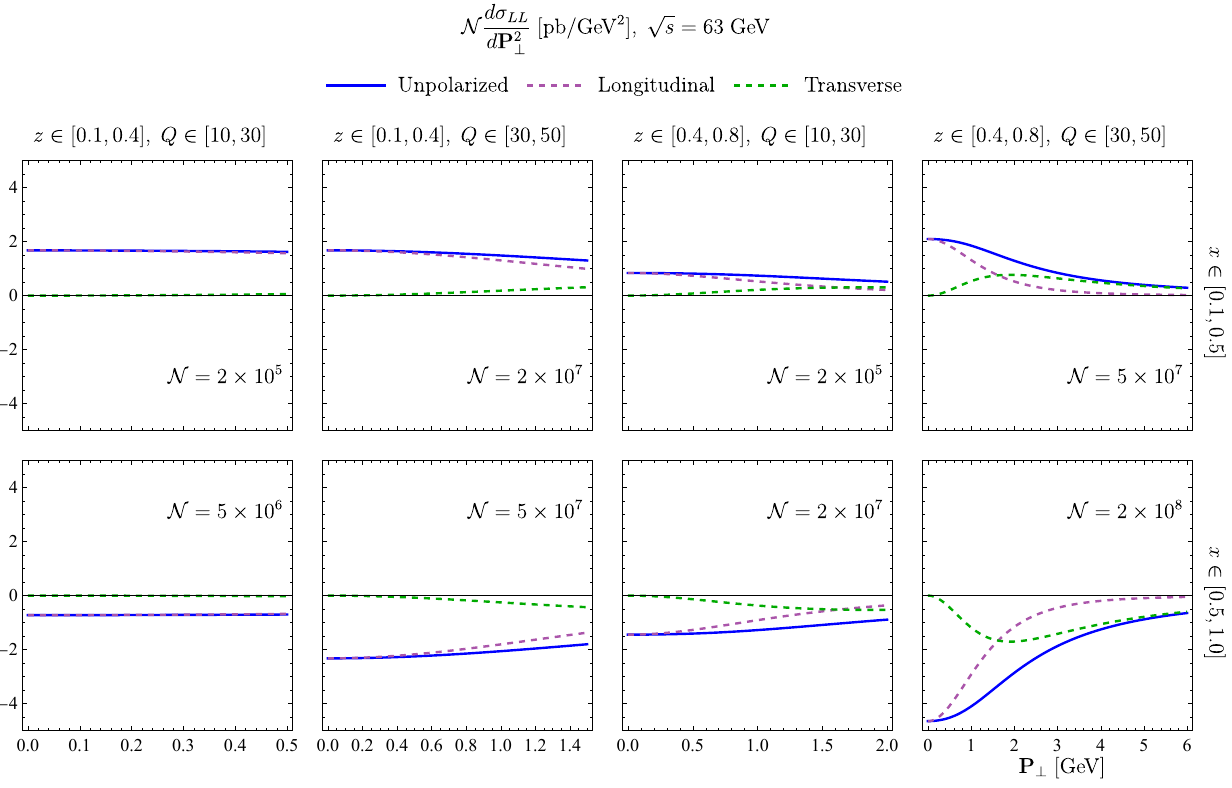}
    \caption[SIDIS cross section for a longitudinally polarized beam and target, producing different polarizations of $J/\psi$.]{SIDIS cross section for a longitudinally polarized beam and target, producing different polarizations of $J/\psi$.  Figure from Ref.~\cite{Copeland:2023wbu}.}
    \label{fig: A cross sections LL}
\end{figure}

\end{landscape}
\chapter{Competing production mechanisms for polarized $J/\psi$ in SIDIS}
\label{chap:JPsiProduction}

We\footnote{The work presented in this chapter was initially published in Ref.~\cite{Copeland:2023qed}. The contributions of each author are listed below. \begin{itemize}
    \item R.~Hodges, M.~Copeland, S.~Fleming: analysis and writing
    \item R.~Gupta: analysis
    \item T.~Mehen: checking calculations and editing manuscript
\end{itemize}}  established in the previous chapter that PDFs and FFs are fundamental objects regarding the internal structure of nucleons and hadrons.  Furthermore, the dependence of PDFs and FFs on the transverse momentum of the partons is a research area where much is yet to be understood.  We derived, for the first time, the polarized TMDFFs for $J/\psi$ production in SIDIS.  Fragmentation is not the only production mechanism for $J/\psi$, however: there can also be direct production via photon-gluon fusion (PGF).  This mechanism is sensitive to gluon PDFs, and also to different NRQCD LDMEs than light quark fragmentation (LQF).  Due to the fact that little is known about spin-dependent gluon TMDs \cite{AbdulKhalek:2021gbh}, and the uncertainty surrounding the NRQCD polarization puzzle and the LDME values (discussed in Sec.~\ref{sec: uncertainties in NRQCD}), we are motivated to find kinematic regimes where different $J/\psi$ production mechanisms dominate the cross section.  Identifying these regimes would allow for the uncertain quantities to be extracted more readily in future experiments.  

TMD direct production of quarkonium has been studied in many papers \cite{Lee:2021oqr,Catani:2014qha,Ma:2014svb,Kang:2014tta,Sun:2012vc,Catani:2010pd,Mukherjee:2016cjw,Mukherjee:2015smo,Boer:2012bt,Echevarria:2019ynx,Fleming:2019pzj,DAlesio:2021yws,Boer:2020bbd,Bor:2022fga,Kishore:2021vsm,Scarpa:2019fol,DAlesio:2019qpk,Bacchetta:2018ivt,Mukherjee:2016qxa,Rajesh:2018qks,Godbole:2013bca,Godbole:2012bx,denDunnen:2014kjo,Kang:2014pya,Zhu:2013yxa,Echevarria:2020qjk}.  However, all but one \cite{Echevarria:2020qjk} of these papers neglect LQF, which we find to have a significant contribution, and only a couple \cite{DAlesio:2021yws,Kang:2014pya} consider polarized $J/\psi$.  Our research, initially presented in Ref.~\cite{Copeland:2023qed}, is intended to extend the analysis of Ref.~\cite{Echevarria:2020qjk} to account for polarized $J/\psi$ and color octet PGF.

\begin{figure}
    \centering
    \includegraphics[width = \linewidth]{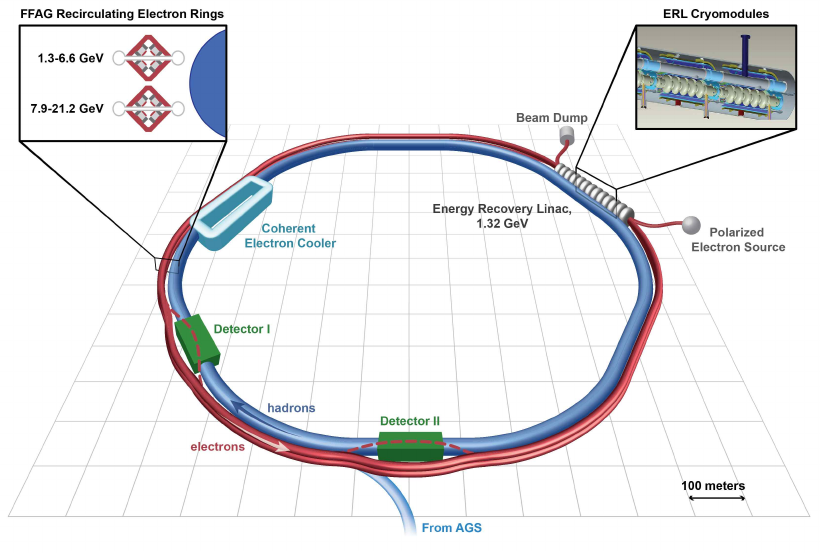}
    \caption[Schematic of the EIC at BNL.]{Schematic of the EIC at BNL. Red is the new electron beam facility, and blue is the existing RHIC hadron beam. Figure from Ref.~\cite{Accardi:2012qut}.}
    \label{fig: EIC schematic}
\end{figure}

\begin{figure}
    \centering
    \includegraphics[width = \linewidth]{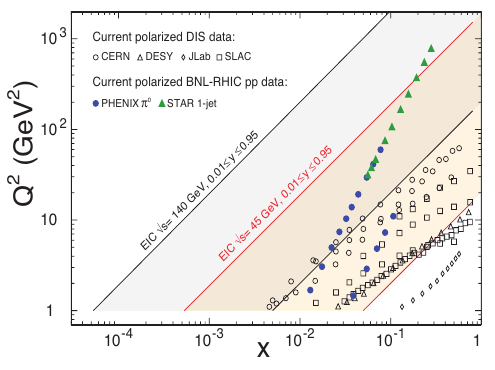}
    \caption[Plot showing the range of $Q^2$ and $x$ the future EIC will be able to access in $e+p$ collisions, compared to existing experiments.]{Plot showing the range of $Q^2$ and $x$ the future EIC will be able to access in $e+p$ collisions, compared to existing experiments.  Figure from Ref.~\cite{Accardi:2012qut}.}
    \label{fig: EIC kinematic range}
\end{figure}

The energies we consider will be accessible at the future Electron-Ion Collider (EIC) at Brookhaven National Lab (BNL), which will make use of the existing facilities of the Relativistic Heavy Ion Collider (RHIC), including its tunnel and polarized proton and nuclear beams, along with a new electron beam facility. A schematic of the EIC design is shown in Fig.~\ref{fig: EIC schematic}.  According to the EIC White Paper \cite{Accardi:2012qut}, some of the most relevant characteristics the new EIC will aim to have are: 
\begin{itemize}
    \item highly polarized ($\sim 70$\%) electron and nucleon beams,
    \item center of mass energies from $20$ to $100$ GeV, upgradable to $140$ GeV,
    \item collision luminosity of $\sim 10^{33-34} \; {\rm cm}^{-2} \, {\rm s}^{-1}$.
\end{itemize}
The EIC will be able to access both higher $Q^2$ and smaller $x$ than previous experiments at CERN, DESY, JLab, SLAC, PHENIX, and STAR (Fig.~\ref{fig: EIC kinematic range}).  Relevant to this chapter is that it will access $Q^2 > 100 \; {\rm GeV}^2$, for which there is no data for polarized DIS.  This will allow us to distinguish between different $J/\psi$ production mechanisms, which have different hierarchies as $Q^2$ increases.

\section{Uncertainties in NRQCD}
\label{sec: uncertainties in NRQCD}

\begin{figure*}[t]
\centering
\begin{minipage}{0.5\textwidth}
\centering
\subfloat[]{\includegraphics[scale=1.3]{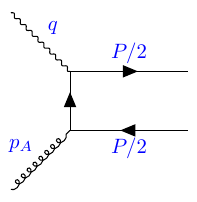}\label{octet}}
\end{minipage}%
\begin{minipage}{0.5\textwidth}
\centering
\subfloat[]{\includegraphics[scale=1.3]{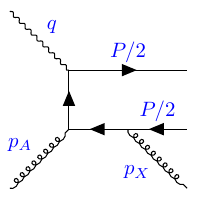}\label{singlet}}
\end{minipage}
\caption[Diagrams at LO and NLO in $\alpha_s$ contributing to $J/\psi$ production via photon-gluon fusion.]{Diagrams at LO and NLO in $\alpha_s$ contributing to $J/\psi$ production via photon-gluon fusion. There is a crossed diagram for (a), and permuting the vertices yields five other graphs like (b).  Figures from Ref.~\cite{Copeland:2023qed}.}
\label{fig: photon-gluon fusion}
\end{figure*}

In addition to giving an avenue to access polarized TMD PDFs, a key motivation behind this research project is to provide an opportunity to shed light on various uncertainties surrounding the NRQCD factorization formalism.  While LQF is only sensitive to the color octet $^3S_1$ LDME, at the same order in $\alpha_s$ and $v$, PGF (whose leading contributions are shown in Fig.~\ref{fig: photon-gluon fusion}) has contributions from three more LDMEs: $\Otsosing$, $\Otpz$, and $\Oosz$.  According to the velocity scaling rules, these scale as $v^3$, $v^7$, and $v^7$, respectively, while $\Otsooct$ also scales as $v^7$.  The contribution from $\Otsosing$ is at roughly the same order as the others in our double power counting, since an extra factor of $\alpha_s$ (as in Fig.~\ref{singlet}) is required to produce a $c\bar{c}$ with those quantum numbers.

\subsection{Polarization puzzle}
\label{sec: polarization puzzle}

One of the key points of tension between the NRQCD factorization formalism and the experimental results is that NRQCD predicts that the $J/\psi$ is transversely polarized at high $P_T$, but the data indicates it is unpolarized in that region.  This is called the ``polarization puzzle''.  

\begin{figure*}[t]
\centering
\includegraphics[scale=1.1]{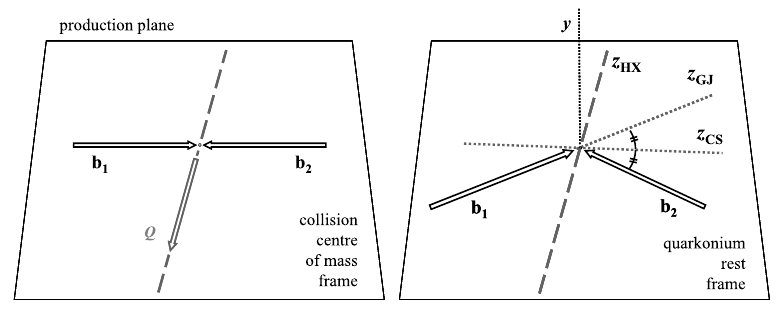} \\
\includegraphics[scale=1.1]{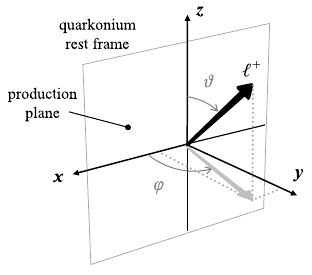}
\caption[The conventions for the $z$ axis and angles $\theta$, $\phi$.]{The conventions for the $z$ axis (we use the helicity convention, $z_{\rm HX}$) and angles $\theta$, $\phi$.  The $b_i$ are the incident beams and $Q$ is the quarkonium.  Figures from Ref.~\cite{Faccioli:2010kd}.}
\label{fig: Jpsi coordinates}
\end{figure*}

The polarization of the final state $J/\psi$ can be measured by looking at the angular distribution of the decay to a lepton pair $\ell^+ \ell^-$.  The coordinate system for this distribution can be confusing and there are a number of different conventions for the $z$ axis.  We use the helicity convention, which can be defined as follows.  
\begin{itemize}
    \item Start in the center of mass frame of the two beams, and let the direction of the $J/\psi$ momentum be the $z$ axis.
    \item Define the $x$ axis such that the $x$-$z$ plane is the production plane, i.e., the plane formed by the incident beams.
    \item Boost to the $J/\psi$ rest frame.  The $x$-$z$ plane is preserved, but now the momenta of the beams are no longer back-to-back.
\end{itemize}
This coordinate system is illustrated in Fig.~\ref{fig: Jpsi coordinates}.  Define the angles according to the $\ell^+$ momentum vector's orientiation. The polar angle $\theta$ is the angle the $\ell^+$ momentum makes with the $z$ axis, and the azimuthal angle $\phi$ is with respect to the production plane.  Then the probability distribution for the $\ell^+\ell^-$ decay is: \cite{Faccioli:2010kd}
\begin{equation}
    W(\theta,\phi) = \frac{1}{3+\lambda_\theta}(1+\lambda_\theta \cos^2 \theta + \lambda_\phi \sin^2\theta \cos 2\phi + \lambda_{\theta\phi}\sin 2\theta \cos \phi) \; .
\end{equation}
Most studies average over $\phi$ and measure only the $\theta$ dependence.  The parameter of interest is the polar anisotropy parameter $\lambda_\theta$, which can be written in terms of the cross sections for unpolarized and longitudinally polarized $J/\psi$ production:
\begin{equation}
    \lambda_\theta \sim \frac{1-3\frac{\sigma_L}{\sigma_U}}{1+\frac{\sigma_L}{\sigma_U}} \; .
\end{equation}
From this one can clearly see that pure longitudinal polarization corresponds to $\lambda_\theta=-1$ and pure transverse polarization to $\lambda_\theta=+1$.

\begin{figure}
    \centering
    \includegraphics[width = \linewidth]{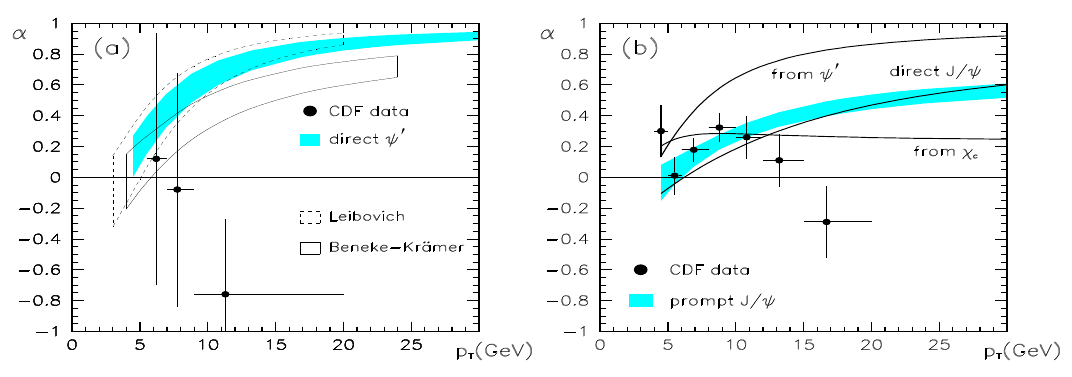}
    \caption[Plot showing the disagreement between NRQCD and data for the polarization of the $\psi^\prime$ and $J/\psi$ at high transverse momentum.]{Plot showing the disagreement between NRQCD and data for the polarization of the (a) $\psi^\prime$ and (b) $J/\psi$ at high transverse momentum.  Here, the parameter $\alpha$ is our $\lambda_\theta$.  Figure reused with permission from Ref.~\cite{Braaten:1999qk}; experimental data from Ref.~\cite{CDF:2000pfk}.}
    \label{fig: pol puzzle}
\end{figure}

To see why NRQCD predicts transverse polarization for the $J/\psi$ at high $P_T$, consider the following facts:
\begin{enumerate}
    \item Prevailing wisdom indicates that the color octet $^3S_1$ mechanism is dominant for $J/\psi$ production.\footnote{This will be discussed in more detail in Sec.~\ref{sec: determining dominant prod}.}
    \item The color octet $^3S_1$ configuration of the $c\bar{c}$ pair must undergo two chromoelectric transitions ${\bf A}\cdot \nabla$ to attain the quantum numbers of the $J/\psi$.  This operator preserves heavy quark spin symmetry, i.e., it is insensitive to the transformations of the spin of the heavy quark, unlike the chromomagnetic transition ${\bf B} \cdot \boldsymbol{\sigma}$.
    \item On-shell gluons must be transversely polarized, i.e., $p\cdot \varepsilon =0$.
\end{enumerate}
These support the following argument: at high $P_T$, the virtual gluon that produces the $c\bar{c}$ is almost on-shell, and therefore is mostly transversely polarized.  Because the operators that take the $c\bar{c}$ to the $J/\psi$ quantum numbers preserve heavy quark spin symmetry, the $J/\psi$ inherits the polarization of the virtual gluon.  Therefore, at high $P_T$, the $J/\psi$ should be mostly transversely polarized.  However, this is not at all consistent with data from the Collider Detector at Fermilab (CDF); see Fig.~\ref{fig: pol puzzle}.  This figure shows the polar anisotropy parameter $\alpha=\lambda_\theta$ as a function of $P_T$ for the production of the charmonium states $\psi^\prime$ and $J/\psi$.  For the $J/\psi$ plot, it shows several curves predicted by NRQCD.  The blue band is prompt $J/\psi$ production, which is all the production that does not arise via the decay of a $B$ meson.  This is further divided into direct production and feed-down production, the latter of which is e.g., production from the decay of a higher-mass charmonium bound state like the $\psi^\prime$ or $\chi_{c0}$.  The partonic processes we consider in this dissertation are an example of direct production.

\subsection{LDME fits}

The LDMEs are nonperturbative objects that need to be fit to experiment; however, there is significant disagreement between the three fits most commonly quoted in the literature \cite{Butenschoen:2011yh,Butenschoen:2012qr,Chao:2012iv,Bodwin:2014gia}; see Table~\ref{tab:LDMEs}.  The color singlet $^3S_1$ has a mostly consistent central value between the three fits, but the rest can vary by as much as a factor of 5, and the uncertainties are large in all cases.  

References \cite{Butenschoen:2011yh,Butenschoen:2012qr}, by Butensch\"on and Kniehl, are global fits to the world's data at NLO in NRQCD.  These authors were the first to show that the color octet mechanisms are crucial for aligning the NRQCD predictions for the $J/\psi$ production cross sections to data from many collaborations \cite{ZEUS:2002src,H1:2002voc,H1:2010udv,CDF:2004jtw,PHENIX:2009ghc,CMS:2010nis,CDF:1997ykw,CDF:1997uzj,Scomparin:2011zzb,LHCb:2011zfl}, and that the color singlet model alone is an underprediction.  Refer to Fig.~1 of Ref.~\cite{Butenschoen:2011yh} to see the agreement.  The main difference between the work of Butensch\"on and Kniehl and the fits of Refs.~\cite{Chao:2012iv,Bodwin:2014gia} is that the latter attempt to account for the NRQCD polarization puzzle, and only fit to the larger values of $P_T$, arguing that NRQCD factorization is not valid at small $P_T$.  It is worth noting that despite their disagreement between fits, the relative magnitudes of the different LDMEs are consistent with their $v$ scaling predicted by NRQCD.

By deriving new observables that are sensitive to all these matrix elements and the $J/\psi$ polarization, we hope that comparison to experiments at the future EIC can help resolve the NRQCD polarization puzzle and further constrain the LDMEs. Before delving into which production mechanisms are dominant in which regions of phase space, first we derive the cross sections themselves.

\begin{table*}
\caption{Fits of NRQCD LDMEs.}
\begin{tabular}{l||r|r|r|r|}
\cline{2-5}
 & $\begin{aligned} \braket{\mathcal{O}^{J/\psi}(^3S_1^{[1]})} & \\ \times \; {\rm GeV}^3 & \end{aligned}$ & $\begin{aligned} \braket{\mathcal{O}^{J/\psi}(^3S_1^{[8]})} & \\ \times 10^{-2} \; {\rm GeV}^3 & \end{aligned}$ & $\begin{aligned} \braket{\mathcal{O}^{J/\psi}(^1S_0^{[8]})} & \\ \times 10^{-2} \; {\rm GeV}^3 & \end{aligned}$ & $\begin{aligned} \braket{\mathcal{O}^{J/\psi}(^3P_0^{[8]})}/m_c^2 & \\ \times 10^{-2} \; {\rm GeV}^3 & \end{aligned}$ \\ \hline \hline
\multicolumn{1}{|l||}{B \& K \cite{Butenschoen:2011yh,Butenschoen:2012qr}} & $1.32\pm 0.20$ & $0.224\pm 0.59$ & $4.97\pm0.44$ & $-0.72\pm 0.88$ \\ \hline
\multicolumn{1}{|l||}{Chao et al. \cite{Chao:2012iv}} & $1.16\pm 0.20$ & $0.30\pm 0.12$ & $8.9\pm 0.98$ & $0.56\pm 0.21$ \\ \hline
\multicolumn{1}{|l||}{Bodwin et al. \cite{Bodwin:2014gia}} & $1.32 \pm 0.20$ & $1.1\pm 1.0$ & $9.9\pm 2.2$ & $0.49\pm 0.44$ \\ \hline
\end{tabular}
\label{tab:LDMEs}
\end{table*}

\section{Cross section expressions}

In this section we outline the cross sections for $J/\psi$ production via LQF and PGF, before their properties are discussed in the next section.

\subsection{Light quark fragmentation}

The cross section for production via light-quark fragmentation was discussed in Chap.~\ref{chap:FFs}.  We quote the results here for convenience.  For an unpolarized beam and target, and considering only unpolarized and longitudinally polarized $J/\psi$, we found
\begin{equation}
\begin{aligned}
\label{eq: fac cross}
    \frac{d \sigma_{UU}(l + H \to l' + J/\psi + X)}{dx ~dz ~dQ^2 ~d^2 {\bf P_\perp}} 
    = &\frac{4\pi \alpha^2}{Q^4} \left(1 - y +\frac{y^2}{2}\right) \bigg\{{\bf I}[f_1 D_1] + S_{LL} {\bf  I}[f_1 D_{1LL}]\bigg\} \; .\\
\end{aligned}
\end{equation}
The convolution integral is:
\begin{equation}
    \begin{aligned}
        {\bf I}[f~D] = \int d^2 \pv_T~d^2\kv_T~\delta^{(2)}(\pv_T-\kv_T+\Pv_\perp/z)~f(\pv_T)D(\kv_T) \; ,
    \end{aligned}
\end{equation}
and the relevant FFs are:
\begin{equation}
    \begin{aligned}
        D_1 = & \; \frac{2\alpha_s^2}{27\pi z M^3} \frac{z^2 \kv_T^2(z^2-2z+2)+2M^2(z-1)^2}{[z^2\kv_T^2+M^2(1-z)]^2} \Otsooct \; , \\
        D_{1LL} =  & \; \frac{2\alpha_s^2}{27\pi z M^3} \frac{z^2 \kv_T^2(z^2-2z+2)-4M^2(z-1)^2}{[z^2\kv_T^2+M^2(1-z)]^2} \Otsooct \; . \\
    \end{aligned}
\end{equation}
Recall that for unpolarized $J/\psi$, one sums over $S_{LL}\in \{1/2,1/2,-1\}$, and for longitudinally polarized $J/\psi$, $S_{LL}=-1$.  

\subsection{Photon-gluon fusion}

Collinear $J/\psi$ production via PGF has been studied in the NRQCD factorization formalism has been studied in previous literature \cite{Fleming:1997fq}.  Here we extend the analysis to have transverse momentum dependence.

The reference frame we use for this production mechanism is different than the one we used to derive the fragmentation cross sections.  Define $p$, $p_A$, $q$, and $P$ to be the proton, parton, virtual photon, and $J/\psi$ momenta, respectively.  We work in a reference frame where the photon has zero transverse momentum, and the photon and proton are back-to-back.
\begin{equation}
\begin{aligned}
    q^\mu = & \; (0,0,0, -Q)\; , \\
    p_A^\mu = & \; \frac{Q}{2x}(1, 0, 0, 1)\;, \\
    P^\mu = & \; \frac{1}{2}\left(\frac{\Pv_T^2+M^2}{z Q} + zQ, 2{\bf P_\perp}, 0,\frac{\Pv_T^2+M^2}{z Q} - zQ\right) \; .
\end{aligned}
\label{eq: breit frame}
\end{equation}
This is therefore a quasi-TMD framework, where the transverse momentum dependence of the parton is taken to be negligibly small, and so the transverse momentum dependence of the final state $J/\psi$ arises in the hadronization process.

Recall from the previous chapter that the $J/\psi$ longitudinal polarization vector is:
\begin{equation}
    \epsilon_L^\mu = \frac{1}{M}\left(|{\bf P}|, P^0 \hat{\bf P} \right) \; .
\end{equation}
This is frame-independent, and allows us to easily project out the desired polarizations of the $J/\psi$.

We consider two diagrams contributing to PGF: the $\mathcal{O}(\alpha_s)$ diagram in Fig.~\ref{octet} and the $\mathcal{O}(\alpha_s^2)$ diagram in Fig.~\ref{singlet}.  The former is proportional to two color octet NRQCD LDMEs, $\Oosz$ and $\Otpz$, while the latter is proportional to the color singlet LDME $\Otsosing$.  Here we state the results for the cross sections in the limit $Q^2 \gg M^2, \Pv_T^2$.  The full expressions are lengthy, and are provided in the Github repository  \cite{Copeland_JPsi_Production_NRQCD} associated with the initial publication of these results \cite{Copeland:2023qed}.
\begin{equation}
    \begin{aligned}
        \frac{d\sigma_U(^1S_0^{[8]})}{dx \, dz \, dQ^2 \, d\Pv_T^2} = & \; \frac{32\pi^3 z \alpha_{\rm em}^2 \alpha_s}{9MQ^6\zt^3}\left(1-y+\frac{y^2}{2}\right) \\
        & \times \delta(\zb)\delta^{(2)}(\Pv_T)f_g(x\zt) \Oosz + \mathcal{O}\bigg(\frac{M^2}{Q^2},\frac{\Pv_T^2}{Q^2}\bigg)  \; ,
    \end{aligned}
\end{equation}
\begin{equation}
    \begin{aligned}
        \frac{d\sigma_L(^1S_0^{[8]})}{dx \, dz \, dQ^2 \, d\Pv_T^2} = \frac{1}{3} \frac{d\sigma_U(^1S_0^{[8]})}{dx \, dz \, dQ^2 \, d\Pv_T^2} + \mathcal{O}\bigg(\frac{M^2}{Q^2},\frac{\Pv_T^2}{Q^2}\bigg) \; , 
    \end{aligned}
\end{equation}
\begin{equation}
    \begin{aligned}
        \frac{d\sigma_U(^3P_0^{[8]})}{dx \, dz \, dQ^2 \, d\Pv_T^2} =& \; \frac{64\pi^3 z \alpha_{\rm em}^2 \alpha_s}{9M^3 Q^6 \zt^5}\left[ y^2 \left(8+z(3z-8) \right) +8(1-y)\zt -2(1-y)z\zt^2 \right] \\
        & \times \delta(\zb)\delta^{(2)}(\Pv_T)f_g(x\zt) \Otpz + \mathcal{O}\bigg(\frac{M^2}{Q^2},\frac{\Pv_T^2}{Q^2}\bigg)  \; ,
    \end{aligned}
\end{equation}
\begin{equation}
    \begin{aligned}
        \frac{d\sigma_L(^3P_0^{[8]})}{dx \, dz \, dQ^2 \, d\Pv_T^2} =& \; \frac{64\pi^3 z^3 \alpha_{\rm em}^2 \alpha_s}{9M^3 Q^6 \zt^5}\left[(2-y)^2 -2z(1-y) \right] \\
        & \times \delta(\zb)\delta^{(2)}(\Pv_T)f_g(x\zt) \Otpz  \, .
    \end{aligned}
\end{equation}
\begin{equation}
    \begin{aligned}
        \frac{d\sigma_U(^3S_1^{[1]})}{dx \, dz \, dQ^2 \, d\Pv_T^2} =& \; \frac{512\pi \zb \alpha_{\rm em}^2 \alpha_s^2}{243M Q^6 z \zt^2(\Pv_T^2+\zb^2M^2)^2}\left(1-y+\frac{y^2}{2} \right)\left[\Pv_T^2+\zb^2M^2(2-z\zt) \right] \\
        & \times f_g(x) \Otsosing + \mathcal{O}\bigg(\frac{M^2}{Q^2},\frac{\Pv_T^2}{Q^2}\bigg) \; , 
    \end{aligned}
\end{equation}
\begin{equation}
    \begin{aligned}
        \frac{d\sigma_L(^3S_1^{[1]})}{dx \, dz \, dQ^2 \, d\Pv_T^2} =& \; \frac{512\pi \zb \alpha_{\rm em}^2 \alpha_s^2}{243M Q^6 z \zt^2(\Pv_T^2+\zb^2M^2)^2}\left(1-y+\frac{y^2}{2} \right)\Pv_T^2 \\
        & \times f_g(x) \Otsosing + \mathcal{O}\bigg(\frac{M^2}{Q^2},\frac{\Pv_T^2}{Q^2}\bigg)  \; .
    \end{aligned}
\label{eq: singlet L cs}
\end{equation}
Here, $\bar{z}\equiv 1-z$ and $\tilde{z} \equiv 2-z$.  The color octet and color singlet PGF contributions can clearly be differentiated according to their dependence on $z$.  Recall that in the proton rest frame, $z$ is the ratio of the energy of the $J/\psi$ to the energy of the photon.  Therefore $z\rightarrow 1$ is the limit where none of the $c\bar{c}$'s energy is lost via the emission of a hard gluon, and so we would expect Fig.~\ref{octet} to dominate.  Away from $z=1$, Fig.~\ref{singlet} should dominate.  This is reflected in the delta function in $(1-z)$ present in the color octet cross sections. 

When plotting the color octet PGF cross sections, we need to decide how to handle the delta functions. The delta function in $\Pv_T$ arises because we are doing an expansion in $P_T$ to leading order, and we replace it with a Gaussian centered at $\Pv_T=0$ in the plots.  Similarly, we can identify the delta function in $(1-z)$ as the leading order term of a shape function \cite{Neubert:1993ch,Beneke:1997qw}.  Effects from higher order in the $\alpha_s$ and $v$ expansions would yield a more complicated $z$ dependence, having the effect of ``smearing out'' this delta function.  The resulting $z$ dependence is contained in a so-called shape function that describes the physics near the endpoint.  We model these higher order effects as a Gaussian.
\begin{equation}
    \begin{aligned}
        \delta^{(2)}(\Pv_T) \rightarrow & \; \frac{1}{\pi \braket{P_T^2}}e^{-\Pv_T^2/\braket{P_T^2}} \; , \\
        \delta(1-z) \rightarrow & \; \frac{1}{\sqrt{\pi \braket{\bar{z}}}} e^{-(1-z)^2/\braket{\bar{z}}} \; .
    \end{aligned}
\end{equation}
We choose the parameters to be $\braket{P_T^2} = 0.25$ GeV$^2$ and $\braket{\bar{z}}=0.04$. The former corresponds to the smallest of the $|\Pv_T|$ upper bounds for the bins we consider in the next section.  The latter ensures we have non-negligible contributions from $z\in[0.7,1]$, which is a range we choose in reference to $J/\psi$ production in $e^+e^-$ collisions, where the NLO contributions have logarithms $(1-z)^{-1}\log{(1-z)}$ that become large around $z\approx 0.7$ \cite{Fleming:2003gt}.  In practice these parameters are somewhat arbitrary and would need to be extracted from experiment.

\section{Determining dominant production mechanisms}
\label{sec: determining dominant prod}

Echevarria et al.~\cite{Echevarria:2020qjk} were the first to compare the relative contribution to $J/\psi$ production of LQF and PGF, using TMD factorization. Our unpolarized $J/\psi$ cross sections presented in the preceding two sections are consistent with theirs.  However, Echevarria et al.~notably do not consider color octet PGF in their plots, as they work in the region $z \ll 1$, and do not take into account shape function effects as described above.  In the $Q^2 \gg M^2, \Pv_T^2$ limit, they compare their expressions for the $\Otsosing$ PGF and LQF cross sections, denoted by $d\sigma(\gamma^* g)$ and $d\sigma(\gamma^*q)$, respectively, and estimate the relative contribution of one to the other: \cite{Echevarria:2020qjk}
\begin{equation}
    \frac{d\sigma(\gamma^*g)}{d\sigma(\gamma^*q)} \sim \bigg( \frac{M}{Qv^2} \bigg)^2 \; .
\label{eq: relative scaling}
\end{equation}
For charmonium, $v^2 \sim 0.3$ and $M\sim 3$ GeV, so this reduces to:
\begin{equation}
    \frac{d\sigma(\gamma^*g\rightarrow c\bar{c})}{d\sigma(\gamma^*q\rightarrow c\bar{c})} \sim \frac{100 \; {\rm GeV}^2}{Q^2} \; .
\end{equation}
In other words, after about $Q\sim 10$ GeV the contribution from PGF should rapidly diminish compared to LQF.  This is indeed reflected in the plots of $d\sigma/d\Pv_T^2$ when only considering the $\Otsosing$ contribution to PGF.  However, in our results we find that the $\Oosz$ and $\Otpz$ contributions are non-negligible when replacing the delta function in $(1-z)$ with a Gaussian to account for shape function effects.  This is to be expected from the power counting.  The cross sections for production via PGF are evaluated at $\alpha_s(\mu^2=Q^2)$, and for the large $Q^2$ accessible at the EIC, $\alpha_s(Q^2) \sim 0.1$.  Since the $\Oosz$ and $\Otpz$ LDMEs scale as $v^7$ and $\Otsosing$ scales as $v^3$, the diagrams in Fig.~\ref{fig: photon-gluon fusion} are roughly the same order in the double power counting. Therefore, Eq.~(\ref{eq: relative scaling}) should still hold for the color octet process, albeit with an exponential suppression away from $z=1$.

As in the previous chapter, we plot the differential cross sections in eight bins to be consistent with Ref.~\cite{Echevarria:2020qjk}, for unpolarized and longitudinally polarized $J/\psi$.\footnote{For the LDMEs, we will use the fits from Ref.~\cite{Chao:2012iv}.  The fits of Ref.~\cite{Bodwin:2014gia} would increase the LQF curves by a factor of about four.
}  For unpolarized $J/\psi$ (Fig.~\ref{fig: cross sections 1}), PGF is most significant in the smaller $x$ bin, $x\in[0.1,0.5]$.  In particular, PGF is most dominant over LQF in the bin $x \in[0.1,0.5]$, $z\in[0.4,0.8]$, and $Q \in [10,30]$ GeV.  This fact would be unchanged if using the LDME fits that increase the LQF contribution by a factor of four.  The color octet PGF contribution only becomes noticeable at larger $z$, which is to be expected due to the presence of the $\delta(1-z)$ in those cross sections.

These observations about which mechanisms are dominant in which bins are mostly unchanged when looking at longitudinally polarized $J/\psi$ (Fig.~\ref{fig: cross sections long}). However, it is notable that the color singlet mechanism now goes to zero as $\Pv_T \rightarrow 0$, which is clear from the expression in Eq.~(\ref{eq: singlet L cs}).    

\begin{landscape}

\begin{figure}
    \centering
    \includegraphics[width = \linewidth]{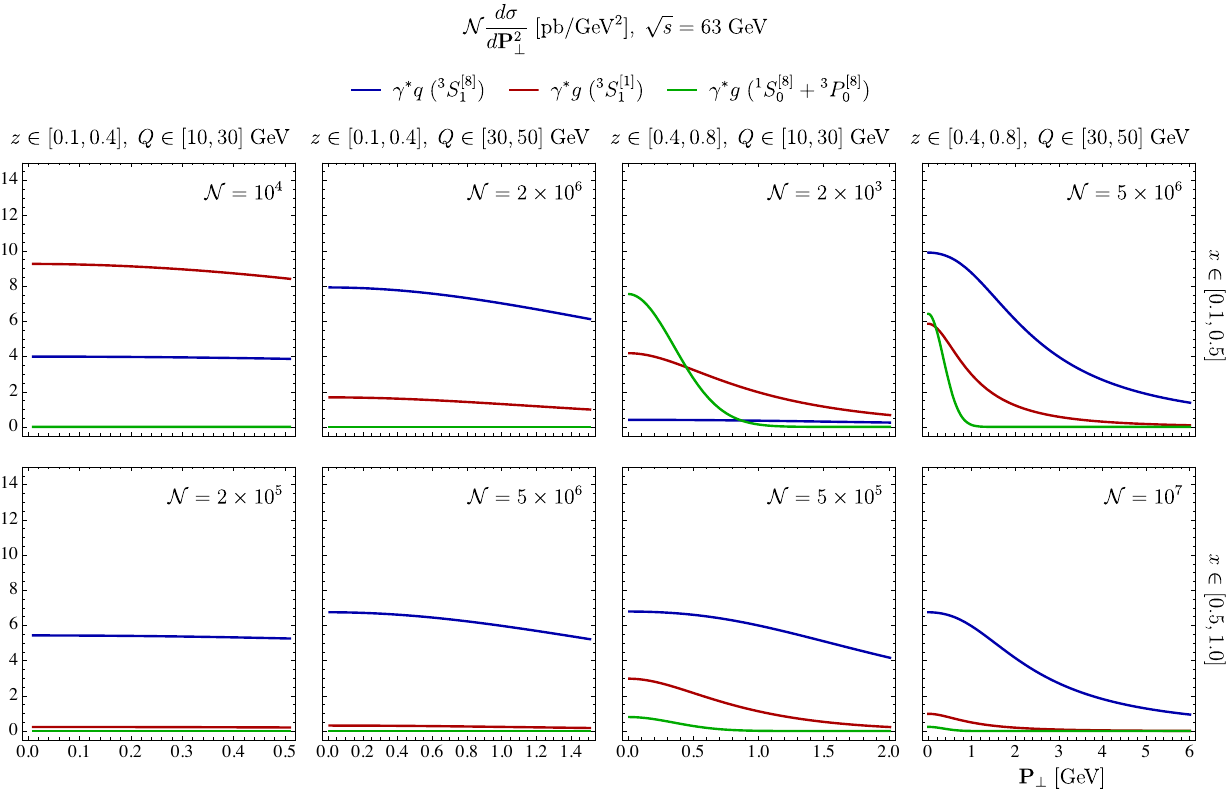}
    \caption[Contributions to the cross section for the production of unpolarized $J/\psi$ from three production mechanisms.]{Contributions to the cross section for the production of unpolarized $J/\psi$ from three production mechanisms.  Each plot is a bin integrated over the subset of the integration domains indicated. Figure from Ref.~\cite{Copeland:2023qed}.}
    \label{fig: cross sections 1}
\end{figure}

\begin{figure}
    \centering
    \includegraphics[width = \linewidth]{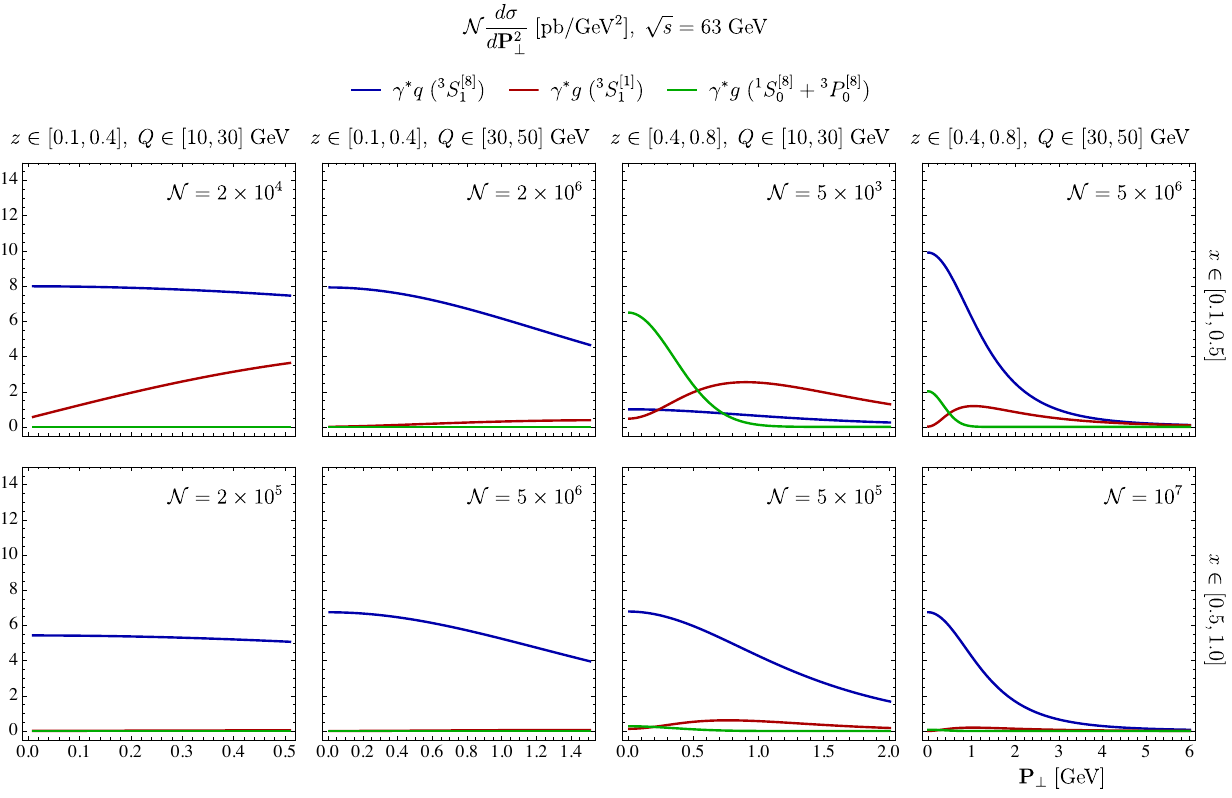}
    \caption[Contributions to the cross section for the production of longitudinally polarized $J/\psi$ from three production mechanisms.]{Contributions to the cross section for the production of longitudinally polarized $J/\psi$ from three production mechanisms.  Each plot is a bin integrated over the subset of the integration domains indicated. Figure from Ref.~\cite{Copeland:2023qed}.}
    \label{fig: cross sections long}
\end{figure}

\end{landscape}

\section{Angular distributions \& asymmetries}

Earlier we mentioned that the final state leptons have an angular distribution according to \cite{Faccioli:2010kd}
\begin{equation}
    W(\theta,\phi) \sim \frac{1}{3+\lambda_\theta}(1+\lambda_\theta \cos^2 \theta + \lambda_\phi \sin^2\theta \cos 2\phi + \lambda_{\theta\phi}\sin 2\theta \cos \phi) \; .
\label{eq: ang dist}
\end{equation}
Recall that the polar anisotropy parameter $\lambda_\theta$ is a measure the polarization of the $J/\psi$.  We can plot the contribution to $\lambda_\theta$ from each of the bins (Fig.~\ref{fig: lambda theta}), and we see that most contribute to a longitudinally polarized $J/\psi$.  However, at larger $\Pv_T$ the contribution tends towards transverse polarization, which is consistent with previous conclusions from NRQCD.  

The $\cos{2\phi_h}$ azimuthal asymmetry is a frequent subject of research \cite{Bor:2022fga,DAlesio:2021yws,Kishore:2021vsm,Boer:2020bbd,Scarpa:2019fol,DAlesio:2019qpk,Bacchetta:2018ivt,Mukherjee:2016qxa}.  This angle $\phi_h$ is distinct from the $\phi$ in Eq.~(\ref{eq: ang dist}); $\phi_h$ is the angle between the $J/\psi$ production plane and the lepton plane.  The asymmetry is defined by projecting out the $\cos 2\phi_h$ contribution from the cross section:
\begin{equation}
    A_{\cos 2\phi_h}(\Pv_T) = \frac{\int d\phi \; d\sigma \cos{2\phi_h}}{\int d\phi \; d\sigma} \; .
\end{equation}
The LQF cross sections have no azimuthal dependence, so they do not contribute to the numerator of this expression.  This suggests that $A_{\cos 2\phi_h}$ is a good avenue to extract gluon TMDs.  However, the inclusion of LQF in the denominator greatly suppresses the asymmetry (Fig.~\ref{fig: Asym}).  The suppression is smallest in the bins where LQF is less dominant compared to PGF, and the bin with the largest magnitude asymmetry is $z\in[0.4,0.8]$, $Q\in [10,30]$ GeV, $x\in[0.1,0.5]$.    

\begin{landscape}

\begin{figure}
    \centering
    \includegraphics[width = \linewidth]{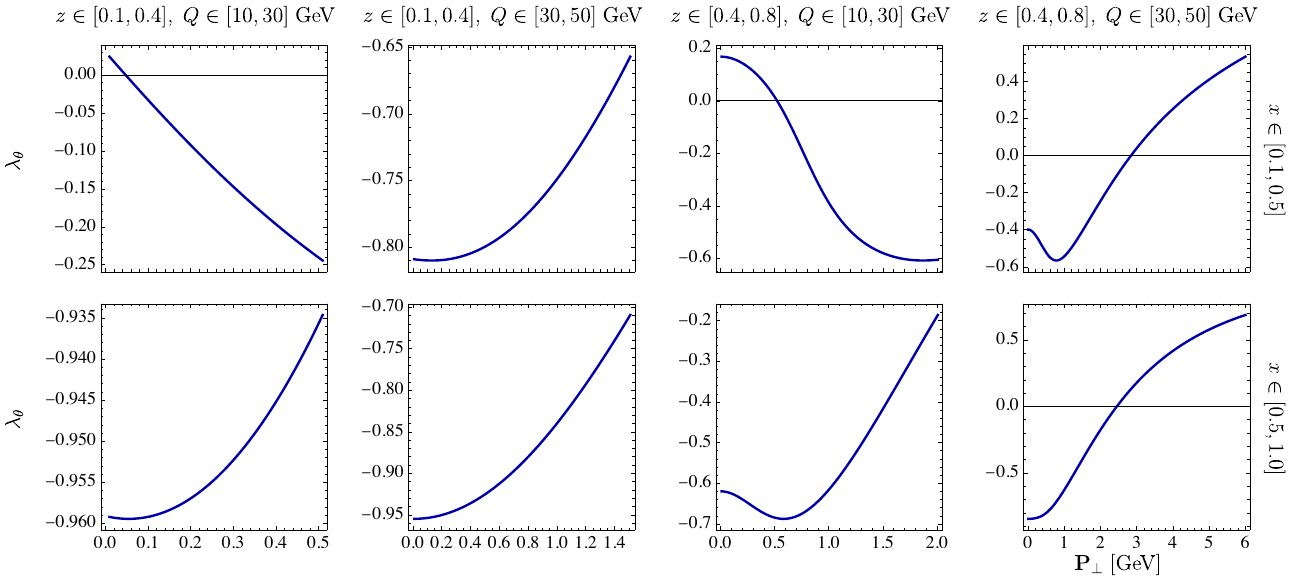}
    \caption[Contributions to the decay distribution parameter $\lambda_\theta$.]{Contributions to the decay distribution parameter $\lambda_\theta$. Figure from Ref.~\cite{Copeland:2023qed}.}
    \label{fig: lambda theta}
\end{figure}

\begin{figure}
    \centering
    \includegraphics[width = \linewidth]{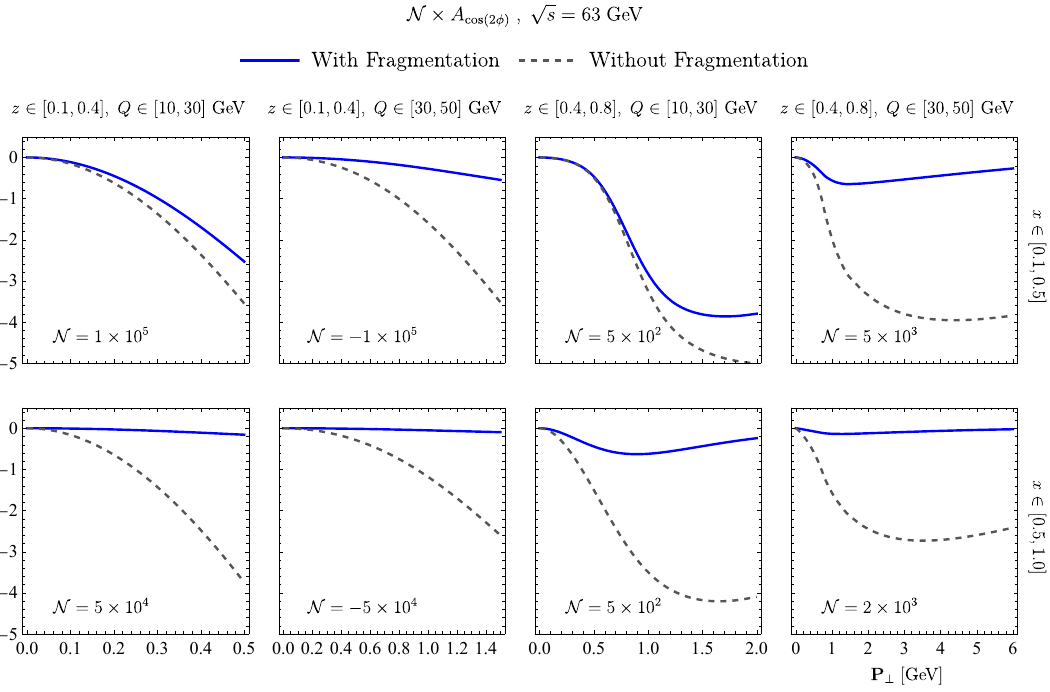}
    \caption[Contributions to the $\cos(2\phi_h)$ azimuthal asymmetry for unpolarized $J/\psi$ production.]{Contributions to the $\cos(2\phi_h)$ azimuthal asymmetry for unpolarized $J/\psi$ production. Figure from Ref.~\cite{Copeland:2023qed}.}
    \label{fig: Asym}
\end{figure}

\end{landscape}
\chapter{Conclusion}
\label{chap:conclusion}

Quantum chromodynamics is a sector of the Standard Model that has plentiful interesting research areas.  Beyond the everyday realm of protons and neutrons, it also describes bound states with heavier quarks, like the $T_{cc}^+$ exotic meson and $J/\psi$.  A crucial element of learning about these systems is the interplay between theory and experiment.  This dissertation has focused on theoretical methods that attempt to predict the outcomes of experiments at particle colliders -- including both explaining past experiments and motivating new ones.  In this way we hope to gain greater insight into QCD and nuclear matter.

In Chap.~\ref{chap:intro} we framed QCD in the context of the Standard Model that describes most of what we know about particle physics.  We also gave the background on effective field theory: a formalism that can make calculations easier by taking what we know about a system, and re-organizing it in a systematic way that highlights the physics we care about.  Furthermore, we discussed how the parton model and factorization can be useful tools in describing scattering experiments that probe the inner structure of nucleons.

Chapter \ref{chap:Tcc} delved into an EFT for the $T_{cc}^+$ developed out of an earlier theory for the $\chi_{c1}(3872)$.  We showed how the theory correctly predicts the total decay width and differential decay distributions, as measured by the LHCb collaboration.  The findings reinforce the view of the $T_{cc}^+$ as a weakly bound molecule of a $D^*D$ pair, and showcase the remarkable utility of EFT models in explaining experimental results.  Future directions in this research could be to perform statistical analyses of the data to further constrain the NLO couplings and thereby make the predictions of the effective theory more precise.

Chapters \ref{chap:FFs} and \ref{chap:JPsiProduction} both dealt with transverse momentum dependent $J/\psi$ production in the formalism of NRQCD factorization.  Chapter \ref{chap:FFs} was mostly a theoretical endeavor that explained our efforts to write down, for the first time, all of the polarized TMD quark and gluon fragmentation functions.  These FFs were then used in Chap.~\ref{chap:JPsiProduction} to look at production via both light quark fragmentation and photon gluon fusion for large $Q^2$. No current measurements exist in this regime, but we made a case for $J/\psi$ production to be studied at the future Electron Ion Collider as a way to learn about gluon TMDs and extract the NRQCD LDMEs.  Further constraining the TMDs and LDMEs would allow for more precise theoretical calculations in nuclear physics.  Future directions in this research could be to compute the TMD FFs to further orders in $\alpha_s$ and $v$, and to develop a TMD version of NRQCD factorization.


\titleformat{\chapter}
[display] 
{\raggedleft\LARGE\bfseries\usefont{T1}{ppl}{b}{n}\fontsize{18pt}{\baselineskip}} 
{\thechapter} 
	{.5em} 
{\normalbaselines} 

\begin{appendices}
	\titleformat{\chapter}[display]
	{
		\raggedleft \LARGE \bfseries \usefont{T1}{ppl}{b}{n}\fontsize{18pt}{\normalbaselines}
		\normalbaselines
		\usefont{T1}{ppl}{b}{n}\fontsize{18pt}{\normalbaselineskip}}
	{Appendix \thechapter}{.5em}{\normalbaselines}

    \chapter{Supplementary material: $T_{cc}^+$ decays}

\section{$D$ meson-pion interaction terms}
\label{app: interaction terms}

Here we see how to write down terms in the $T_{cc}^+$ interaction Lagrangian by constructing isospin invariants.  The charm mesons are an isospin-$1/2$ field, and the pions are an isospin-$1$ field; represent them with $\psi$ and $v$, respectively.
\begin{equation}
    \psi_\alpha = \begin{pmatrix} D^+ \\ D^0 \end{pmatrix} \; , \qquad v_i =  \frac{1}{\sqrt{2}}\begin{pmatrix} \pi_1 \\ \pi_2 \\ \pi_0 \end{pmatrix}  \; .
\end{equation}
By simple addition of angular momentum, to construct interaction terms between one charm meson and one pion, we can get either isospin-$1/2$ or isospin-$3/2$ terms.  For the isospin-$1/2$ terms, we need to have one free spinor index.  We can write down
\begin{equation}
    O_{1/2} = {\bf v} \cdot \boldsymbol{\sigma}\psi = v_i \sigma^i_{\alpha \beta}\psi_\beta \; .
\end{equation}
For the isospin-$3/2$ term, we need to have a free vector index.
\begin{equation}
    O^i_{3/2} = v^i \psi - \frac{1}{3} \sigma^i O_{1/2} \; ,
\end{equation}
where the second term is necessary to make $O_{3/2}^i$ orthogonal to $O_{1/2}$.  The appropriate terms in the Lagrangian are then
\begin{equation}
    \mathcal{L}_{C_\pi} = c_{1/2} O^\dagger_{1/2} O_{1/2} + c_{3/2} O^{\dagger i}_{3/2}O^i_{3/2} \; .
\end{equation}
We can then multiply out these operators to get the interaction terms in terms of the charm meson and pion fields themselves.  Redefining the couplings as linear combinations of the $c_{1/2}$ and $c_{3/2}$, we get:
\begin{equation}
\begin{aligned}
    {\mathcal L}_{C_\pi} \rightarrow & \; C_\pi^{(1)}D^{0\dagger}\pi^{0\dagger} D^+ \pi^- - C_\pi^{(1)} D^{+\dagger}\pi^{0\dagger} D^0 \pi^+ + {\rm H.c.}   \\
    & + C_\pi^{(2)}D^{0\dagger}\pi^{0\dagger} D^0 \pi^0 + C_\pi^{(2)} D^{+\dagger}\pi^{0\dagger} D^+ \pi^0  \\
    & + C_\pi^{(3)}D^{0\dagger}\pi^{+\dagger} D^0 \pi^+ \; ,
\end{aligned}
\end{equation}
where
\begin{equation}
    \begin{aligned}
        C_\pi^{(1)} = & \; \frac{\sqrt{2}}{3}c_{3/2} - \frac{\sqrt{2}}{3}c_{1/2} \; , \\
        C_\pi^{(2)} = & \; \frac{2}{3}c_{3/2} + \frac{1}{3}c_{1/2} \; , \\
        C_\pi^{(3)} = & \; \frac{1}{3}c_{3/2} + \frac{2}{3}c_{1/2} \; .
    \end{aligned}
\end{equation}
For the $B_1$ terms we proceed similarly, but with two pseudoscalar charm mesons and a pion on one side of the vertex, and a vector and pseudoscalar charm meson on the other.
\begin{equation}
    \begin{aligned}
        \mathcal{L}_{B_1} = & \; b_0 \varepsilon_{\alpha \beta} (\psi_\alpha^* \psi_\beta)^\dagger \psi_\rho (\sigma^2 \sigma^i)_{\rho \tau} \psi_\tau v^i \\
        & + b_1 [\psi_\alpha^* (\sigma^2 \sigma^k)_{\alpha \beta} \psi_\beta]^\dagger \varepsilon^{ijk} \psi_\rho (\sigma^2 \sigma^i)_{\rho \tau} \psi_\tau v^j + {\rm H.c.} \; ,
    \end{aligned}
\label{eq: B1 Lag}
\end{equation}
which in terms of the charm meson and pion fields becomes
\begin{equation} 
\begin{aligned}
    {\mathcal L}_{B_1} \rightarrow & \; B_1^{(1)}(D^+D^{*0})^\dagger(D^+D^0\nabla\pi^0) +B_1^{(2)}(D^0D^{*+})^\dagger(D^+D^0\nabla\pi^0)  \\
    & +\frac{B_1^{(3)}}{2}(D^0D^{*+})^\dagger(D^0D^0\nabla\pi^+) + \frac{B_1^{(4)}}{2}(D^+D^{*0})^\dagger(D^0D^0\nabla\pi^+) + \, {\rm H.c.}\; .  \\ 
\end{aligned}
\end{equation}
Here there are four $B_1$ couplings instead of the two in Eq.~(\ref{eq: B1 Lag}); this is because for the $T_{cc}^+$, isospin is only an approximate symmetry, so we need isospin-breaking terms in the Lagrangian to properly renormalize. In the isospin limit we would have
\begin{equation}
\begin{aligned}
    B_1^{(1)} =&\; - B_1^{(2)} = -\sqrt{2}b_0 \; ,  \\
    B_1^{(3)} =&\; 2(b_1+b_0) \; ,  \\
    B_1^{(4)} =&\; 2(b_1-b_0) \; . 
\end{aligned}
\end{equation}

The $C_{0D}$ terms merely need two pseudoscalar charm meson fields.
\begin{equation}
    \mathcal{L}_{C_{0D}} = c_0 (\varepsilon_{\alpha \beta} \psi_\alpha \psi_\beta)^\dagger \varepsilon_{\rho \tau}\psi_\rho \psi_\tau + c_1 [\psi_\alpha (\sigma^2 \sigma^k)_{\alpha \beta} \psi_\beta]^\dagger \psi_\rho (\sigma^2 \sigma^k)_{\rho \tau} \psi_\tau \; .
\end{equation}
In terms of the charm meson and pion fields:
\begin{equation}
\begin{aligned}
    {\mathcal L}_{C_{0D}} \rightarrow & \; \frac{C_{0D}^{(1)}}{2}(D^0D^0)^\dagger(D^0D^0) + C_{0D}^{(1)}(D^+ D^0)^\dagger (D^+ D^0) \; ,
\end{aligned}
\end{equation}
where $C_{0D}^{(1)} = 4c_1$.

\subsection{Obtaining numerical values for the $C_\pi$}

This analysis was first presented in Ref.~\cite{Dai:2019hrf}.

We can write down the charm mesons and pions as vectors n the isospin $\ket{I,m_I}$ basis.
\begin{equation}
\begin{aligned}
    \ket{\pi^+} =&\; -\ket{1,1} \, , \quad \ket{\pi^0} = \ket{1,0} \, ,  \\
    \ket{D^+} =&\; \Ket{\frac{1}{2},\frac{1}{2}} \, , \quad \ket{D^0} = \Ket{\frac{1}{2},-\frac{1}{2}} \, . 
\end{aligned}
\end{equation}
Then, using Clebsch-Gordan coefficients, we can write
\begin{align}
    \ket{D^0\pi^0} =&\; \sqrt{\frac{2}{3}}\Ket{\frac{3}{2},-\frac{1}{2}} + \frac{1}{\sqrt{3}}\Ket{\frac{1}{2},-\frac{1}{2}} \, ,  \nonumber \\
    \ket{D^+\pi^0} =&\; \sqrt{\frac{2}{3}}\Ket{\frac{3}{2},\frac{1}{2}} + \frac{1}{\sqrt{3}}\Ket{\frac{1}{2},\frac{1}{2}} \, ,  \\
    \ket{D^0\pi^+} =&\; -\sqrt{\frac{2}{3}}\Ket{\frac{1}{2},\frac{1}{2}} - \frac{1}{\sqrt{3}}\Ket{\frac{3}{2},\frac{1}{2}} \nonumber \, .
\end{align}
This implies that the scattering lengths are related by:
\begin{equation} \label{scattlen}
\begin{aligned}
    a_{D^0\pi^0} =&\; a_{D^+\pi^0} = \frac{2}{3}a_{D_\pi}^{3/2}+\frac{1}{3}a_{D\pi}^{1/2} \, ,  \\
    a_{D^0\pi^+} =&\; \frac{1}{3}a_{D\pi}^{3/2}+\frac{2}{3}a_{D\pi}^{1/2} \, .
\end{aligned}
\end{equation}
Lattice calculations in Ref.~\cite{Liu:2012zya} give $a_{D\pi}^{1/2} = 0.37_{-0.02}^{+0.03}\, {\rm fm}$ and $a_{D\pi}^{3/2} = -(0.100\pm0.002)\, {\rm fm}$.  We can then use $C_\pi = 4\pi(1+m_\pi/m_D)a_{D\pi}$ \cite{Dai:2019hrf} to get values for $C_\pi^{(2)}$ and $C_\pi^{(3)}$, and from there solve for $c_{1/2}$ and $c_{3/2}$ to obtain $C_\pi^{(1)}$.
\begin{equation}
\begin{aligned}
    C_\pi^{(1)} =&\; -3.0_{-0.40}^{+0.32}\, {\rm fm} \, ,  \\
    C_\pi^{(2)} =&\; -0.76_{-0.09}^{+0.14}\, {\rm fm} \, ,  \\
    C_\pi^{(3)} =&\; 2.9_{-0.2}^{+0.3}\, {\rm fm} \, . 
\end{aligned}
\end{equation}

\section{Transition magnetic moments} \label{app: transition magnetic moments}

At tree level the emission of a photon by a $D^*$ meson is given by the amplitude
\begin{equation}
    \mathcal{A} = i\mu \boldsymbol{\varepsilon}_{D} \cdot {\bf k}_\gamma \times \boldsymbol{\varepsilon}_\gamma \; ,
\end{equation}
for transition magnetic moment $\mu$.  The corresponding decay width is then
\begin{equation}
    \Gamma(D^* \rightarrow D \gamma) = \frac{1}{3\pi} \frac{m_D}{m_{D^*}}|\mu|^2 |{\bf k}_\gamma|^3 \; .
\end{equation}
We can then use the kinematics for a two-body decay to write
\begin{equation}
    |{\bf k}_\gamma| = \frac{m_{D^*}^2 - m_D^2}{2m_{D^*}} \; ,
\end{equation}
and solve for $\mu$ using known or derived values for the decay widths. The value $\Gamma(D^{*+}\rightarrow D^+\gamma) = 1.33$ keV is known from the Particle Data Group (PDG) \cite{Zyla:2020zbs}, and $\Gamma(D^{*0}\rightarrow D^0 \gamma)=19.9$ keV can be obtained by using isospin symmetry to relate $\Gamma[D^{*0}\to D^0 \pi^0]$ to $\Gamma[D^{*+}\to D^+ \pi^0]$, then writing in terms of the branching ratios $\text{Br}[D^{*0}\to D^{0}\gamma]$ and $\text{Br}[D^{*0}\to D^{0}\pi^0]$ from the PDG.   We get:
\begin{equation}
    \mu_{D^{0}} = \frac{2.80\times 10^{-4}}{\rm MeV} \; , \quad \mu_{D^{+}} = - \frac{7.34 \times 10^{-5}}{\rm MeV} \; .
\end{equation}
In HH$\chi$PT, these couplings have the form \cite{Amundson:1992yp,Stewart:1998ke}
\begin{equation}
    \mu_{D^0} = \frac{2e}{3}\beta +\frac{2e}{3m_c} , \quad  \mu_{D^+} = -\frac{e}{3}\beta +\frac{2e}{3m_c} \, .
\end{equation}
Our values are consistent with these in sign and relative magnitude.

\section{Power divergence subtraction scheme} \label{app: PDS scheme}
These integrals were first evaluated in Ref.~\cite{Guo:2010ak}.

The primary integral for which we use the PDS scheme is the following integral over a Cartesian momentum:
\begin{equation}
    \begin{aligned}
        \Sigma_{\rm dim \; reg}(c-i\epsilon) = \int \frac{d^{d-1}\lv}{(2\pi)^{d-1}} \frac{1}{\lv^2+c-i\epsilon} \; .
    \end{aligned}
\end{equation}
Here the integral is regularized using dimensional regularization.  The integral clearly has a linear divergence in $d=4$, since the radial integrand goes as $d|\lv|\;\lv^2/(\lv^2+c-i\epsilon)$.  However, using dimensional regularization yields a finite result for $d=4$:
\begin{equation}
    \Sigma_{\rm dim \; reg}(c-i\epsilon)|_{d=4} = - \frac{\sqrt{c-i\epsilon}}{4\pi} \, .
\end{equation}
It can be a useful to ensure that all divergences cancel as a consistency check that the effective theory is properly renormalized.  We therefore use the PDS scheme to make the linear divergence explicit.  However, the result is not finite in $d=3$.  We start by introducing a new dimensionless scale to keep track of the divergence.
\begin{equation}
    \begin{aligned}
        \Sigma_{\rm PDS}(c-i\epsilon) = \bigg(\frac{\Lambda_{\rm PDS}}{2}\bigg)^{4-d}\int \frac{d^{d-1}\lv}{(2\pi)^{d-1}} \frac{1}{\lv^2+c-i\epsilon} \; .
    \end{aligned}
\label{eq: PDS sigma}
\end{equation}
Evaluating in $d$ dimensions, we see the integral has a pole at $d=3$.  Expanding the result around the pole yields:
\begin{equation}
    \begin{aligned}
        \Sigma_{\rm PDS}(c-i\epsilon) = - \frac{\Lambda_{\rm PDS}}{4\pi(d-3)} + {\rm finite} \; .
    \end{aligned}
\end{equation}
We now redefine $\Sigma_{\rm PDS}$ to have a counterterm which cancels this divergence in $d=3$, $\Sigma_{\rm PDS} \rightarrow \Sigma_{\rm PDS} + \Lambda_{\rm PDS}/4\pi(d-3)$.
This redefined integral, when then evaluated in $d=4$, gives
\begin{equation}
    \begin{aligned}
        \Sigma_{\rm PDS}(c-i\epsilon)|_{d=4} = \frac{1}{4\pi}\big(\Lambda_{\rm PDS} - \sqrt{c-i\epsilon}\big) \; .
    \end{aligned}
\end{equation}
There is now a linear divergence in $\Lambda_{\rm PDS}$ which must cancel in our final answers for observables.

The other integrals in our calculation for which the PDS scheme is relevant are as follows.
\begin{equation}
    \begin{aligned}
        I({\bf p}) = \int \frac{d^{d-1}{\bf l}}{(2\pi)^{d-1}} \frac{1}{{\bf l}^2+c_1-i\epsilon} \frac{1}{{\bf l}^2-2b{\bf l}\cdot{\bf p}+c_2-i\epsilon} \; ,
    \end{aligned}
\end{equation}
\begin{equation}
    \begin{aligned}
        \pv^iI^{(1)}({\bf p}) =  \int \frac{d^{d-1}{\bf l}}{(2\pi)^{d-1}} \lv^i \frac{1}{{\bf l}^2+c_1-i\epsilon} \frac{1}{{\bf l}^2-2b{\bf l}\cdot{\bf p}+c_2-i\epsilon} \; ,
    \end{aligned}
\label{eq: linear tensor PDS}
\end{equation}
\begin{equation}
    \begin{aligned}
        \pv^i\pv^jI^{(2)}_0({\bf p})+\delta^{ij}\pv^2 I^{(2)}_1({\bf p}) = \int \frac{d^{d-1}{\bf l}}{(2\pi)^{d-1}} \lv^i \lv^j \frac{1}{{\bf l}^2+c_1-i\epsilon} \frac{1}{{\bf l}^2-2b{\bf l}\cdot{\bf p}+c_2-i\epsilon} \; .
    \end{aligned}
\end{equation}
The scalar integral $I(\pv)$ is finite in both $d=3$ and $d=4$, and so no counterterm is needed.
\begin{equation}
    \begin{aligned}
        I({\bf p}) = \frac{1}{8\pi}\frac{1}{\sqrt{b^2\pv^2}}\bigg[\tan^{-1}\bigg( \frac{c_2-c_1}{2\sqrt{b^2\pv^2c_1}}\bigg) + \tan^{-1}\bigg(\frac{2b^2\pv^2+c_1-c_2}{2\sqrt{b^2\pv^2(c_2-b^2\pv^2)}} \bigg) \bigg] \; .
    \end{aligned}
\end{equation}
To evaluate the linear tensor integral $I^{(1)}({\bf p})$, one contracts both sides of Eq.~(\ref{eq: linear tensor PDS}) with $\pv^i$, and algebraic manipulation of the numerator on the right-hand side yields two integrals of the form of Eq.~(\ref{eq: PDS sigma}).  The two have an opposite sign for the linear divergence in $\Lambda_{\rm PDS}$, and so $I^{(1)}({\bf p})$ is also finite.
\begin{equation}
    \begin{aligned}
        \pv^2I^{(1)}({\bf p})= \frac{1}{2b}\bigg[\frac{1}{4\pi}\sqrt{c_1-i\epsilon}-\frac{1}{4\pi}\sqrt{c_2-b^2\pv^2-i\epsilon}+(c_2-c_1)I(\pv)\bigg] \; .
    \end{aligned}
\end{equation}
The quadratic tensor integrals $I^{(2)}_0({\bf p})$ and $I^{(2)}_1({\bf p})$ have a subtlety in the PDS scheme.  After using Feynman parameters to combine the propagator denominators, it is necessary to perform the numerator algebra in $d=4$ before evaluating the momentum integral.  Otherwise, one obtains a different coefficient for the subtraction scale $\Lambda_{\rm PDS}$, and its dependence does not cancel in the final expression for the $T_{cc}^+$ decay width.  Likewise, algebraic manipulation of the numerator, like the method used to solve for $I^{(1)}(\pv)$, leads to a distinct incorrect coefficient for $\Lambda_{\rm PDS}$.  The correct procedure yields the following intermediate steps after evaluating the momentum integral:
\begin{equation}
    \begin{aligned}
        I_0^{(2)}(\pv) = & \; \frac{b^2}{8\pi} \int_0^1 dx \, \frac{x^2}{\sqrt{\Delta (x)}} \; , \\
        \pv^2I_1^{(2)}(\pv) =&\; \frac{1}{8\pi}\bigg[\frac{2}{3}\Lambda_{\rm PDS} - \int_0^1 dx\, \sqrt{\Delta (x)}\bigg] \; , 
    \end{aligned}
\label{eq: quad int weight}
\end{equation}
In our effective theory for the $T_{cc}^+$, all dependence on $\Lambda_{\rm PDS}$ cancels in the limit $\mu_0 = \mu_+$, which is an approximation we make in the cutoff-dependent terms. 

The final answers we obtain are:
\begin{equation}
    \begin{aligned}
        \pv^2 I_0^{(2)} =&\; -\frac{1}{16\pi}\sqrt{c_2-b^2\pv^2-i\epsilon} + \frac{c_1}{2}I(\pv)  +\frac{3}{4}\frac{c_2-c_1}{b}I^{(1)}(\pv) \; , \\
        \pv^2 I_1^{(2)} =&\; \frac{\Lambda_{\rm PDS}}{12\pi} -\frac{1}{16\pi}\sqrt{c_2-b^2\pv^2-i\epsilon} - \frac{c_1}{2}I(\pv)-\frac{1}{4}\frac{c_2-c_1}{b}I^{(1)}(\pv) \; .
    \end{aligned}
\end{equation}
%

\section{Expressions for $T_{cc}^+$ amplitudes and differential decay widths} \label{app: Tcc diagram expressions}

Here we give the expressions for the diagrams on the first two lines on the right-hand side of Fig.~\ref{bubbleDiagram}, and the decay diagrams in Fig.~\ref{DecayDiagrams}.  We have neglected terms in the propagators that go as $\pv^4/m_H^2$ or $(\delta m)\pv^2/m_H$, where $H$ is a charm meson and $\delta m \sim m_\pi$, since they are small compared to $\pv^2$.
\begin{equation}
    \begin{aligned}
            -i\Sigma_1(m,m^*) =&\; -\frac{i\mu(m,m^*)}{2\pi}[\Lambda_{\rm PDS}-\gamma(m,m^*)] \, , \\
    \end{aligned}
\end{equation}
\begin{equation}
    \begin{aligned}
        & -i\Sigma_2(m_1,m_1^*,m_2,m_2^*,m_\pi,g_1,g_2) \\
        =&\; -\frac{4ig_1 g_2}{3}\mu(m_1,m_1^*)\mu(m_2,m_2^*)   \bigg\{\frac{1}{16\pi^2}[\Lambda_{\rm PDS}-\gamma(m_1,m_1^*)]  [\Lambda_{\rm PDS}-\gamma(m_2,m_2^*)]  \\
        & + \frac{(m_2^*-m_1)^2-m_\pi^2}{(8\pi)^2} \bigg[\frac{1}{\epsilon}+2 -4 \log \bigg( \gamma(m_1,m_1^*)+\gamma(m_2,m_2^*)  \\
        & -i(m_2^*-m_1)^2+im_\pi^2 \bigg) -4\log \mu \bigg] \bigg\} \, , \\
    \end{aligned}
\end{equation}
\begin{equation}
    \begin{aligned}
        -i\Sigma_3(m_1,m_1^*,m_2,m_2^*,C_2) =&\; -\frac{i}{4\pi^2}C_2[\gamma^2(m_1,m_1^*)+\gamma^2(m_2,m_2^*)]\mu(m_1,m_1^*) \\ &  \times \mu(m_2,m_2^*) 
        [\Lambda_{\rm PDS}-\gamma(m_1,m_1^*)] \\
        & \times [\Lambda_{\rm PDS}-\gamma(m_2,m_2^*)] \, .
    \end{aligned}
\end{equation}
The integrals involved in $\Sigma_2$ and $\Sigma_3$ can be evaluated using a Fourier transform method described in Ref.~\cite{Braaten:1995cm}.  The $1/\epsilon$ pole that appears in $\Sigma_2$ is irrelevant to the $T_{cc}^+$ decay width because it will drop out when taking the derivative with respect to $E$. The reduced mass is defined as $\mu(m_1,m_2)\equiv m_1 m_2/(m_1+m_2)$ and the binding momenta are defined as $\gamma^2(m_1,m_2)=2\mu(m_1,m_2)(m_1+m_2-m_T)$. 

The combinations of these self-energy diagrams that we need in the expression for the $T_{cc}^+$ decay width are:
\begin{equation}
\begin{aligned}
    {\rm Re}\, {\rm tr}\,\Sigma^{\prime \rm LO} =&\; {\rm Re}\,\Sigma_1^\prime(m_0,m_+^*) + {\rm Re}\,\Sigma_1^\prime(m_+,m_0^*) \, ,  \\
    {\rm Re}\, \Sigma_0^{\prime \rm NLO} |_{C_2\rightarrow 0} =&\; {\rm Re} \bigg[
    \Sigma_2^\prime(m_+,m_0^*,m_+,m_0^*,m_{\pi^+},g/f_\pi,g/f_\pi)  \\
    & + \Sigma_2^\prime(m_0,m_+^*,m_0,m_+^*,m_{\pi^+},g/f_\pi,g/f_\pi)  \\
    & + \Sigma_2^\prime(m_+,m_0^*,m_0,m_+^*,m_{\pi^0},-g/\sqrt{2}f_\pi,g/\sqrt{2}f_\pi)  \\
    & + \Sigma_2^\prime(m_0,m_+^*,m_+,m_0^*,m_{\pi^0},g/\sqrt{2}f_\pi,-g/\sqrt{2}f_\pi) \bigg]
\end{aligned}
\end{equation}

The amplitudes $\mathcal{A}$ for the decay diagrams have a subscript that labels to which subdiagram of Fig.~\ref{DecayDiagrams} they correspond. If there is a single pion/charm meson vertex in a diagram, its coupling is labeled $g_\pi$, and if there are more than one such vertex, the couplings are indicated with a numeric subscript. The basis integrals are defined in App.~\ref{app: PDS scheme}. The parameters $c_1$ and $c_2$ are provided where appropriate, $b=1$ unless otherwise specified, and the momentum arguments for the integrals are $\pv$ unless otherwise specified. 
\begin{equation}
    \begin{aligned}
        i\mathcal{A}_{{\rm(\ref{figX1a})}}(\pv,m,m^*,g_\pi) =&\; \frac{2ig_\pi {\boldsymbol \epsilon}_T \cdot \pv_\pi \mu(m,m^*)}{\pv^2+\gamma^2(m,m^*)} \, . \\
    \end{aligned}
\end{equation}
\begin{equation}
    \begin{aligned}
        i\mathcal{A}_{{\rm(\ref{figX1b})}}(\pv,m,m_{\rm ext},m_\pi,m_1^*,m_2^*,g_1,g_2,g_3) =&\;  \frac{4i \mu(m,m_1^*)\mu(m_{\rm ext},m_2^*)}{\pv^2+\gamma^2(m_{\rm ext},m_2^*)}  \\
        & \times g_1g_2g_3 \bigg[ {\boldsymbol \epsilon}_T\cdot\pv \,\pv_\pi\cdot\pv \\
        & \times \big(I_0^{(2)}-2I^{(1)}+I\big)  \\
        &+{\boldsymbol \epsilon}_T\cdot\pv_\pi\pv^2 I_1^{(2)}\bigg] \, ,  \\
        c_1 =&\; \gamma^2(m,m_1^*)\, ,  \\
        c_2 =&\; \pv^2-(m_T-m-m_{\rm ext})^2+m_\pi^2 \, .  \\
    \end{aligned}
\end{equation}
\begin{equation}
    \begin{aligned}
        i\mathcal{A}_{{\rm(\ref{figX1d})}}(m,m_{\rm ext},m_\pi,m^*,g_\pi,C_\pi) =&\; 2i\mu(m,m^*)g_\pi C_\pi {\boldsymbol \epsilon}_T\cdot\pv[I^{(1)}-I] \, , \\
        c_1 =&\; \gamma^2(m,m^*)\, ,  \\
        c_2 =&\; \pv^2-(m_T - m-m_{\rm ext})^2+m_\pi^2 \, .  \\
    \end{aligned}
\end{equation}
\begin{equation}
    \begin{aligned}
        i\mathcal{A}_{{\rm(\ref{figX1e})}}(m,m_{\rm ext},m_1^*,m_2^*,g_\pi,C_2) =&\; \frac{1}{\pi}iC_2g_\pi{\boldsymbol \epsilon}_T\cdot\pv_\pi\mu(m,m_1^*)\mu(m_{\rm ext},m_2^*)  \\
        &\times \frac{\pv^2-\gamma^2(m,m_1^*)}{\pv^2+\gamma^2(m_{\rm ext},m_2^*)} [\gamma(m,m_1^*)-\Lambda_{\rm PDS}] \, . \\
    \end{aligned}
\end{equation}
\begin{equation}
    \begin{aligned}
        i\mathcal{A}_{{\rm(\ref{figX1f})}}(m,m^*,B_1) =&\; -\frac{iB_1}{2\pi} {\boldsymbol \epsilon}_T\cdot\pv_\pi \mu(m,m^*)[\gamma(m,m^*)-\Lambda_{\rm PDS}] \, . \\
    \end{aligned}
\end{equation}
\begin{equation}
    \begin{aligned}
        i\mathcal{A}_{{\rm(\ref{figX1g})}}(m_1,m_2,m^*,p_\pi^0,g_\pi,C_{0D}) =&\; 4i\mu(m_1,m_2)\mu(m_2,m^*) \\
        & \times g_\pi C_{0D}{\boldsymbol \epsilon}_T\cdot\pv_\pi I(\pv_\pi) \, , \\
        c_1 =&\; \gamma^2(m_2,m^*) \, ,  \\
        c_2 =&\; -2\mu(m_1,m_2)\bigg(m_T-m_1 \\
        & \; -m_2 - p_\pi^0 - \frac{\pv_\pi^2}{2m_1}\bigg) \, ,  \\
        b =&\; \frac{\mu(m_1,m_2)}{m_1} \, . 
    \end{aligned}
\end{equation}

Using the expressions defined therein, the differential decay widths for the two strong decays of $T_{cc}^+$ are below. The subscripts on $\mu$ and $\gamma$ indicate which charm meson is a pseudoscalar in that channel, e.g., $\mu_0 = \mu(m_0,m_+^*)$. 
\begin{equation}
\begin{aligned}
    & \frac{d\Gamma_0^{\rm NLO}(T_{cc}^+\rightarrow D^+D^0\pi^0)}{d\pv_0^2d\pv_+^2} \\
    =&\; \frac{2}{{\rm Re}\,{\rm tr} \, \Sigma^{\prime \rm LO}(-E_T)}{\rm Re}\, \bigg[ {\mathcal A}_{(\ref{figX1a})}(\pv_+,m_+,m_0^*,-g/\sqrt{2}f_\pi)  \\
    & \times \bigg( \mathcal{A}_{(\ref{figX1b})}(\pv_0,m_+,m_0,m_{\pi^0},m_0^*,m_+^*,-g/\sqrt{2}f_\pi,g/\sqrt{2}f_\pi,g/\sqrt{2}f_\pi)  \\
    & + {\mathcal A}_{(\ref{figX1b})}(\pv_+,m_+,m_+,m_{\pi^-},m_0^*,m_0^*,g/f_\pi,g/f_\pi,-g/\sqrt{2}f_\pi)  \\
    & - {\mathcal A}_{(\ref{figX1b})}(\pv_0,m_0,m_0,m_{\pi^+},m_+^*,m_+^*,g/f_\pi,g/f_\pi,g/\sqrt{2}f_\pi)  \\
    & - {\mathcal A}_{(\ref{figX1b})}(\pv_+,m_0,m_+,m_{\pi^0},m_+^*,m_0^*,g/\sqrt{2}f_\pi,-g/\sqrt{2}f_\pi,-g/\sqrt{2}f_\pi)  \\
    & + {\mathcal A}_{(\ref{figX1d})}(\pv_0,m_+,m_0,m_{\pi^0},m_0^*,-g/\sqrt{2}f_\pi,C_\pi^{(2)})  \\
    & - {\mathcal A}_{(\ref{figX1d})}(\pv_0,m_0,m_0,m_{\pi^+},m_+^*,g/f_\pi,C_\pi^{(1)})  \\
    & +{\mathcal A}_{(\ref{figX1g})}(m_0,m_+,m_0^*,-g/\sqrt{2}f_\pi,C_{0D}^{(1)})  \\
    & -{\mathcal A}_{(\ref{figX1g})}(m_+,m_0,m_+^*,g/\sqrt{2}f_\pi,C_{0D}^{(1)}) \bigg)^* + (D^0 \leftrightarrow D^+, \pi^+ \leftrightarrow \pi^-) \bigg]  \\
    & - \frac{1}{{\rm Re}\,{\rm tr} \, \Sigma^{\prime \rm LO}(-E_T)} \bigg[[\beta_1(\pv_+^2+\gamma_+^2)+\beta_2] \\ 
    & \times \big(\big|{\mathcal A}_{(\ref{figX1a})}(\pv_+,m_+,m_0^*,-g/\sqrt{2}f_\pi)\big|^2  \\
    & - {\mathcal A}_{(\ref{figX1a})}(\pv_0,m_0,m_+^*,g/\sqrt{2}f_\pi){\mathcal A}_{(\ref{figX1a})}^*(\pv_+,m_+,m_0^*,-g/\sqrt{2}f_\pi)\big)  \\
    &+[\beta_3(\pv_0^2+\gamma_0^2)+\beta_4]\big(\big|{\mathcal A}_{(\ref{figX1a})}(\pv_0,m_0,m_+^*,g/\sqrt{2}f_\pi)\big|^2  \\
    & - {\mathcal A}_{(\ref{figX1a})}(\pv_+,m_+,m_0^*,-g/\sqrt{2}f_\pi){\mathcal A}_{(\ref{figX1a})}^*(\pv_0,m_0,m_+^*,g/\sqrt{2}f_\pi)\big) \bigg]  \\ 
    & - \frac{d\Gamma_0^{\rm LO}(T_{cc}^+ \rightarrow D^+ D^0 \pi^0)}{d\pv_0^2d\pv_+^2} \frac{{\rm Re}\,\Sigma_0^{\prime \rm NLO}}{{\rm Re}\,{\rm tr} \, \Sigma^{\prime \rm LO}} \bigg|_{C_2 \rightarrow 0,E=-E_T} \; ,
\end{aligned}
\end{equation}
\begin{equation}
\begin{aligned}
    & \frac{d\Gamma_0^{\rm NLO}(T_{cc}^+\rightarrow D^0D^0\pi^+)}{d\pv_1^2d\pv_2^2} \\ =&\; \frac{1}{{\rm Re}\,{\rm tr} \, \Sigma^{\prime \rm LO}(-E_T)}{\rm Re}\, \bigg[ {\mathcal A}_{(\ref{figX1a})}(\pv_2,m_0,m_+^*,g/f_\pi)  \\
    & \times \bigg( \mathcal{A}_{(\ref{figX1b})}(\pv_1,m_0,m_0,m_{\pi^+},m_+^*,m_+^*,g/f_\pi,g/f_\pi,g/f_\pi)  \\
    & + {\mathcal A}_{(\ref{figX1b})}(\pv_2,m_0,m_0,m_{\pi^+},m_+^*,m_+^*,g/f_\pi,g/f_\pi,g/f_\pi)  \\
    & - {\mathcal A}_{(\ref{figX1b})}(\pv_1,m_+,m_0,m_{\pi^0},m_0^*,m_+^*,-g/\sqrt{2}f_\pi,g/\sqrt{2}f_\pi,g/f_\pi)  \\
    & - {\mathcal A}_{(\ref{figX1b})}(\pv_2,m_+,m_0,m_{\pi^0},m_0^*,m_+^*,-g/\sqrt{2}f_\pi,g/\sqrt{2}f_\pi,g/f_\pi)  \\
    & + {\mathcal A}_{(\ref{figX1d})}(\pv_1,m_0,m_0,m_{\pi^+},m_+^*,g/f_\pi,C_\pi^{(3)})  \\
    & - {\mathcal A}_{(\ref{figX1d})}(\pv_1,m_+,m_0,m_{\pi^0},m_0^*,-g/\sqrt{2}f_\pi,C_\pi^{(1)})  \\
    & +{\mathcal A}_{(\ref{figX1g})}(m_0,m_0,m_+^*,g/f_\pi,C_{0D}^{(1)}/2)\bigg)^* + (\pv_1 \leftrightarrow \pv_2)   \\
    & -\bigg(\frac{2g\mu_0}{f_\pi}\bigg)^2 \frac{\pv_\pi^2}{3} \beta_5\bigg(\frac{1}{\pv_1^2+\gamma_0^2}+\frac{1}{\pv_2^2+\gamma_0^2}\bigg)\bigg]  \\
    & - \frac{d\Gamma_0^{\rm LO}(T_{cc}^+ \rightarrow D^0 D^0 \pi^+)}{d\pv_1^2d\pv_2^2}\bigg(\beta_4+\frac{{\rm Re}\,\Sigma_0^{\prime \rm NLO}}{{\rm Re}\,{\rm tr} \,\Sigma^{\prime \rm LO}} \bigg|_{C_2 \rightarrow 0,E=-E_T}\bigg) \; .
\end{aligned}
\end{equation}
The parameters $\beta_i$ depend on the $C_2$ and the $B_1$, and are defined in App.~\ref{app: beta expressions}.

\section{$\beta_i$ expressions} \label{app: beta expressions}

Here we give the expressions for the $\beta_i$ parameters.
\begin{equation}
\begin{aligned}
    \beta_1 =&\; (\Lambda_{\rm PDS}-\gamma_+ )\bigg(\frac{f_\pi}{\sqrt{2}\pi g}B_1^{(1)}+\frac{1}{\pi}C_2^{(+)}\mu_+-\frac{1}{\pi}C_2^{(-)}\mu_0 \frac{\Lambda_{\rm PDS}-\gamma_0}{\Lambda_{\rm PDS}-\gamma_+} \bigg) \, , \\
\end{aligned}
\end{equation}
\begin{equation}
\begin{aligned}
    \beta_2 =&\; \bigg[\frac{1}{\pi}C_2^{(+)}\mu_+(-2\gamma_+^2)(\Lambda_{\rm PDS}-\gamma_+)-\frac{1}{\pi}C_2^{(-)}\mu_0(-\gamma_0^2-\gamma_+^2)(\Lambda_{\rm PDS}-\gamma_0)  \\
    & +2\pi\bigg(\frac{\mu_0^2}{\gamma_0}+\frac{\mu_+^2}{\gamma_+}\bigg)^{-1}\bigg[-\frac{1}{\pi^2}C_2^{(+)}\mu_+^3(\gamma_+-\Lambda_{\rm PDS})(2\gamma_+-\Lambda_{\rm PDS})  \\
    & -\frac{1}{\pi^2}C_2^{(+)}\mu_0^3(\gamma_0-\Lambda_{\rm PDS})(2\gamma_0-\Lambda_{\rm PDS})  \\
    & -\frac{C_2^{(-)}(\gamma_+^2+\gamma_0^2)\mu_+\mu_0}{2\pi}\bigg(\frac{\mu_+}{\gamma_0}(\Lambda_{\rm PDS}-\gamma_0)+\frac{\mu_0}{\gamma_+}(\Lambda_{\rm PDS}-\gamma_+)\bigg)  \\
    & +\frac{C_2^{(-)}\mu_+\mu_0(\mu_++\mu_0)}{\pi^2}(\Lambda_{\rm PDS}-\gamma_+)(\Lambda_{\rm PDS}-\gamma_0)\bigg]\bigg] \, , \\
\end{aligned}
\end{equation}
\begin{equation}
\begin{aligned}
    \beta_3 =&\; (\Lambda_{\rm PDS}-\gamma_0)\bigg(-\frac{f_\pi}{\sqrt{2}\pi g}B_1^{(2)}+\frac{1}{\pi}C_2^{(+)}\mu_0-\frac{1}{\pi}C_2^{(-)}\mu_+ \frac{\Lambda_{\rm PDS}-\gamma_+}{\Lambda_{\rm PDS}-\gamma_0} \bigg) \, , \\
\end{aligned}
\end{equation}
\begin{equation}
\begin{aligned}
    \beta_4 =&\; \bigg[\frac{1}{\pi}C_2^{(+)}\mu_0(-2\gamma_0^2)(\Lambda_{\rm PDS}-\gamma_0)-\frac{1}{\pi}C_2^{(-)}\mu_+(-\gamma_0^2-\gamma_+^2)(\Lambda_{\rm PDS}-\gamma_+)  \\
    & +2\pi\bigg(\frac{\mu_0^2}{\gamma_0}+\frac{\mu_+^2}{\gamma_+}\bigg)^{-1}\bigg[-\frac{1}{\pi^2}C_2^{(+)}\mu_+^3(\gamma_+-\Lambda_{\rm PDS})(2\gamma_+-\Lambda_{\rm PDS})  \\
    & -\frac{1}{\pi^2}C_2^{(+)}\mu_0^3(\gamma_0-\Lambda_{\rm PDS})(2\gamma_0-\Lambda_{\rm PDS})  \\
    &-\frac{C_2^{(-)}(\gamma_+^2+\gamma_0^2)\mu_+\mu_0}{2\pi}\bigg(\frac{\mu_+}{\gamma_0}(\Lambda_{\rm PDS}-\gamma_0)+\frac{\mu_0}{\gamma_+}(\Lambda_{\rm PDS}-\gamma_+)\bigg)  \\
    & +\frac{C_2^{(-)}\mu_+\mu_0(\mu_++\mu_0)}{\pi^2}(\Lambda_{\rm PDS}-\gamma_+)(\Lambda_{\rm PDS}-\gamma_0)\bigg]\bigg] \, , \\
\end{aligned}
\end{equation}
\begin{equation}
\begin{aligned}
    \beta_5 =&\; \frac{1}{\pi}C_2^{(+)}\mu_0(\Lambda_{\rm PDS}-\gamma_0)-\frac{1}{\pi}C_2^{(-)}\mu_+(\Lambda_{\rm PDS}-\gamma_+)  \\
    & +\frac{B_1^{(3)}f_\pi}{4\pi g}(\gamma_0-\Lambda_{\rm PDS})-\frac{B_1^{(4)}f_\pi}{4\pi g}(\gamma_+-\Lambda_{\rm PDS})\frac{\mu_+}{\mu_0} \, . 
\end{aligned}
\end{equation}
In the isospin limit $m_+ = m_0$ these become:
\begin{equation}
\begin{aligned}
    \beta_1 = \beta_3=\beta_5 = & \; \frac{1}{\pi}(\gamma-\Lambda_{\rm PDS})\bigg(\frac{b_0f_\pi}{g}-2C_2^{(0)}\mu\bigg) \, ,  \\
    \beta_2 = \beta_4 = & \; -\frac{4C_2^{(0)}\mu\gamma}{\pi}(\gamma-\Lambda_{\rm PDS})^2 \, .
\end{aligned}
\label{eq: betas in isospin limit}
\end{equation}
Since we are dropping isospin-breaking interactions in this limit, the isospin-$1$ couplings drop out. Equation (\ref{eq: betas in isospin limit}) is consistent with the dependence on $\Lambda_{\rm PDS}$ in XEFT \cite{Fleming:2007rp}. 
    \chapter{Supplementary material: $J/\psi$ production}

\section{NRQCD matching}
\label{app: NRQCD matching}

The content of this appendix follows the results of a paper by Braaten and Chen \cite{Braaten:1996jt}.

Define the momentum of the $c$ and $\bar{c}$ to be $p=\frac{P}{2}+q$ and $\bar{p} = \frac{P}{2} - q$, respectively, where $q^\mu = {\Lambda^\mu}_j q^j$ is a boosted relative three-momentum.  The boost matrix is: \cite{Braaten:1996jt}
\begin{equation}
    \begin{aligned}
        {\Lambda^0}_j = & \; \frac{1}{2E_q}P_j \; , \\
        {\Lambda^i}_j = & \; {\delta^i}_j - \frac{P^iP_j}{\Pv^2} + \frac{P^0}{2E_q} \frac{P^iP_j}{\Pv^2} \; .
    \end{aligned}
\end{equation}
These satisfy the relations:
\begin{equation}
    \begin{aligned}
        P_\mu {L^\mu}_j = & \; 0 \; , \\
        g_{\mu\nu}{\Lambda^\mu}_i {\Lambda^\nu}_j = &\; -\delta_{ij} \; , \\
        {\Lambda^\mu}_i{\Lambda^\nu}_j = & \; -g^{\mu\nu} + \frac{P^\mu P^\nu}{P^2} \; .
    \end{aligned}
\end{equation}

The nonrelativistic expansions of the QCD spinors, to linear order in $q$, are given by:
\begin{equation}
    \begin{aligned}
        \bar{u}(p)v(\bar{p}) = & \; -2 \xi^\dagger (\qv \cdot \boldsymbol{\sigma})\eta \; , \\
        \bar{u}(p)\gamma^\mu v(\bar{p}) = & \; M {\Lambda^\mu}_j \xi^\dagger \sigma^j \eta \; , \\
        \bar{u}(p)\frac{i}{2}[\gamma^\mu,\gamma^\nu]v(\bar{p}) = & \; i(P^\mu {\Lambda^\nu}_j - P^\nu{\Lambda^\mu}_j)\xi^\dagger \sigma^j \eta - 2{\Lambda^\mu}_j{\Lambda^\nu}_k \epsilon^{jkl}q^l \xi^\dagger \eta \; , \\
        \bar{u}(p)\gamma^\mu \gamma_5 v(\bar{p}) = & \; P^\mu \xi^\dagger \eta - 2i{\Lambda^\mu}_j \xi^\dagger (\qv \times \boldsymbol{\sigma}^j) \eta \; , \\
        \bar{u}(p)\gamma_5 v(\bar{p}) = & \; M \xi^\dagger \eta \; .
    \end{aligned}
\end{equation}
These two-spinors then match onto vacuum matrix elements of NRQCD operators: \cite{Braaten:1996jt}
\begin{equation}
    \begin{aligned}
        M^2 \eta^{\prime\dagger} \sigma_i \xi^\prime \xi^\dagger \sigma_j \eta &\leftrightarrow \braket{\chi^\dagger \sigma_i \psi \, \mathcal{P}_{J/\psi (\lambda)} \, \psi^\dagger \sigma_j  \chi}  \, , \\
        M^2 q^\prime_m q_n \eta^{\prime\dagger} \sigma_i T^a \xi^\prime \xi^\dagger \sigma_j T^a \eta &\leftrightarrow \braket{\chi^\dagger \sigma_i \big(-\frac{i}{2}\overleftrightarrow{{\bf D}}_m \big)T^a \psi \, \mathcal{P}_{J/\psi (\lambda)} \, \psi^\dagger \sigma_j \big(-\frac{i}{2}\overleftrightarrow{{\bf D}}_n \big) T^a \chi}  \, , \\
        M^2 \eta^{\prime\dagger} \sigma_i T^a \xi^\prime \xi^\dagger \sigma_j T^a \eta &\leftrightarrow \braket{\chi^\dagger \sigma_i T^a \psi \, \mathcal{P}_{J/\psi (\lambda)} \, \psi^\dagger \sigma_j T^a \chi}  \, , \\
        M^2 \eta^{\prime\dagger} T^a \xi^\prime \xi^\dagger T^a \eta &\leftrightarrow \braket{\chi^\dagger T^a \psi \, \mathcal{P}_{J/\psi (\lambda)} \, \psi^\dagger  T^a \chi}  \, , \\
    \end{aligned}
\end{equation}
which in turn can be decomposed into the $J/\psi$ polarization vectors and NRQCD LDMEs using rotational symmetry and tensor analysis:
\begin{equation}
    \begin{aligned}
        \braket{\chi^\dagger \sigma_i \psi \, \mathcal{P}_{J/\psi (\lambda)} \, \psi^\dagger \sigma_j  \chi} &= \frac{2M}{3} \epsilon^*_i \epsilon_{j}  \braket{\mathcal{O}^{J/\psi}(^3S_1^{[1]})} \, , \\
        \braket{\chi^\dagger \sigma_i \big(-\frac{i}{2}\overleftrightarrow{{\bf D}}_m \big)T^a \psi \, \mathcal{P}_{J/\psi (\lambda)} \, \psi^\dagger \sigma_j \big(-\frac{i}{2}\overleftrightarrow{{\bf D}}_n \big) T^a \chi} &= 2M \epsilon^*_i \epsilon_{j}  \delta_{mn} \braket{\mathcal{O}^{J/\psi}(^3P_0^{[8]})} \, , \\
        \braket{\chi^\dagger \sigma_i T^a \psi \, \mathcal{P}_{J/\psi (\lambda)} \, \psi^\dagger \sigma_j T^a \chi} &= \frac{2M}{3} \epsilon^*_i \epsilon_{j}  \braket{\mathcal{O}^{J/\psi}(^3S_1^{[8]})} \, , \\
        \braket{\chi^\dagger T^a \psi \, \mathcal{P}_{J/\psi (\lambda)} \, \psi^\dagger  T^a \chi} &= 2M \braket{\mathcal{O}^{J/\psi}(^1S_0^{[8]})} \, . \\
    \end{aligned}
\end{equation}
%

\section{Wilson lines}
\label{app: wilson lines}

To present an introduction to Wilson lines, we will briefly turn to the Abelian case of QED.  Consider an operator with two Dirac fields at different spacetime points: $\bar{\psi}(x) \psi(y)$.  This is not invariant under local gauge transformations $\psi(z) \rightarrow V(z) \psi(z)$, where $V(z) = \exp[ig_{\rm em}\theta(z)]$, since $V^\dagger(x)V(y) \neq 1$.  However, if we could construct an object $W(x,y)$ which transforms as $W\rightarrow V(x)W(x,y)V^\dagger(y)$ and insert it into the product as $\bar{\psi}(x)W(x,y)\psi(y)$, then that object would be gauge invariant.  The gauge field transforms as $A_\mu(z) \rightarrow A_\mu(z) + \partial_\mu\theta(z)$.  Consider, then, the ansatz
\begin{equation}
    W(x,y) = \exp \bigg[ig_{\rm em} \int_P dz^\mu \; A_\mu (z) \bigg] \; ,
\end{equation}
where $P$ is a path connecting $y$ and $x$.  Changing from $z$ to a variable that $s$ that parametrizes the path, we get
\begin{equation}
    W(x,y) = \exp \bigg[ig_{\rm em} \int_{s_y}^{s_x} ds \; \frac{dz^\mu}{ds} A_\mu\big(z(s)\big) \bigg] \; ,
\label{eq: QED wilson line}
\end{equation}
where the bounds of the integration domain have been transformed according to $y=z(s_y)$ and $x=z(s_x)$.  Under the local gauge transformation, this becomes
\begin{equation}
\begin{aligned}
    W(x,y) \rightarrow  & \; \exp \bigg[ig_{\rm em} \int_{s_y}^{s_x} ds \; \frac{dz^\mu}{ds} A_\mu\big(z(s)\big) + ig_{\rm em} \int_{s_y}^{s_x} ds \; \frac{dz^\mu}{ds} \partial_\mu\theta\big(z(s)\big) \bigg] \\
    = & \; \exp \bigg[ig_{\rm em} \int_{s_y}^{s_x} ds \; \frac{dz^\mu}{ds} A_\mu\big(z(s)\big) + ig_{\rm em} \int_{s_y}^{s_x} ds \; \frac{d}{ds} \theta\big(z(s)\big) \bigg] \\
    = & \; \exp \bigg[ig_{\rm em} \int_{s_y}^{s_x} ds \; \frac{dz^\mu}{ds} A_\mu\big(z(s)\big) + ig_{\rm em} \theta(x) - ig_{\rm em}\theta(y) \bigg] \\
    = & \; V(x) W(x,y) V^\dagger(y) \; .
\end{aligned}
\end{equation}
Therefore this Wilson line $W(x,y)$ has the desired transformation property to let the operator be gauge invariant.

The half-staple-shaped Wilson lines in Eq.~(\ref{eq: q TMDFF}) are products of two Wilson lines along straight line segments: 
\begin{equation}
\begin{aligned}
    W{\scalebox{1.7}{$\lrcorner$}} = &\; W_{\hat{b}_T}(+\infty n;b_T,+\infty) W_{n}(b;0,+\infty) \\
    W_{\scalebox{1.7}{$\urcorner$}} = &\; W_{n}(0;+\infty,0) W_{\hat{b}_T}(+\infty n;+\infty,0) \;.
\end{aligned}
\end{equation}
Here, the notation convention used is
\begin{equation}
    W_v (x^\mu; a, b) =  {\cal P} \, {\rm exp}\bigg[i g_s \int_a^b ds~ v\cdot A^c (x + s v) t^c\bigg].
\label{eq: TMDFF wilson line}
\end{equation}
This is a Wilson line for a field that starts at $x$ and travels along the direction specified by $v$.  The path ordering $\mathcal{P}$ is needed because we have returned to the non-Abelian case where the gauge fields are gluons.  The gauge transformation property of the Wilson line holds, so the Wilson line in Eq.~(\ref{eq: TMDFF wilson line}) is analogous to the QED Wilson line in Eq.~(\ref{eq: QED wilson line}) with $z^\mu(s) = x^\mu + sv^\mu$.  We find
\begin{equation}
\begin{aligned}
    W_v (x^\mu; a, b) \rightarrow & \; V\big(z(b)\big) W_v (x^\mu; a, b) V^\dagger\big(z(a)\big) \\
    = &\; V(x+bv) W_v (x^\mu; a, b) V^\dagger(x+av) \; .
\end{aligned}
\end{equation}
The half-staple-shaped Wilson lines, together with the light quark fields, then transform as:
\begin{equation}
\begin{aligned}
    W{\scalebox{1.7}{$\lrcorner$}} \psi(b) \rightarrow & \; V(+\infty n + \infty \hat{b}_T) W_{\hat{b}_T}(+\infty n;b_T,+\infty) V^\dagger({\bf b}_T + \infty n) \\
    & \; \times V(b+\infty n)  W_{n}(b;0,+\infty) V^\dagger(b) V(b) \psi(b) \\
   = & \; V(+\infty n + \infty \hat{b}_T) W{\scalebox{1.7}{$\lrcorner$}} \psi(b) \\
    \bar{\psi} (0) W_{\scalebox{1.7}{$\urcorner$}} \rightarrow & \; \bar{\psi}(0)V^\dagger(0) 
    V(0) W_{n}(0;+\infty,0) V^\dagger(+\infty n) \\
     & \; \times V(+\infty n) W_{\hat{b}_T}(+\infty n;+\infty,0) V^\dagger(+\infty n + \infty \hat{b}_T) \\
    = & \; \bar{\psi}(0)W_{\scalebox{1.7}{$\urcorner$}}  V^\dagger(+\infty n + \infty \hat{b}_T)
\end{aligned}
\end{equation}
The combination of these is then gauge invariant.

\section{Interpretation of J/$\psi$ polarization parameters}
\label{app: pol par interpretation}

This analysis of the polarization parameters for a spin-1 particle follows from Appendix A of a paper by Bacchetta and Mulders \cite{Bacchetta:2000jk}.

The general construction of the density matrix of a spin-1 particle is written in terms of the three-dimensional generalization of the Pauli matrices, $\Sigma^i$.  
\begin{equation}
    \rho = \frac{1}{3}\bigg(\hat{1} + \frac{3}{2}{\bf S}^i \Sigma^i + 3 T^{ij}\Sigma^{ij}\bigg) \; .
\end{equation}
The tensor $\Sigma^{ij}$ is symmetric and has zero trace:
\begin{equation}
    \Sigma^{ij} = -\frac{2}{3}\hat{1}\delta^{ij} + \frac{1}{2}\Sigma^{\{i}\Sigma^{j\}} \; .
\end{equation}
The coefficients of the operators are the spin vector ${\bf S}$ and the spin tensor $T^{ij}$.  In the rest frame of the hadron, Ref.~\cite{Bacchetta:2000jk} parametrizes them as:
\begin{equation}
\begin{aligned}
    {\bf S}= & \;  (S^x_T, S_T^y, S_L) \; , \\
    T_{ij} = & \;
\begin{pmatrix}
-\frac23 S_{LL} + S_{TT}^{xx} & S_{TT}^{xy} & S_{LT}^x\\
S_{TT}^{yx} & -\frac23 S_{LL} - S_{TT}^{xx} & S_{LT}^y\\
S_{LT}^x & S_{LT}^y & \frac43 S_{LL}
\end{pmatrix} \; .
\end{aligned}
\end{equation}
The spin vector operator is:
\begin{equation}
    \boldsymbol{\Sigma}\cdot \hat{\bf n} = \Sigma^x \cos{\theta}\cos{\varphi} + \Sigma^y \cos{\theta} \sin{\varphi} + \Sigma^z\sin{\theta} \; ,
\end{equation}
for polar and azimuthal angles $\theta$ and $\varphi$. Let $\ket{m_{(\theta,\varphi)}}$ be the eigenstates of this operator with magnetic quantum number $m$.  The probability associated with one of these states is 
\begin{equation}
    P(m_{(\theta,\varphi)}) = {\rm Tr}\; \rho \ket{m_{(\theta,\varphi)}} \bra{m_{(\theta,\varphi)}} \; .
\end{equation}
It follows then that the tensor polarization parameters can be written in terms of probabilities:
\begin{equation}
    \begin{aligned}
        S_{LL} = & \; \frac{1}{2}P(1_{(0,0)}) + \frac{1}{2}P(-1_{(0,0)}) - P(0_{(0,0)}) \; , \\
        S_{LT}^x = & \; P(0_{(-\frac{\pi}{4},0)}) - P(0_{(\frac{\pi}{4},0)}) \; , \\
        S_{LT}^y = & \; P(0_{(-\frac{\pi}{4},\frac{\pi}{2})}) - P(0_{(\frac{\pi}{4},\frac{\pi}{2})}) \; , \\
        S_{LT}^{xx} = & \; P(0_{(\frac{\pi}{2},-\frac{\pi}{4})}) - P(0_{(\frac{\pi}{2},\frac{\pi}{4})}) \; , \\
        S_{LT}^{xy} = & \; P(0_{(\frac{\pi}{2},\frac{\pi}{2})}) - P(0_{(\frac{\pi}{2},0)}) \; .
    \end{aligned}
\end{equation}
This puts constraints on the ranges for the parameters.
\begin{equation}
    \begin{aligned}
        -1 & \le S_{LL} \le \frac{1}{2} \; , \\
        -1 & \le S_{LT}^i, S_{TT}^{ij}  \le 1 \; . \\
    \end{aligned}
\end{equation}

\section{Projection operators for TMDFFs}
\label{app: projection operators}

Below are listed the projection operators for the quark TMDFFs \cite{Bacchetta:2000jk,Boer:2016xqr}. The notation $\Delta_{\rm pol}^{[\Gamma]}$ is the part of the quark FF with parton polarization projector $\Gamma$, proportional to the $J/\psi$ polarization parameter $S_{\rm pol}$.
\begin{equation}
    \begin{aligned}
        \Delta_U^{[\gamma^+]}(x,\kv_T) = & \; D_1 \; , \\
        \Delta_L^{[\gamma^+]}(x,\kv_T) = & \; 0 \; , \\
        \Delta_T^{[\gamma^+]}(x,\kv_T) = & \; \frac{1}{M}\epsilon_T^{\mu\nu}S_{T\; \nu} k_{T\;\mu} D_{1T}^\perp \; , \\
        \Delta_{LL}^{[\gamma^+]}(x,\kv_T) = & \; S_{LL} D_{1LL} \; , \\
        \Delta_{LT}^{[\gamma^+]}(x,\kv_T) = & \; \frac{1}{M}{\bf S}_{LT} \cdot \kv_T D_{1LT} \; , \\
        \Delta_{TT}^{[\gamma^+]}(x,\kv_T) = & \; \frac{1}{M^2} \kv_T \cdot {\bf S}_{TT} \cdot \kv_T D_{1TT} \; ,
    \end{aligned}
\label{eq: U quark FFs}
\end{equation}
\begin{equation}
    \begin{aligned}
        \Delta_U^{[\gamma^+ \gamma_5]}(x,\kv_T) = & \; 0 \; , \\
        \Delta_L^{[\gamma^+ \gamma_5]}(x,\kv_T) = & \; S_L G_{1L} \; , \\
        \Delta_T^{[\gamma^+ \gamma_5]}(x,\kv_T) = & \; \frac{1}{M} {\bf S}_T \cdot \kv_T G_{1T} \; , \\
        \Delta_{LL}^{[\gamma^+ \gamma_5]}(x,\kv_T) = & \;  0 \; , \\
        \Delta_{LT}^{[\gamma^+ \gamma_5]}(x,\kv_T) = & \; \frac{1}{M}\epsilon_T^{\mu \nu} S_{LT \; \nu} k_{T \; \mu} G_{1LT} \; , \\
        \Delta_{TT}^{[\gamma^+ \gamma_5]}(x,\kv_T) = & -\frac{1}{M^2}\epsilon_T^{\mu\nu} S_{TT \; \nu \rho} k_T^\rho k_{T\; \mu} G_{1TT} \; ,
    \end{aligned}
\label{eq: L quark FFs}
\end{equation}
\begin{equation}
    \begin{aligned}
        \Delta_U^{[i\sigma^{i+}\gamma_5]}(x,\kv_T) = & \; \frac{1}{M} \epsilon_T^{ij} \kv_{T \; j}H_1^\perp \; , \\
        \Delta_L^{[i\sigma^{i+}\gamma_5]}(x,\kv_T) = & \; \frac{1}{M}S_L \kv_T^i H_{1L}^\perp \; , \\
        \Delta_T^{[i\sigma^{i+}\gamma_5]}(x,\kv_T) = & \; {\bf S}_T^i H_{1T} + \frac{1}{M^2} {\bf S}_T\cdot \kv_T \kv_T^i H_{1T}^\perp \; , \\
        \Delta_{LL}^{[i\sigma^{i+}\gamma_5]}(x,\kv_T) = & \frac{1}{M}S_{LL} \epsilon_T^{ij}\kv_{T \; j} H_{1LL}^\perp \; , \\
        \Delta_{LT}^{[i\sigma^{i+}\gamma_5]}(x,\kv_T) = & \epsilon_T^{ij} {\bf S}_{LT \; j}H_{1LT}^\prime + \frac{1}{M^2} {\bf S}_{LT} \cdot \kv_T \epsilon_T^{ij}\kv_{T \; j} H_{1LT}^\perp \; , \\
        \Delta_{TT}^{[i\sigma^{i+}\gamma_5]}(x,\kv_T) = & \frac{1}{M} \epsilon_T^{ij}S_{TT \; jl} \kv_T^l H_{1TT}^\prime + \frac{1}{M^3} \kv_T \cdot {\bf S}_{TT} \cdot \kv_T \epsilon_T^{ij} \kv_{T \; j} H_{1TT}^\perp \; .
    \end{aligned}
\label{eq: T quark FFs}
\end{equation}
The gluon FF $\Delta_{g\rightarrow J/\psi}^{\alpha \alpha'}$ has no $\kv_T$ dependence at leading order in $\alpha_s$, so the only non-vanishing FFs at this order are:
\begin{equation}
    \begin{aligned}
        \Delta_U^{\alpha \beta} = & \; - \frac{1}{2}g_T^{\alpha\beta} D_1^g \; , \\
        \Delta_L^{\alpha \beta} = & \; \frac{i}{2}\epsilon_T^{\alpha \beta} S_L G_1^g \; , \\
        \Delta_T^{\alpha \beta} = & \; 0 \; , \\
        \Delta_{LL}^{\alpha \beta} = & \; - \frac{1}{2}g_T^{\alpha\beta} S_{LL} D_{1LL}^g \; , \\
        \Delta_{LT}^{\alpha \beta} = & \; 0 \; , \\
        \Delta_{TT}^{\alpha \beta} = & \; \frac{1}{2}S_{TT}^{\alpha \beta}H_{1TT}^g \; . \\
    \end{aligned}
\end{equation}
\end{appendices}

\cleardoublepage
\normalbaselines 
\addcontentsline{toc}{chapter}{Bibliography} 
\printbibliography

\biography
Reed Hodges graduated with a major in physics and minor in mathematics in May 2018 from Georgia Southern University.  While an undergraduate, he worked with his advisor Dr.~Maxim Durach on projects related to the photonic properties of metamaterials.  He began graduate studies at Duke University in August 2018, where he and his advisor Dr.~Thomas Mehen and their collaborators completed projects applying effective field theories to various heavy quark systems.

\end{document}